\documentclass[suppldata]{interact}

\usepackage{epstopdf}
\usepackage{graphicx}
\usepackage{framed}
\usepackage[normalem]{ulem}
\usepackage{amsmath}
\usepackage{amsthm}
\usepackage{amssymb}
\usepackage{mathtools}
\usepackage{multirow}
\usepackage{amsfonts}
\usepackage{hyperref}
\usepackage{isotope}
\usepackage{caption}
\usepackage{subcaption}
\usepackage{comment}
\usepackage[numbers,sort&compress]{natbib}
\usepackage{color}                  
\usepackage{lscape}
\usepackage{multicol}
\usepackage{fancyhdr}
\usepackage{apxproof}

\DeclarePairedDelimiter\ket{\lvert}{\rangle}
\DeclarePairedDelimiterX\braket[2]{\langle}{\rangle}{#1 \delimsize\vert #2}

\usepackage{longtable}
\usepackage{datetime}

\theoremstyle{plain}

\theoremstyle{definition}

\theoremstyle{remark}

\definecolor{purple}{rgb}{0.6,0,1}
\definecolor{orange}{rgb}{0.9,0.32,0}
\definecolor{cyan}{rgb}{0.1,0.63,0.9}
\definecolor{pink}{rgb}{0.9,0.45,0.85}
\definecolor{crimson}{rgb}{0.64,0,0.14}
\definecolor{green}{rgb}{0,0.6,0}
\definecolor{grey}{rgb}{0.75,0.75,0.75}

\begin{document}

\title{QEYSSat~2.0 - White Paper on Satellite-based Quantum Communication Missions in Canada}

\author{
\name{Thomas Jennewein\textsuperscript{a}\thanks{$\dagger$thomas.jennewein@uwaterloo.ca}$\dagger$, Christoph Simon\textsuperscript{b}, Andr\'{e} Foug\`{e}res\textsuperscript{c}, Fran\c{c}ois Babin\textsuperscript{c}, Faezeh Kimiaee Asadi\textsuperscript{b}, Katanya B. Kuntz\textsuperscript{a}, Mathieu Maisonneuve\textsuperscript{c}, Brian Moffat\textsuperscript{a}, Kimia Mohammadi\textsuperscript{a}, and Denis Panneton\textsuperscript{c}}
\affil{\textsuperscript{a}Institute for Quantum Computing, Department of Physics and Astronomy, University of Waterloo, Ontario, Canada N2L 3G1; \\ \textsuperscript{b}Institute for Quantum Science and Technology, Department of Physics and Astronomy, University of Calgary, Alberta, Canada T2N 1N4; \\ \textsuperscript{c}Institut National d'Optique, 2740 rue Einstein, Qu\'{e}bec, Canada G1P 4S4}
}

\maketitle

\begin{abstract}
We present the white paper developed during the QEYSSat~2.0 study, which was undertaken between June 2021 and March 2022. The study objective was to establish a technology road-map for a Canada-wide quantum network enabled by satellites. We survey the state-of-art in quantum communication technologies, identify the main applications and architectures, review the technical readiness levels and technology bottlenecks and identify a future mission scenario. We report  the findings of a dedicated one-day workshop that included Canadian stakeholders from government, industry and academia to gather inputs and insights for the applications and technical road-map. We also provide an overview of the Quantum EncrYption and Science Satellite (QEYSSat) mission expected to launch in 2024-2025 and its anticipated outcomes. One of the main outcomes of this study is that developing the main elements for a Canada-wide quantum internet will have the highest level of impact, which includes Canada-wide entanglement distribution and teleportation.  We present and analyze a  possible future mission  (`QEYSSat~2.0') that would enable long range quantum teleportation across Canada as an important step towards this vision.  
\end{abstract}


\newpage

\tableofcontents



\newpage

\section{Executive Summary}
 The QEYSSat~2.0 study was contracted by the National Research Council of Canada (NRC) together with Defence Research and Development Canada (DRDC) with the objective to identify future opportunities and directions for a Canada-wide quantum network enabled by satellites. The duration of this study was less than a year, kicking off in June 2021, and concluding in March 2022. The contractors were the University of Waterloo (lead), the University of Calgary, and the Institut National d’Optique. The main deliverable of the QEYSSat~2.0 study was a White Paper that provides an overview on the state of art in quantum technologies, identifies the opportunity for quantum networking, and provides the vision for future missions and technology. This article contains the bulk of the white paper that was delivered on 31 March 2022,  with  minor changes including updated references and the correction of typos. Some content was omitted as noted.

A virtual workshop was held on 15 February 2022, where a total of 91 attendees from Government, Academia and Industry, came together to discuss the quantum internet and possible future missions. A total of 71 colleagues participated in the applications round table, 61 colleagues in the architectures round table, and 54 colleagues in the technologies round table.

Canadian researchers have historically had a strong leadership role in the development of quantum communications, such as the first quantum key distribution (QKD) protocol developed in 1984, the invention of quantum teleportation in 1993, and blind quantum computing in the early 2000s. And roughly 30 years later since the first protocol was shown, Canada is working on its very own quantum communication satellite mission called QEYSSat, which will advance the science and technology around establishing quantum links between ground and space. The knowledge and science learnt in QEYSSat will provide  valuable insights and baselines for future operational quantum communication satellite that may be developed.

\begin{figure*}[ht]
\centering\includegraphics[width=0.7\linewidth]{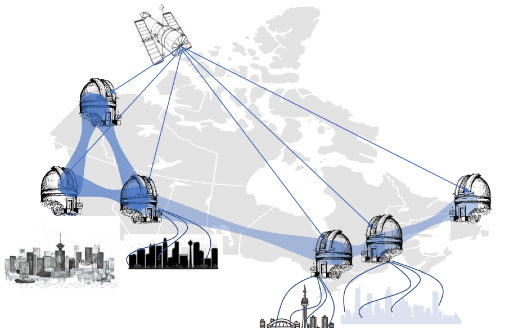}
\caption{QEYSSat 2.0 mission: Teleportation Across Canada. \textit{A mari usque ad mare}.} (Image by Khabat Heshami)\label{fig:QEYSSat2.0}
\end{figure*}

However, while the QEYSSat mission will be an important stepping stone towards making this technology viable, it is intended to be only the first quantum communication mission. This QEYSSat~2.0 project was initiated in order to establish a road-map of objectives and scopes of future quantum communication mission(s). As the main outcome of this study, it is found that developing the main elements for a Canada-wide quantum internet will have the highest level of impact in the intermediate term. This includes Canada-wide entanglement distribution and teleportation. 

Drawing a parallel with the today's (classical) internet, the end goal of the quantum internet is to connect two remote end-users (Alice and Bob) via their respective network access points, where the  fundamental resource of the quantum internet is the entanglement of quantum states of qubits. 


As the distance between end-users strongly affects the performance of the network link due to channel losses, classical networks use a series of distributed amplifiers and repeaters to restore the signal along the optical fibre path. For quantum systems, amplification is not possible due to the `no-cloning' theorem. Therefore so-called quantum repeaters based on entanglement swapping between a series of nodes are required to create a long-distance quantum entanglement link between the two end users' Alice and Bob's access point. The entanglement swapping operation establishes  entanglement of one node with the next. Once the full entanglement link is completed, Alice will let her qubit state coherently interact with the entangled state at her access point, and effectively teleport the qubit state to Bob. Fiber-based quantum repeaters can be used over moderate distances (i.e., 500–2,000 \:km). On the other hand, long-reach systems (i.e., those with a range of more than 2,000 \:km) will likely have to make use of free-space links via satellites, either alone or in conjunction with repeater nodes. 

Due to Canada's  large size, it is critical to develop the required satellite quantum technology for long-range quantum communication within the country.  Canada's  strong capability in academic research, as well as related industry and startups, means it has the opportunity to build Canadian quantum technology solutions, rather than depend on overseas vendors, and become an international leader in this technology, rather than playing catch-up with other nations.

\subsection{The Quantum Internet across Canada}
\begin{figure}[ht]
    \centering
    \includegraphics[width=1\textwidth]{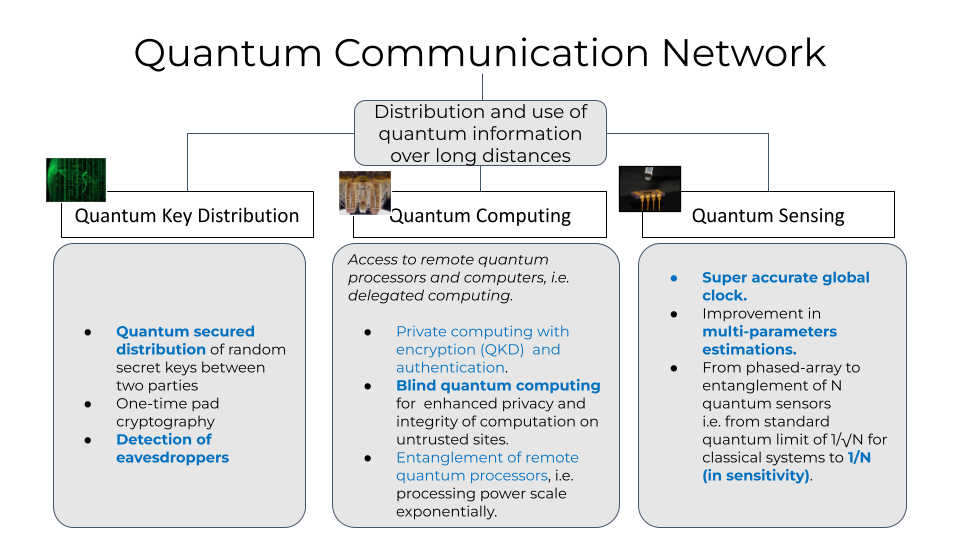}
    \caption{Outline of the use-cases for the `Quantum Internet.'}
    \label{fig:quantum_network_use_cases}
\end{figure}

The Quantum Internet will be a network that allows for the seamless transmission of quantum information between various users and devices. It is an extension of the classical internet, and uses quantum teleportation to ultimately enable quantum information transfer between many devices and users -- across the world. Applications include highly secure communications, authentication of documents, dense coding of information, oblivious transfer of information, blind quantum computing, and ultimately distributed quantum computing, see Fig.~\ref{fig:quantum_network_use_cases}.  This quantum network could also enable distributed quantum sensing and metrology, with applications such as quantum enhanced telescopes or improved clock synchronisation.

One of the most established use cases for quantum communication is the need for secure communications. As the global development of quantum computers is progressing very rapidly and with large investments, it is only a matter of time until the key exchange algorithms used in today's internet could become vulnerable. Quantum key distribution (QKD) is a possible solution, and commercial systems have been available for several years. However, these systems are only point-to-point (P2P) links, and their main-stream deployment will require multi-user access and networking, similar to today's internet, where billions of end-nodes can talk to each other. The best approach to reach the multi-user capability is a quantum internet, which intrinsically enables the coherent transfer and routing of quantum information through a network. 

Quantum teleportation is the main enabling protocol for a general, multi-user and scalable quantum network. An underlying requirement for a quantum internet is the long-distance distribution of quantum information, and therefore the QEYSSat~2.0 team recommends that the next quantum satellite mission should involve entanglement distribution across Canada with the goal to enable {\bf Canada-wide quantum teleportation}. This proposed mission is of high relevance as such a mission would be conceived as a science and technology demonstration, many of its technologies, including deterministic photon sources and quantum memories, are critical elements for a future quantum internet both on the ground and in space, and will therefore help establish Canada's world leadership in this field.

\begin{figure}[ht]
    \centering
    \includegraphics[ width=0.9\textwidth]{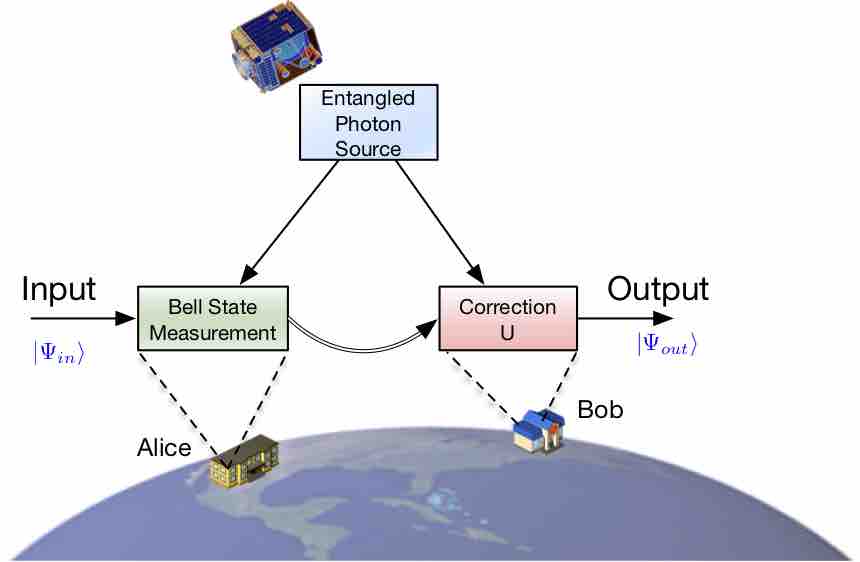}
    \caption{General overview of the quantum teleportation scheme proposed for a future QEYSSat~2.0 mission, where the two ground parties are spread across Canada.}
    \label{fig:teleport_across_Canada}
\end{figure}

\subsection{Technical Feasibility of Long-range Quantum Teleportation}\label{Technical-Feasibility}
We performed a preliminary performance and data transfer rate estimation for several quantum repeater and quantum teleportation scenarios, and determined that it could be feasible to implement long range teleportation  even with established technology. This  includes state of the art  tracking telescopes, single-photon detectors, weak coherent pulse sources, and spontaneous parametric down-conversion (SPDC) sources.  While it is reassuring to see that even the relatively inefficient original teleportation scheme from 1997 (\cite{Bouwmeester1997}) could be scaled up to operate from a high-altitude medium Earth orbit (MEO) satellite, a future QEYSSat 2.0 mission would need to utilize advanced quantum technologies such as deterministic photon sources and quantum memories as its functionality and scalability would  greatly benefit from such emerging technologies.

Specifically, we identified the following  areas as \textbf{critical technologies} to enable and improve the performance of a long-range teleportation mission:

\begin{itemize}
    \item \textbf{Deterministic and high-rate quantum sources and emitters of single and entangled photons}, such as quantum dot or memory-based systems. This could also involve SPDC-based sources with multiplexing features.
    \item \textbf{Heralded quantum memory and quantum non-demolition detection} are required for fully scalable quantum communication for its capability to determine if a photon successfully arrived through the channel, while leaving its quantum information intact. 
    \item \textbf{Bell-state measurement (BSM)} between photons sent from satellite to ground, as well as ground-based BSMs.  Quantum teleportation inherently relies on a BSM operation. The main challenges are the rapidly varying links between ground and  satellites which requires unique solutions to establish real-time stabilisation and synchronisation of photon arrival times. Another challenge will be to compensate for the distortion of spatial photon modes caused by turbulence in atmospheric propagation.   
    \item  \textbf{Adaptive optics (AO) for wave-front correction} for the ground-based systems are required for various applications, including coupling received optical signals to a single-mode system for improved coupling for two-photon interference, quantum memories and better single photon detectors (SPDs) (e.g. superconducting nanowire single photon detectors (SNSPDs)). AO could also be useful for improving the up-link beam pointing. 
    \item \textbf{Multiplexed quantum memories} are needed in order to enhance the channel transfer rate, and could involve temporal, spatial or spectral multiplexing.
    
\end{itemize}

\subsection{The Opportunity for Canada}
As part of the discussions for this project, it is clear that Canada has a great opportunity to step-up its efforts in this realm. Essentially, all the required technology and scientific expertise exists in Canada, be it with academic researchers or through industrial and startup capabilities. For instance, some of the world-leading single photon detectors are built by Excelitas in Canada, and are already used in space. New detector designs are being worked on at Sherbrooke University.  Quantum memories are an essential element for this project, and are in development by several researchers across Canada, including the Universities of Calgary, Alberta,  Ottawa, as well as Simon Fraser University.  Several Canadian startups are working on quantum memories or quantum repeater technologies, including  Quantum Bridge Technologies, Aurora Quantum Technologies, Photonic, and Quantized Technologies.

Quantum communication research groups and startups are developing end-to-end secure communication solutions, and are also considering the space segment, including the Universities of Waterloo, Toronto, Montreal, Ottawa, Calgary, and startups including  QEYNet, EvolutionQ, and ISARA.

The underlying theory of quantum networking and its applications are being researched at the Universities of Calgary and Ottawa, together with the National Research Council of Canada (NRC).

Quantum emitters for deterministic single photons and entangled photons are crucial elements for success, and are currently being built and studied by researchers at the NRC and the Universities of Waterloo and Calgary. There is also some technological development being made in several Canadian companies and industrial research centers that can be used as a stepping stone for entangled sources.

Furthermore, the Canadian photonics industry is very strong, and there are several companies and startups in the domain of optics that could benefit from such a project, including OZOptics and Iridian Spectral Technologies. Notably, the strong Canadian space industry has experience with the implementation and operation of satellite payloads and missions, including UTIAS, Honeywell-Canada, MDA, ABB, INO, Telesat and others. We provide a non-exhaustive overview of relevant expertise in Section~\ref{appendix:CanadaList}

\subsection{Outlook}
The question is not if the Quantum Internet will happen, but when. Canadian expertise and know-how mean that its entities and researchers are well positioned to take a leading role. The proposed mission on {\bf Teleportation Across Canada} emerged from this study  as an ideal mid-term goal. The development of critical quantum technologies and components, such as   photon emitters,  quantum memories or novel space-platforms for applications both on the ground and in space are very well represented within the Canadian ecosystem.  The  researchers, industry and startups in this domain are already well prepared and moving towards development and research in these areas. Given the rather large size and unique population distribution of Canada, it is critical that the technology for long-range quantum communication networks are developed and available from within the country.

\clearpage

\section{Introduction}

\subsection{The Quantum Internet - Overview}
A global quantum network or `Quantum Internet' would allow for the exchange of quantum information between spatially separated parties connected by quantum communication channels \cite{kimble2008quantum,simon2017towards}, and assisted by classical channels. This network, in which the laws of quantum physics govern the transferred information, will boost our capabilities of performing computation and communication tasks, and accommodate new functionalities with no classical counterpart. One of the most prominent examples of such an advantage is securely transmitting quantum information between distant users using quantum key distribution (QKD) \cite{alleaume2014using,gisin2002quantum}. The security of QKD is robust against advances in computing technologies and mathematical algorithms. Secure execution of quantum computation tasks, where both computational commands and results are hidden from the computer that performs the computation, is also possible using blind quantum computing \cite{barz2012demonstration}. The correlation of quantum sensors across a quantum network would also enhance the precision and sensitivity of the resulting system. This, in particular, can lead to long-baseline telescopes with improved angular resolution and/or sensitivity \cite{gottesman2012longer, moore2017astrometric, khabiboulline2019quantum}, more precise clock synchronisation and therefore global timekeeping \cite{komar2014quantum}, and high sensitivity magnetometers \cite{otterstrom2014nonlinear}.

\begin{figure}[ht]
    \centering
    \includegraphics[ width=0.6\textwidth]{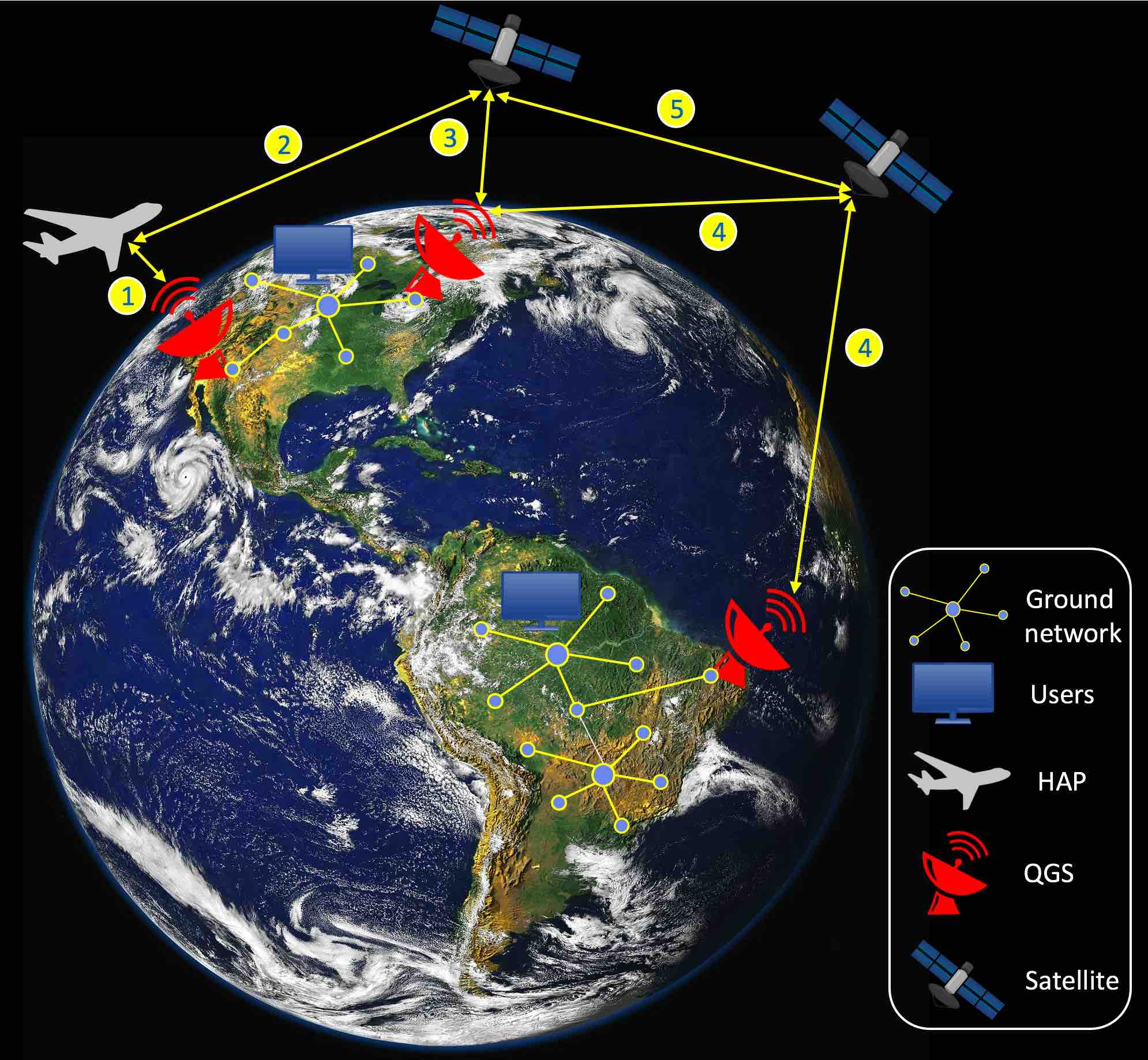}
    \caption{General conceptual overview of the Quantum Internet. Link 1: QGS to HAP, Link 2: HAP to satellite, Link 3: Single link QGS to satellite (like QEYSSat 1.0 mission, trusted-node QKD), Link 4: Double link QGS to satellite (entanglement distribution, untrusted-node QKD), Link 5: Intersatellite links. QGS = Quantum ground station, HAP = High altitude platform.}
    \label{fig:QuantumInternet}
\end{figure}

Figure \ref{fig:QuantumInternet} is a conceptual overview of the main components that would form a quantum internet. Ground based networks transfer quantum information via optical fibers,  quantum ground stations (QGS) establish a space-to-ground link, while satellite or airborne systems enable long distances and large sections without ground links.


Photons are individual particles of light that are naturally suited to carry quantum information because of their long lifetime and ease of distribution. However, unavoidable transmission losses present in every communication channel (e.g., terrestrial free-space or optical fibres) are a bottleneck for quantum communication over more than a few hundred kilometers. We cannot use the usual techniques to overcome the impact of transmission losses because unlike classical communication, which is free to use amplifiers to copy and regenerate classical signals, according to the no-cloning theorem, it is impossible to amplify an unknown quantum state perfectly. As a result, quantum networks will be limited in scale unless alternative solutions are employed.

In principle, a quantum repeater (i.e., quantum counterpart to a classical amplifier) allows for continental-scale quantum networks by circumventing the effects of exponential loss \cite{briegel1998quantum, sangouard2011quantum, simon2007quantum}. To date, technical challenges such as limited memory efficiency and storage time have hindered the practical usefulness of repeaters that are actually better than the direct transmission of photons. In addition, distances beyond a few thousand-kilometer range are out of reach with fibre-based quantum repeaters. An alternative approach to overcome this limitation is by incorporating quantum satellites and space-based technologies. The key benefit of using satellite links for quantum channels is that beam diffraction, rather than absorption, is the dominant transmission loss effect, which means the channel losses scale quadratically with distance for a satellite link, rather than exponentially as for a ground-based link. Besides, most weather phenomena occur only in the troposphere (i.e., the lowest layer of the atmosphere with a total average height of $\sim 20$ km).  Of course,  the use of satellites to transmit quantum signals comes with its own set of challenges. These include additional losses due to atmospheric extinction and scattering, background thermal noise, and turbulence-induced beam wandering.

\subsection{Terrestrial-based Quantum Communication}
Over the last decade, there have been several impressive experimental advances of quantum communication using terrestrial communication links. Notably, long-range entanglement distribution was demonstrated over an optical free-space link between the Canary Islands separated by 144\:km \cite{ursin2007entanglement}. In addition, QKD was demonstrated through very long optical fibre in laboratory settings such as 509\:km \cite{chen2020sending}, 421\:km \cite{boaron2018secure}, and 404\:km \cite{yin2016measurement}. A fibre network containing 700 QKD links with a total length of $\sim20,000$\:km between Beijing and Shanghai in China has been demonstrated \cite{chen2021integrated}, as well as secure QKD over a 23\:km free-space link between two mountain tops \cite{kurtsiefer2002step}. Other notable demonstrations include the entanglement distribution between the Mediterranean islands over a 96\:km submarine optical fibre \cite{wengerowsky2019entanglement}.
Most recently, secure QKD has been realized over 511\:km of optical fibre between two remote metropolitan cities \cite{chen2021twin}. 

To date, heralded entanglement generation over an elementary link of a quantum repeater has been demonstrated using different platforms, including atomic ensembles \cite{yu2020entanglement,yuan2008experimental, chou2007functional} and rare-earth doped crystals \cite{usmani2012heralded}, and also single systems such as quantum dots \cite{delteil2016generation}, trapped ions \cite{moehring2007entanglement} and atoms \cite{hofmann2012heralded}, and defects in diamond \cite{hensen2015loophole}. Nevertheless, none of these demonstrations could address all of the requirements for an efficient quantum network (e.g., memories with high multimode, large bandwidth, long lifetime capabilities, and compatibility with telecom fibres). However, most recently, entanglement generation between two remote quantum memories has made substantial experimental progress. To be more precise, using rare-earth doped crystals, telecom-heralded entanglement generation between distant multimode quantum memories \cite{lago2021telecom}, as well as heralded-entanglement distribution between memories that can support multimode operations \cite{liu2021heralded} have been demonstrated. Considering these progresses, designing a simple quantum repeater with only two elementary links seems accessible in the near future.

Using distributed entangled pairs, quantum teleportation of unknown quantum states has also been realized between remote locations. In particular, quantum teleportation between the Canary Islands over 143\:km \cite{ma2012quantum}, as well as entanglement distribution over two free-space links of total length of $\sim100$\:km, including quantum teleportation over 97\:km of free-space channel \cite{yin2012quantum} was demonstrated utilising terrestrial free-space links. Entanglement swapping has also been reported across 100\:km of optical fibre \cite{sun2017entanglement}. Quantum teleportation over metropolitan fibre networks with independent entanglement sources in Geneva~\cite{gisin_natphys_2007}, Calgary \cite{valivarthi2016quantum} and Hefei \cite{sun2016quantum}, have been realised employing optical fibres, as well as teleportation of telecom qubits across 22\:km \cite{valivarthi2020teleportation}. However, none of these schemes actually utilised the stored, pre-distributed entanglement as this would require quantum memories, but rather these demonstrations consumed the entanglement resources in real-time.

\subsection{Satellite-based Quantum Communication: The QEYSSat Mission}
\label{sec:qeyssat1.0}
\begin{figure*}[ht]
\centering\includegraphics[width=0.8\linewidth]{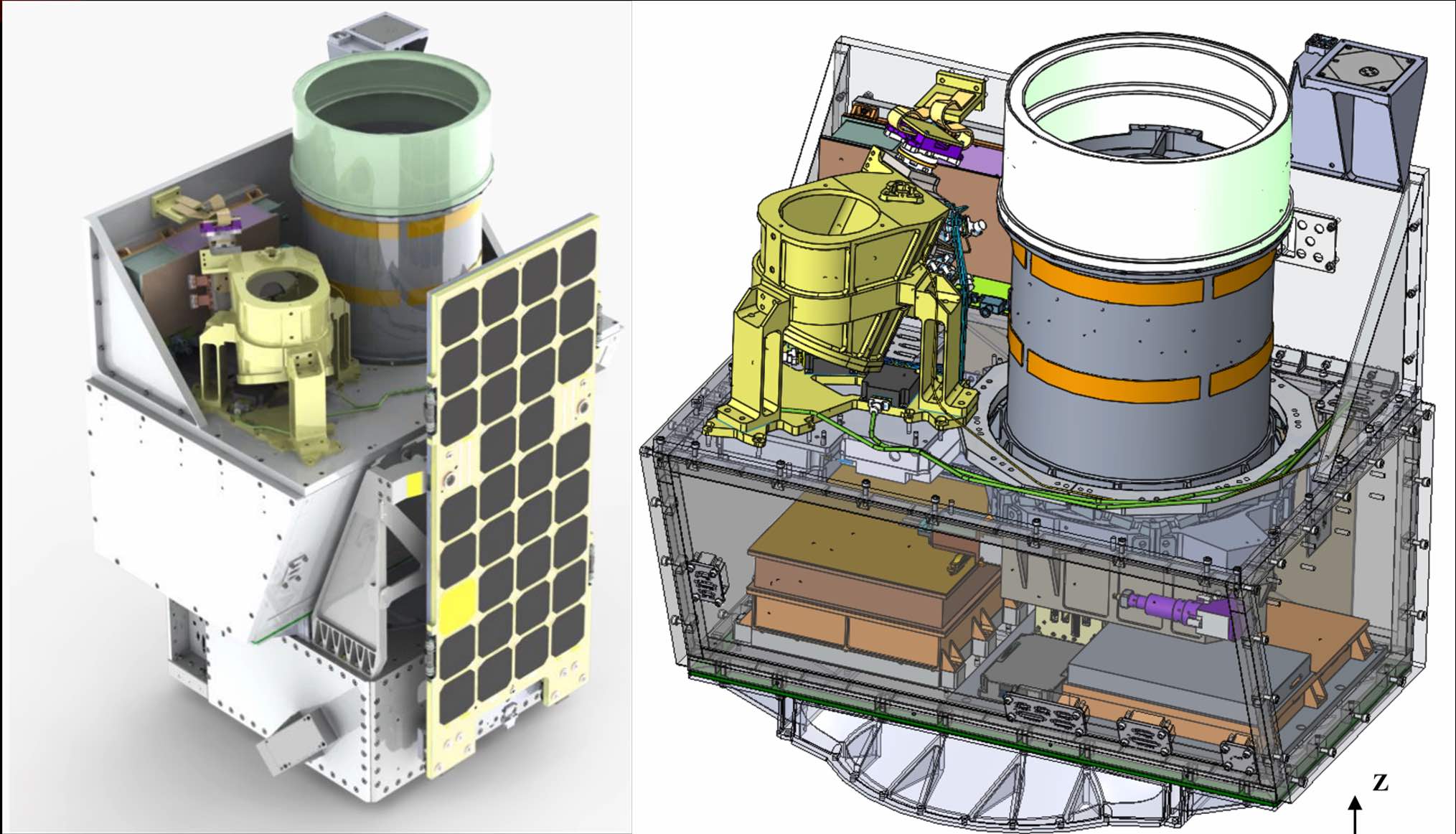}
\caption{Renderings of the QEYSSat space craft. The main aperture is a 25\:cm telescope. The primary function of the QEYSSat payload is to receive and analyse photons \cite{Jennewein2014}. It will also include an additional WCP source module, which will emit photons with random polarisations to be received at the quantum ground station. (Image Credit Honeywell)} \label{fig:QEYSSatModel}
\end{figure*}

The Canadian Quantum EncrYption and Science Satellite (QEYSSat) mission is intended to be a scientific platform and demonstrator, offering a unique opportunity for the Canadian scientific community to implement satellite-based quantum communication, and perform cutting edge scientific experiments. QEYSSat will provide both scientific and technological advances to Canada, with a clear benefit to the governmental and industrial sectors, and society in general. A technical schematic and model drawings of the QEYSSat design are shown in Figure~\ref{fig:QEYSSatModel}.

\subsubsection{QEYSSat Mission Overview}
The QEYSSat mission was green-lit in 2017 during a federal funding announcement at the Canadian Space Agency (CSA). In brief, the QEYSSat mission aims to advance the technical readiness of quantum communication using ground-to-space links, and study the science related to such a quantum channel. The primary functionality of the payload is a receiver for photons using a 25\:cm aperture telescope, and the photons will travel along an up-link path. The benefit of the up-link is we can use various different photon emitters at the QGS. This mission is the result of eight years of extensive studies and research on the feasibility, concept and prototyping of the system led by the IQC team at the University of Waterloo~\cite{Rideout2012,Bourgoin2013, Holloway2013, MeyerScott2011, Bourgoin2015a, Bourgoin2015b, Anisimova2017}. Specifically, QEYSSat's primary mission objective is to develop and demonstrate the capability to distribute highly secure encryption keys from an optical ground station to a QKD satellite.  The mission will also conduct fundamental science such as tests of long-distance quantum entanglement at distance and velocity combinations not possible on the ground, and explore additional applications of the quantum channel \cite{Rideout2012}.

In 2016, the IQC team demonstrated the mission's viability by flying a quantum payload prototype on an NRC Twin Otter Airborne Research aircraft, and receiving quantum signals from the IQC QGS~\cite{pugh2017airborne}. The IQC team also led the development of the mission's scientific concepts and objectives~\cite{Rideout2012, Jennewein2014}, which foremost aims to demonstrate a quantum up-link from ground-to-space, utilising multiple different ground station configurations, including different quantum sources and quantum networks. 

\subsubsection{What is Space-based QKD?}
QKD establishes highly secure keys between distant parties by using single photons to transmit each bit of the key. Since single photons behave according to the laws of quantum mechanics, they cannot be tapped, copied or directly measured without detection. Ground-based QKD systems are commercially available today, however, current systems can only cover distances of a few hundred kilometers due to photon absorption in fibre optic cables. 

\subsubsection{Current Mission Timeline}
\begin{figure*}[ht]
\centering\includegraphics[width=\linewidth]{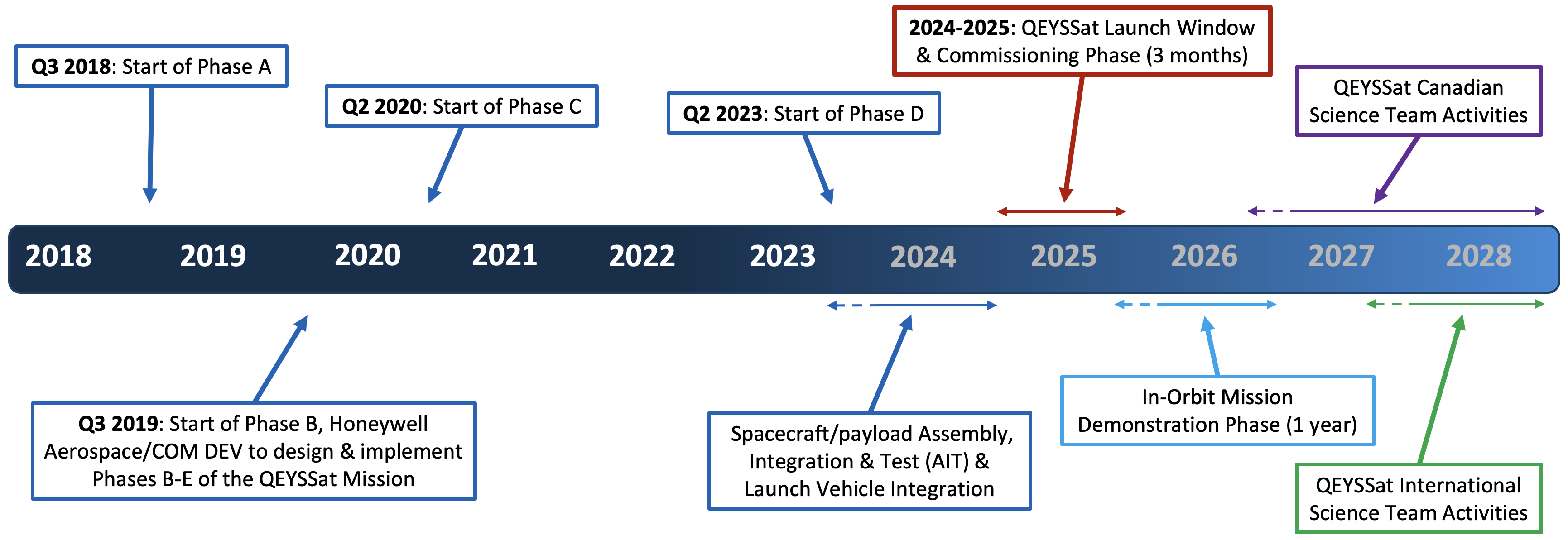}
\caption{Current QEYSSat mission timeline (Updated Q1 2023).} \label{fig:QEYSSatTimeline}
\end{figure*}

The current QEYSSat mission timeline is shown in Figure~\ref{fig:QEYSSatTimeline}. The project is currently in Phase C with Phase D due to start in Q2 2023 (updated Q1 2023). 

\subsubsection{Mission Requirements and Goals}
The QEYSSat mission will be a technology demonstration platform, primarily intended to study the quantum up-link channel (ground-to-space) using a space-based quantum receiver including a photon polarisation analyser and single photon detectors. A recent addition to the mission is a quantum source module on QEYSSat to demonstrate QKD downlink channels using a novel reference frame independent protocol \cite{QeyssatRFI}. The mission duration will be for at least 1 year, with the possibility of an extended phase that includes Science Team member activities for several years to maximize use of QEYSSat.  The mission is designed so that it shall distribute at least 100\:kbit of secure key between at least two ground stations with separation of at least 400\:km, with the goal of many thousands of kilometers in collaboration with several QGS partners.  QEYSSat will also demonstrate re-keying the satellite over consecutive periods of time to study the reliability of the system. 

QEYSSat also has several fundamental science studies in addition to the QKD demonstration. For example, the mission shall perform a Bell test for entangled photons separated over a distance of more than 400\:km, with the goal of 1000\:km or more. QEYSSat will also study the performance of ground-satellite quantum links with at least two different quantum sources, which are a WCP source (see Section~\ref{sec:WCP}) and an EPS (see Section~\ref{sec:EPS}). The data collected by the mission shall be used to improve the knowledge of the environmental parameters and assumptions used in quantum link budgets in order to facilitate the design of future systems. 

The QEYSSat Science Collaborator Teams  (see Sections~\ref{sec:QEYSSatPhase1} and \ref{sec:QEYSSatPhase2}) will also assist with demonstrating an interface between ground-based quantum links and the satellite, as well as testing other novel single photon sources such as quantum dots.

\subsection{Advances towards a large-scale Quantum Internet}

Quantum networks have the potential for a wide range of applications, from secure quantum communication to distributed quantum computation. As a result, many efforts are underway throughout the world to demonstrate certain building blocks of a quantum network which include long-range satellite links and ground-based quantum technologies.

\subsubsection{Quantum Satellite Missions - Past, Present, Future}
To demonstrate space-based quantum communication, the Chinese Academy of Sciences has launched the world’s first major quantum communication low Earth orbit (LEO) satellite known as {\it Micius}. Using this 631\:kg quantum satellite, direct distribution of entangled photon pairs over 1200\:km \cite{yin2017satellite}, entanglement-based QKD over a satellite link \cite{yin2017(b)satellite}, secure QKD over intercontinental distances between ground stations in China and Austria \cite{liao2017satellite}, and ground-to-satellite quantum teleportation of independent single photon qubits over distances up to 1400\:km \cite{ren2017ground} have been demonstrated. More recently, using an existing trusted node link between Beijing and Shanghai in combination with satellite-to-ground free-space links, an intercontinental scale hybrid quantum communication network over a total distance of 4600\:km was demonstrated~\cite{chen2021integrated}. In light of demonstrating these key milestones toward a global quantum network, China is now leading the quantum communication sciences.

Employing a 50\:kg-class microsatellite, Japan has also demonstrated a satellite-to-ground quantum transmission of polarisation states \cite{takenaka2017satellite}. Successful tests of quantum communication between a ground station and a mobile receiver to emulate the motion of a satellite have also been achieved \cite{nauerth2013air, wang2013direct}. 

Another important step towards full quantum communication using satellites is having the ability to generate correlated entangled photon pairs on small satellites in a cubesat format (i.e. nanosatellite) \cite{tang2016generation}, which was demonstrated by a team from Singapore. Several reviews on the recent advances in space quantum communication are available \cite{Sidhu2021,Clark2021,Kaltenbaek2021}.


\subsubsection{Canadian Context}
Several groups in Canada have made significant progress in different building blocks of a quantum network, which includes experimental demonstration of quantum memories \cite{saglamyurek2011broadband, saglamyurek2018coherent, saglamyurek2021storing, sinclair2014spectral, saglamyurek2015quantum}, microwave-to-optical photons conversion \cite{arnold2020converting}, quantum teleportation  \cite{valivarthi2020teleportation, valivarthi2016quantum}, quantum cryptography \cite{pinheiro2018eavesdropping}, free-space QKD \cite{bourgoin2015free, sajeed2015security, pugh2017airborne, sit2017high}, theoretical studies on quantum transduction \cite{lauk2020perspectives}, quantum repeaters \cite{han2010quantum, asadi2018quantum, wu2020near, kumar2019towards, asadi2020protocols, ji2020proposal}, satellite-based quantum communication \cite{pugh2020adaptive, jennewein2018towards}, and global quantum networks \cite{boone2015entanglement, simon2017towards}. 

\begin{figure*}[ht]
\centering\includegraphics[width=0.8\linewidth]{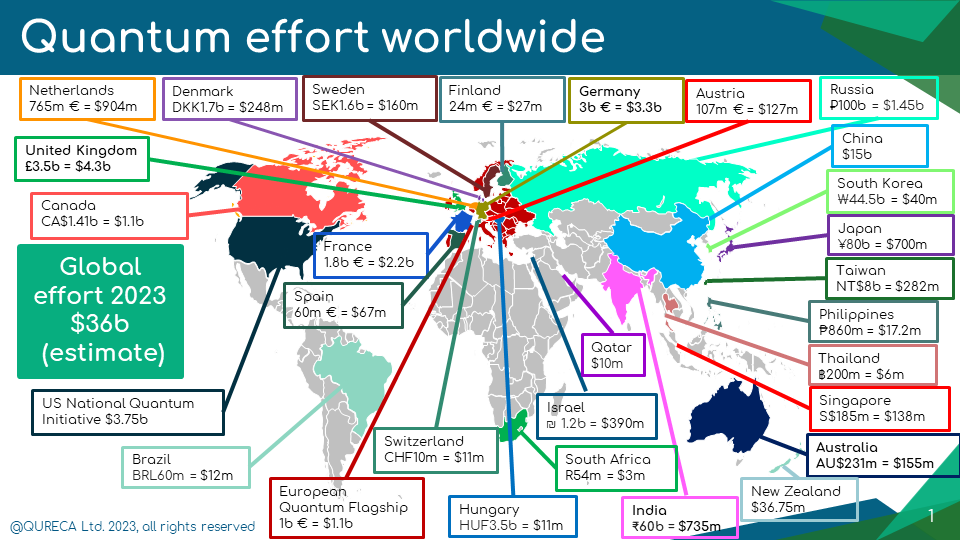}
\caption{World investment into Quantum Technology (May 2023). https://qureca.com/quantum-initiatives-worldwide-update-2023/} \label{fig:WorldFig}
\end{figure*}

\subsubsection{International Investment in Quantum Technologies}

The worldwide investment into quantum technologies has reached very substantial levels. Figure~\ref{fig:WorldFig} provides a rough overview of  quantum technologies at a global scale as of May 2023. It is expected that the level of investments will increase going forward.

\clearpage

\section{Applications and Use-Cases of Quantum Communication Satellites}
In this section we present high-level summaries of the main applications and possible use-cases for quantum satellite technologies for Canada wide, and even global scale, deployment. Table~\ref{tab:use-cases_for_quantum_sat} presents an overview of such applications and use cases, and categorizes them in their respective domain of application. Since the aim of this paper is to establish the right priorities for the Canadian Roadmap in quantum technologies, the authors have selected several applications and use cases relevant to the possible QEYSSat 2.0 mission.

\begin{table}[h]
\centering\includegraphics[width=1\linewidth]{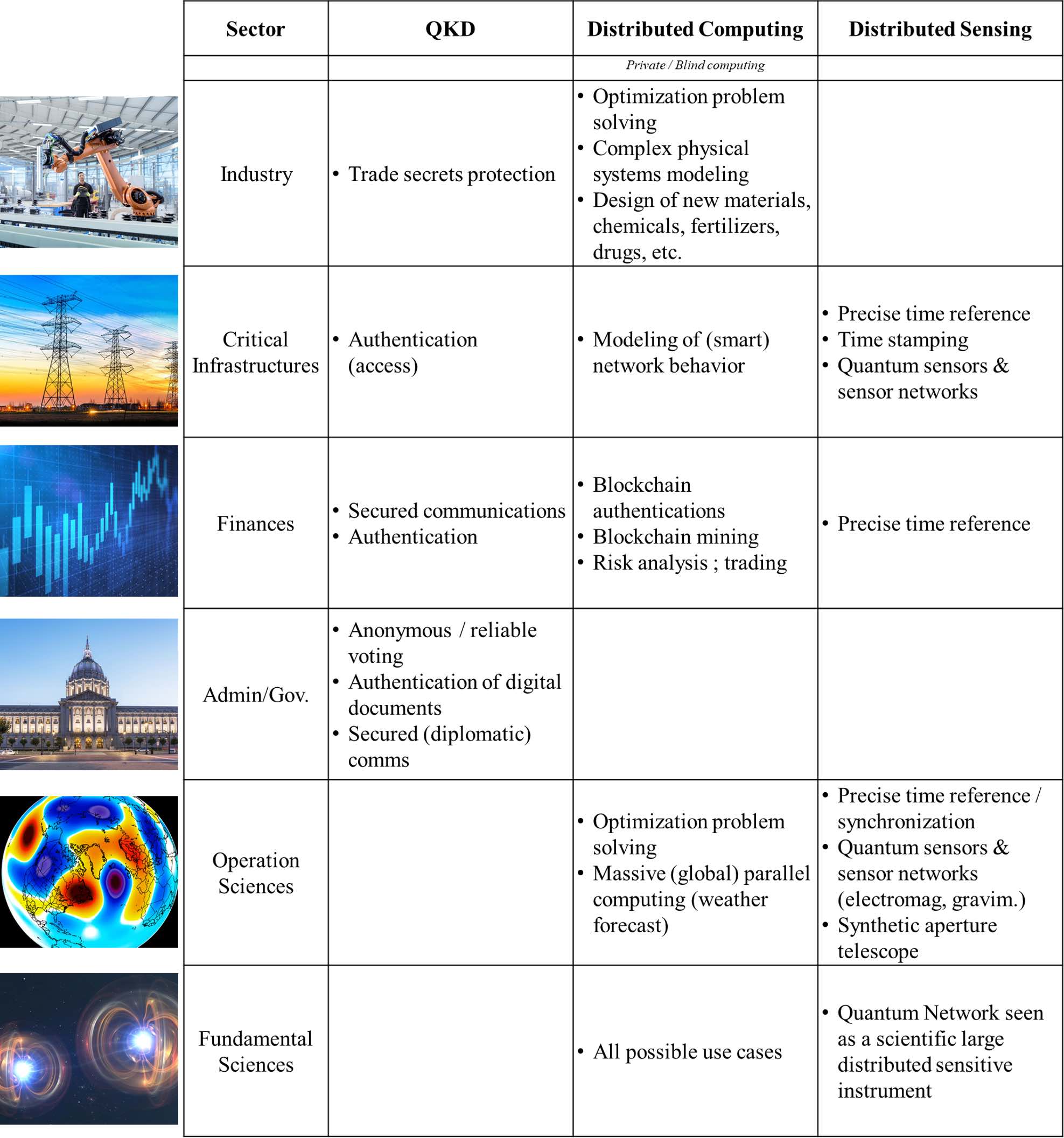}
\caption{\label{tab:use-cases_for_quantum_sat}Use Case, Application and Domain of Application for the quantum satellite technologies \cite{arXiv:2202.01817v2}.}
\end{table}

\subsection{Quantum Key Distribution}
The secure communication of information between distant parties using via quatnum key distribution (QKD) is one of the more developed applications of quantum technology. While there are security advantages to using quantum-based techniques over algorithmic and non-quantum physical layer methods, there are still technical limitations to achieving high key rates over long distances that need to be overcome to make this technology feasible for worldwide deployment. Several countries are working on QKD deployments that consist of terrestrial ground networks involving optical fibre and/or free-space links, and satellite networks with LEO, MEO or GEO satellites \cite{Travagnin2019}. The  QEYSSat mission is Canada's first step into long-range QKD, and will demonstrate the science and technology of ground-to-space quantum links (see Section~\ref{sec:qeyssat1.0} for mission summary).

Quantum key distribution is the generation of encryption keys between two users, typically called `Alice' (Sender) and `Bob' (Receiver). The information is stored in individual packets of energy called photons, and the security of the key is based on the principles of quantum physics. Since a single photon cannot be copied or manipulated without being noticed, the users can know when an eavesdropper (`Eve') is trying to learn the key. If an eavesdropper tries to hack the quantum channel, they will disturb the photons, revealing the attack. Furthermore, an encryption key generated from QKD that is secure today will remain secure against advances in computing power, unlike current public-key encryption methods, which are vulnerable to attacks from future technology like quantum computers.

There has been significant worldwide efforts in implementing real-world QKD networks with hardware made in academic facilities or companies such as QuantumCTek, China Quantum Technologies, ID Quantique, Toshiba and Huawei. Deployments over a few hundred kilometers are possible with a ground optical fibre network. However, larger distances require other approaches (quantum repeaters, satellites) because signal transmission in fibre decreases exponentially, and conventional amplification to compensate for the loss does not work for quantum information. Ground-based quantum repeaters (discussed in Sections~\ref{sec:QRepeater} and \ref{sec:Qmemories}) are in development but are unlikely to be deployable for Canada-wide links for the foreseeable future. Satellites with quantum technologies on board, such as single photon detectors (Section~\ref{sec:detectors}) and quantum sources (Section~\ref{sec:Sources}), can enable Canada-wide quantum links to facilitate a QKD network coast-to-coast-to-coast. A Canadian quantum network will insure Canada's sovereignty over the privacy of public, private, and commercial data.

\subsection{Clock Synchronisation}
Numerous sectors in modern life depend on access to a global standard time, from data transfer to telecommunication networks, metrology and long baseline interferometry, to the Global Navigation Satellite System (GNSS) and Global Positioning System (GPS). In order to have universal agreement across all these networks, it is necessary to achieve clock synchronisation across the entire network. This challenge is currently met with classical optical-based techniques via measuring time of arrival of light pulses. By expanding these protocols to include quantum technologies, such as frequency-entangled or squeezed light pulses, we can improve the accuracy and performance of clock synchronisation and positioning systems by several orders of magnitude \cite{Giovannetti2001,Okeke2018,Jozsa2010}.

The increasing demand for high precision remote clocks is adding to the push for space-based quantum technologies that can enable long distance transfer with enhanced security. High precision clocks in space will open the door to having a secure and independent time base for global time keeping, as well as allowing for a global quantum network when combined with space-space and space-ground optical links. However, there are several technology bottlenecks/challenges left to overcome, as highlighted in Section~\ref{sec:TechBottlenecks}.

\subsubsection{Global Time Standards}
There are two fundamental technology elements to time standards and frequency transfer: precision time standards (clocks) and the ability to transfer frequency (i.e. the phase of the clock) over a large distance with high precision. Current atomic clock technologies have the ability to operate at radio frequencies (GHz range) and achieve accuracies on the order of $10^{-15}$. However, there is a fundamental limit to how much improvement can be achieved with this technology due to the operation frequency. Operating in the optical frequency range dramatically increases the signal bandwidth. Including quantum technologies, such as quantum light sources, has significantly improved optical clock development, with laboratory demonstrations having uncertainty values below $10^{-18}$ and continuing to improve \cite{Brewer2019}. Frequency transfers using lasers in a ground-based 920\:km fibre optical network at an accuracy better than $4\times10^{-19}$ was achieved.

\subsubsection{Optical Clocks in Space}
The European Space Agency (ESA) has studied plans to launch a sequel mission to the Atomic Clock Ensemble in Space (ACES) in the early 2020s that includes optical atomic clocks and optical links called the Space Optical Clock on the International Space Station (ISS, I-SOC) \cite{Origlia2016}.

\subsection{Sensing}
Quantum satellite technologies encompass sensing capabilities that can be used for the measurement of various parameters such as:

\subsubsection{Gravitational Wave Sensing}
Space-based optical clocks open several other applications beyond global time keeping, such as measuring the geopotential difference between two distant locations based on their relative gravitational red shift, which is on the order of $10^{-18}$ per cm of geopotential height \cite{McGrew2018}. The absolute geopotential on Earth's surface can be characterised relative to a satellite's location in orbit (geodesy application). Gravitational wave detection can also be achieved between two optical clocks on distant satellites \cite{Shen2020, Kolkowitz2016, Tino2019}, which is complementary to the ESA/NASA Laser Interferometer Space Antenna (LISA), which has a very high sensitivity to gravitational waves at lower frequencies.

\subsubsection{Large-baseline Telescopes}
Another sensing application that would benefit from quantum technologies is an optical synthetic-aperture telescope, which is analogous to the radio-frequency synthetic-aperture observation technique. A set of fully synchronised optical clocks enables a phase measurement of an incident light wave from a distant celestial object at several locations that are spatially distant. This network creates a telescope with an optical synthetic aperture size comparable to the separation of measurement locations (potentially thousands of km) \cite{gottesman2012longer} . The added sensitivity could potentially allow for the direct observation of extrasolar planets.

\subsubsection{Remote Sensing and Earth Observation}
Improving our understanding of climate change, hydro- and biosphere evolution, and more accurate detection of tectonics and earthquake predictions can be possible through gravity field mapping using space-based quantum technologies. Atomic systems have already been implemented as sensors in space, such as using Bose-Einstein condensates to create atomic interferometers \cite{Muntinga2013}. Quantum gravity sensors use coherent quantum matter waves as the test masses, which leads to more precise and sensitive instruments.

Others application such as altimetry - which provides insightful data on precision sea level, sea surface height, and large wave height could use technologies such as quantum LIDAR with possible improvement in the measurement precision, as well as possibilities to range surfaces that are currently difficult with radar waves. Moreover, LIDAR are also used for wind speed measurements (mission Aeolus from the ESA), and could also use technologies such as quantum LIDAR.

\subsection{Foundations of Quantum Networks}

\subsubsection{Bell test}
A scientifically  interesting long-distance experiment is testing Bell’s inequality. Bell’s theorem states that under the assumption of locality and reality and for any physics theory, the resulting correlations from measurement outcomes performed by remote observers must obey Bell’s inequality. Ultra-long range Bell test enables us to address the locality and freedom-of-choice loopholes \cite{cao2018bell}. To perform a long-range Bell test, one can consider a setup where the entangled source is placed in the middle of two receivers. It is also possible to use an asymmetric setup where the source and one of the receivers are in the same site, e.g., inside a quantum satellite. In this case, employing quantum memories can ensure space-like separation between the detection events. 

Using the Micius quantum satellite, violation of the Bell inequality across 1200\:km has been demonstrated \cite{yin2017satellite}. Testing Bell’s inequality over longer distances, for instance, between Earth and the moon, could be the next step that can improve our understanding of the quantum gravity theories.

\begin{figure*}[ht]
\centering\includegraphics[width=\linewidth]{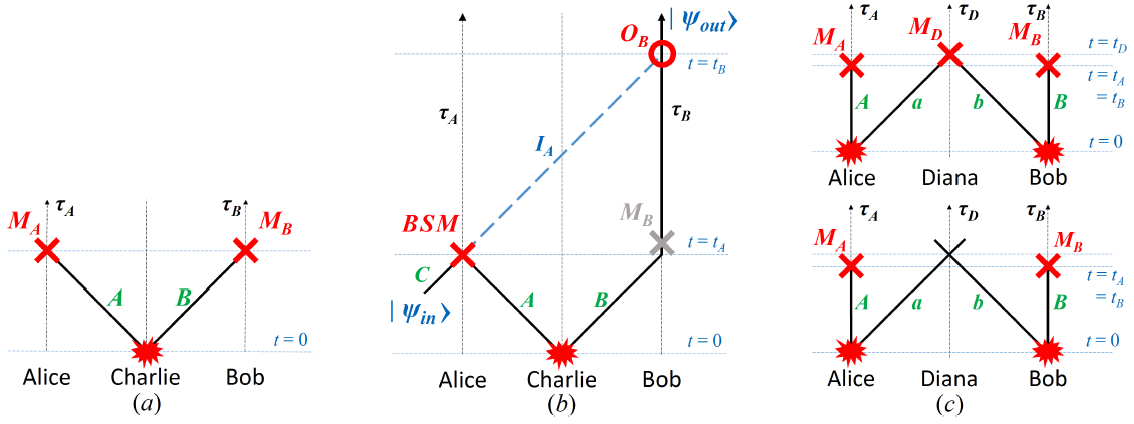}
\caption{Spacetime diagram of Bell test (a), quantum teleportation (b) and delayed choice entanglement swapping (c) \cite{Jennewein2021}.} \label{fig:SpacetimeBell}
\end{figure*}

\subsubsection{Quantum Teleportation}
Quantum state teleportation \cite{bennett1993teleporting} is a uniquely non-classical concept, as it perfectly transfers an unknown quantum state from one system to another by using two channels: a maximally entangled state and a classical signal. The first step is to establish long-distance distribution of entangled photons, see Figure~\ref{fig:SpacetimeBell}, over large distances in space. Quantum teleportation utilises such remote entanglement as follows \cite{Bouwmeester1997}: first, an entangled photon pair is generated by Charlie (photons A and B in Figure~\ref{fig:SpacetimeBell}(a)), and A is sent to Alice, and B to Bob. Alice performs a BSM \cite{Weinfurter1994, Mattle1996, Calsamiglia2001} on photon A jointly with the unknown quantum state carried by another photon, C, thereby projecting her two photons into an entangled state (Figure~\ref{fig:SpacetimeBell}(b)). This BSM will project Bob’s photon B onto one of four possible states depending on the BSM outcome. Bob, in the meanwhile must retain photon B after its arrival in a quantum memory until he receives Alice’s BSM result via the classical channel, which he then uses to apply a unitary operation to fully recover the original input state. Note that neither Alice, Charlie nor Bob obtain any knowledge on the input state, and the final unitary transformation depends only on the (random) BSM result, and thus the protocol fully obeys quantum-no-cloning \cite{Wootters1982}.


Currently, the longest range over which entanglement teleportation has been demonstrated is around 1400\:km between a ground station and a satellite, obtained by the Micius spacecraft \cite{ren2017ground}. In this experiment, It has been suggested that for the teleportation without post-selection between the earth and Moon, a memory storage time of at least 1.3\:s is required \cite{gundougan2021topical}.


\subsubsection{Generalized Entanglement Swapping}
\begin{figure}[ht]
    \centering
    \includegraphics[width=6cm]{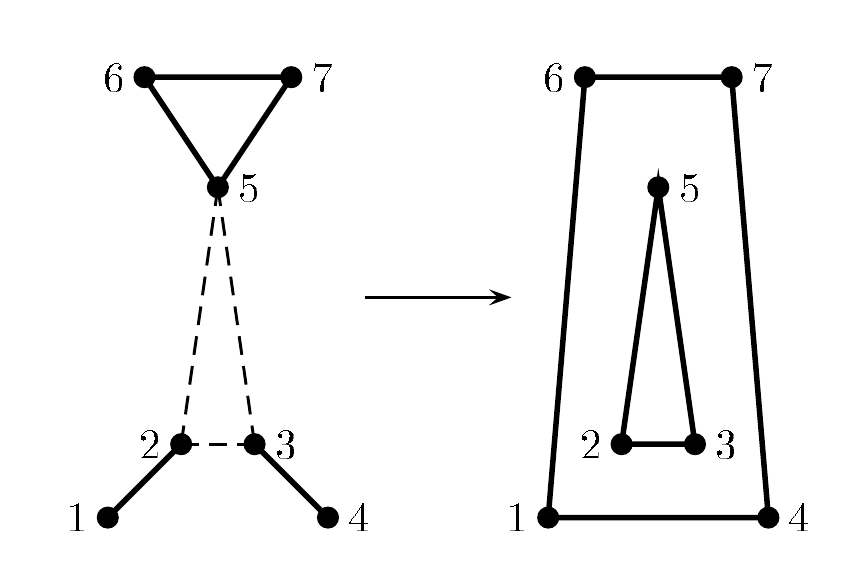}
    \caption{The conversion of two Bell states and a 3-particle GHZ state to a 3 particle GHZ state and a 4-particle GHZ state due to a GHZ state projection on three particles (from \cite{bose1998multiparticle}).}
    \label{fig:knight_generalized_ent_swap}
\end{figure}

The extension of quantum teleportation is entanglement swapping, where the input to a teleportation protocol is an entangled photon itself \cite{zukowski1993event}. To perform entanglement swapping, the two parties (e.g., `Alice' and `Bob') should share two-qubit entangled pairs with a third party (e.g., `Diana'), as shown in Figure~\ref{fig:SpacetimeBell}(c). Diana will then perform a BSM between her portion of the entangled pairs. This, in turn, leaves the state of Alice and Bob qubits in an entangled state. Thereby the entanglement is created between two separated quantum systems that never actually interacted. Entanglement swapping is the core element of a quantum repeater. The extension of entanglement swapping to multiple nodes and systems allows a seamless re-distribution of quantum entanglement across multiple users, see Figure~\ref{fig:knight_generalized_ent_swap}.

The success rate of entanglement swapping based on linear optics is limited to 50$\%$. It is possible to use auxiliary photons to enhance the swapping probability \cite{grice2011arbitrarily}. However, using this scheme, the success probability for entanglement swapping scales as $1-1/2^n$ where $2^n-2$ is the number of auxiliary photons. Hence, to achieve 100$\%$ efficiency, an infinite number of auxiliary photons are necessary. As a result, the practical application of this scheme is limited. Another approach to overcome this problem is by using single-emitter-based systems. In this case, depending on the swapping protocol, it is possible to perform deterministic gates between the quantum systems \cite{asadi2018quantum}.

\subsubsection{Quantum Repeaters}
\label{sec:QRepeater}
A quantum repeater can extend the transmission distance of entangled particles by connecting several short, independent entanglement links together via BSMs. In so-called first-generation quantum repeater protocols, the overall link length is divided into smaller segments with quantum memories at each system node. Next, entanglement is created over these elementary links independently. In this case, a heralding signal can be used to claim the successful implementation of remote entanglement. As entanglement generation is a probabilistic process, it will not be established over all elementary links simultaneously. However, entanglement can be stored using quantum memories until it is established over two adjacent elementary links.  Entanglement is then distributed gradually by performing the entanglement swapping between memories of each node. Although it has not been demonstrated yet, in principle entanglement can be distributed over 1,000-2,000 \:km using fiber-based quantum repeaters.

The repeater performance can be enhanced by using quantum memories capable of storing multiple temporal modes or generating entanglement over many spectral/spatial channels \cite{simon2007quantum, afzelius2009multimode, asadi2018quantum,sinclair2014spectral}. To compensate for losses, quantum repeaters can also employ heralded generation of entanglement and quantum error corrections.

When the individual entanglement distribution is realized with quantum memories, it has been shown that this scheme can clearly outperform the direct transmission of photons. 
To tackle loss and operational errors, the first generation of quantum repeaters use probabilistic error correction protocols \cite{ childress2006fault, briegel1998quantum}. In addition to the probabilistic treatment of loss errors, the second generation of repeaters can overcome operational errors using deterministic protocols \cite{ jiang2009quantum, munro2010quantum}. The third generation, on the other hand, treats all these errors deterministically \cite{ muralidharan2014ultrafast}. However, deterministic error corrections come at the cost of reducing length of elementary links and increasing the number of required qubits. 

So far, several repeater protocols have been proposed \cite{ childress2006fault, sangouard2011quantum, asadi2020protocols, sharman2021quantum}. Quantum repeaters are essential to a global quantum network, and such protocols could form the backbone for a future quantum internet.

\subsection{Distributed Quantum Computing}
Quantum computing makes use of quantum features such as entanglement and superposition to perform calculations. Certain problems are believed to be solvable in a reasonable amount of time only using quantum computers. A building block of a quantum computer is a qubit. Several platforms such as ion trap, rare-earth ions, and transmons can serve as qubits. In principle, increasing the number of qubits can increase the power of the quantum computer. The most important goal of quantum physics is to connect remote quantum devices to form a quantum internet in the long term.

In general, it is not an easy task to increase the number of operating qubits. For instance, as the interaction of qubits with the environment is inevitable, novel error correction strategies should be employed. Errors are not solely caused by decoherence. Any other sources of infidelity, such as imperfect quantum operation, can also cause errors. Hence, several strategies should be employed to resolve these critical errors.

While quantum computers can perform certain quantum tasks a lot more quickly than classical counterparts, not all of clients might have access to quantum computing devices. Blind quantum computation will enable clients to delegate their quantum computing tasks to untrusted devices, yet protecting the privacy against the server \cite{broadbent2009universal, barz2012demonstration}.  So far several blind quantum computing protocols have been implemented. In this regard, the ultimate goal is to make outsourcing computing tasks possible without the need for the clients to perform any quantum tasks or have access to any quantum resources. 

Distributed computing is another application of quantum physics where remote quantum devices are connected to form an arbitrary-scaled environment. The computing power of a quantum computer scales exponentially with the number of qubits. Distributed quantum computing can significantly enhance computing and communication capabilities (by scaling up the number of qubits), and lead to a network of quantum computers or a quantum internet \cite{cacciapuoti2019quantum}. Quantum devices as the building blocks of the system are comprised of quantum memories, transducers, sources and measurement devices.  

Quantum internet is made up of both classical and quantum links. Coherently networking quantum computers relies on generating and distributing entanglement between remote nodes. As stated before, direct transmission of quantum information is limited by the transmission loss in communication channels. Besides, the no-cloning theorem forbids us from copying the state of an unknown quantum system. As a result, the use of quantum repeaters has been suggested. 

\subsection{Ground-based vs. Space-based Solutions}
Quantum communication technology is rapidly advancing. QKD links are commercialized by multiple companies. Furthermore, quantum repeater technologies are being researched and advanced both in academia, and now also some startups are working towards such solutions. Quantum communication satellites could therefore be overtaken by ground-based developments and breakthroughs, such as scalable quantum repeaters, 
or better optical fibres, 
or finally, ultra-long-storage quantum memories that allow physical transport of quantum information \cite{zhong2015optically}.

A historical example of such a development was optical laser communication terminals in space, which were developed in the 1990s in order to achieve global coverage of high-bandwidth communications. However, before laser terminals could be deployed in space, optical fibre technology experienced massive advances and absorbed most of the global data needs. It is only more recently noted that classical laser communication has a role in space, and now multiple terminals have been launched and are in use.

What is unique about quantum communications is that the transmitted power leaving a source is already at the single-photon level, and the signal will only become fainter due to attenuation of the channel (fibre, free-space). 

The currently best  optical fibres have  the  minimal attenuation of  0.16\:dB/km for 1550\:nm signals. This implies that a link of 1000\:km will experience attenuation of 160\:dB. Thus the probability for a single photon to reach the receiver is so low, that it will happen only {\bf once every 100 days}, assuming the source emits photons at 1\:GHz rate. On the other hand, when using the same emitter on a typical 1000\:km space link, the single photon transfer occurs within {\bf 1~micro~seconds (!)}, assuming around 30\:dB of attenuation. More importantly, the scaling of the space link is quadratically with channel length, while the fibre optic link scales exponentially with channel length. 

\begin{figure}[ht]
    \centering
    \includegraphics[width=0.95\textwidth]{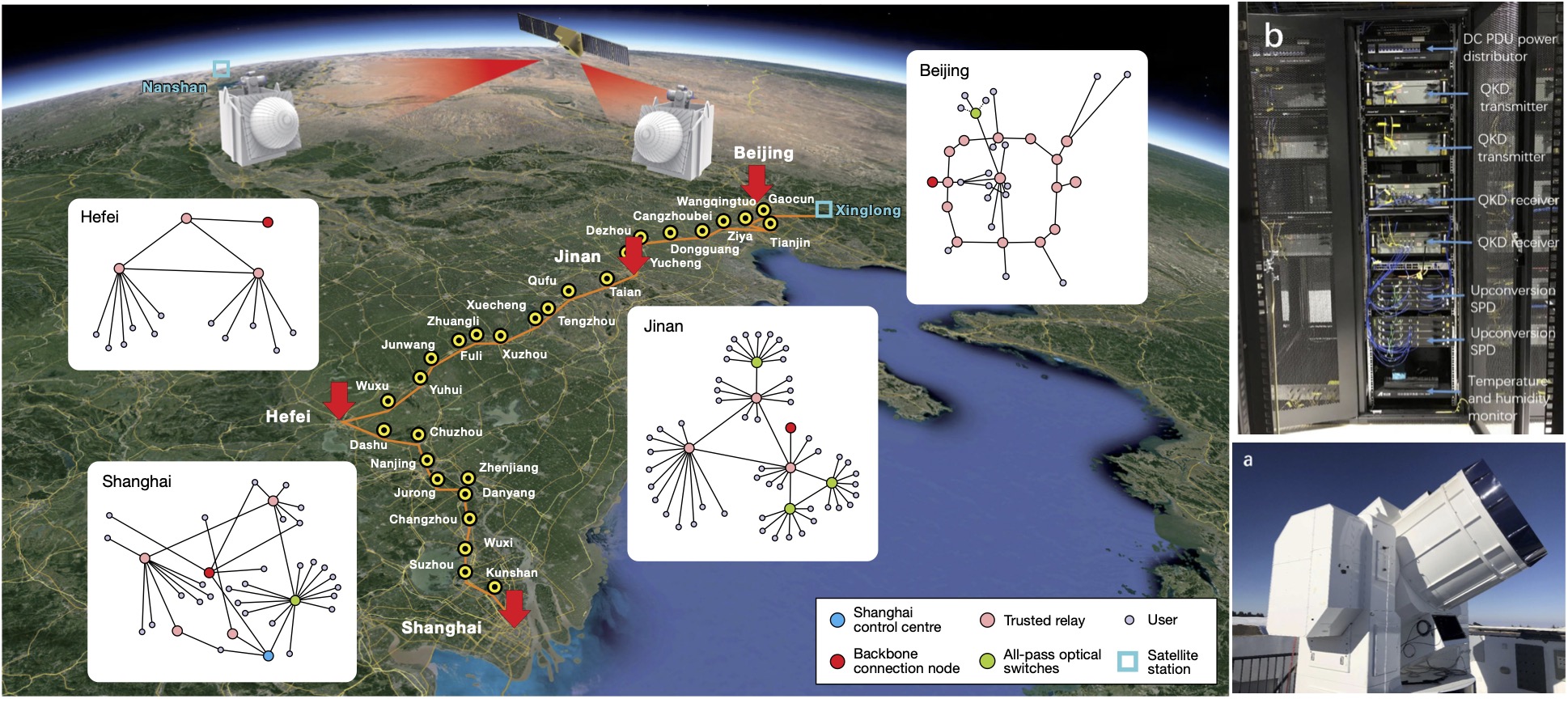}
    \caption{Overview of the Chinese quantum network, covering a total of 4,600 km \cite{chen2021integrated}.}
    \label{fig:china_nework}
\end{figure}

Nevertheless, ground-based quantum networks have been built in several locations, and most importantly Chinese researchers and companies connected multiple cities via ground and space-based quantum links, spanning many thousands of kilometers, see Figure~\ref{fig:china_nework}. However, it is important to note that these nodes are all classical repeaters, so-called 'trusted nodes'. These nodes generate a quantum key with multiple connected links, and generate joint keys via classical processing. Clearly, this classical data could compromise the security of the node if released, and therefore these nodes must be guarded.

Fibre-based quantum repeaters can distribute entanglement over up to a few thousand-kilometer range. Hence, to increase the operating distance, quantum satellites are required. In principle, for a global quantum network, we need to employ either LEO satellites combined with quantum repeaters and memories or GEO satellites \cite{simon2017towards}. Therefore, given today's technology, quantum communication satellites have a clear benefit when it comes to long-range quantum channels. 
 It will require several scientific breakthroughs before a purely ground based solution could make a satellite solution obsolete. And even in that case, given that vast areas of Canada are only sparsely populated and have very little infrastructure, a satellite solution might be the only viable option to connect such locations via quantum communication.

\clearpage

\section{Architectures}

\subsection{Satellites-based Quantum Communication}
    
\subsubsection{Entanglement Distribution with Satellites Comparison - Scenario Overview}
One of the most important prerequisites for a worldwide quantum network is the capability to perform quantum teleportation and entanglement swapping over large baselines \cite{Jennewein2021}.

Extending this truly fundamental quantum protocol to longer distances such as Earth-Moon, will expand validity tests of quantum mechanics and act as a precursor for quantum networks that can be useful for sensing, secure communications, dense-coding and interlinking of quantum computers in the context of Deep-Space missions. To date, only long-baseline passive teleportation \cite{Pirandola2015} has been demonstrated over long distances, including into space \cite{ren2017ground}.

Generally, the following link scenarios are relevant for entanglement distribution envisioned in the context of this paper:

\begin{figure*}[ht]
\centering\includegraphics[width=\linewidth]{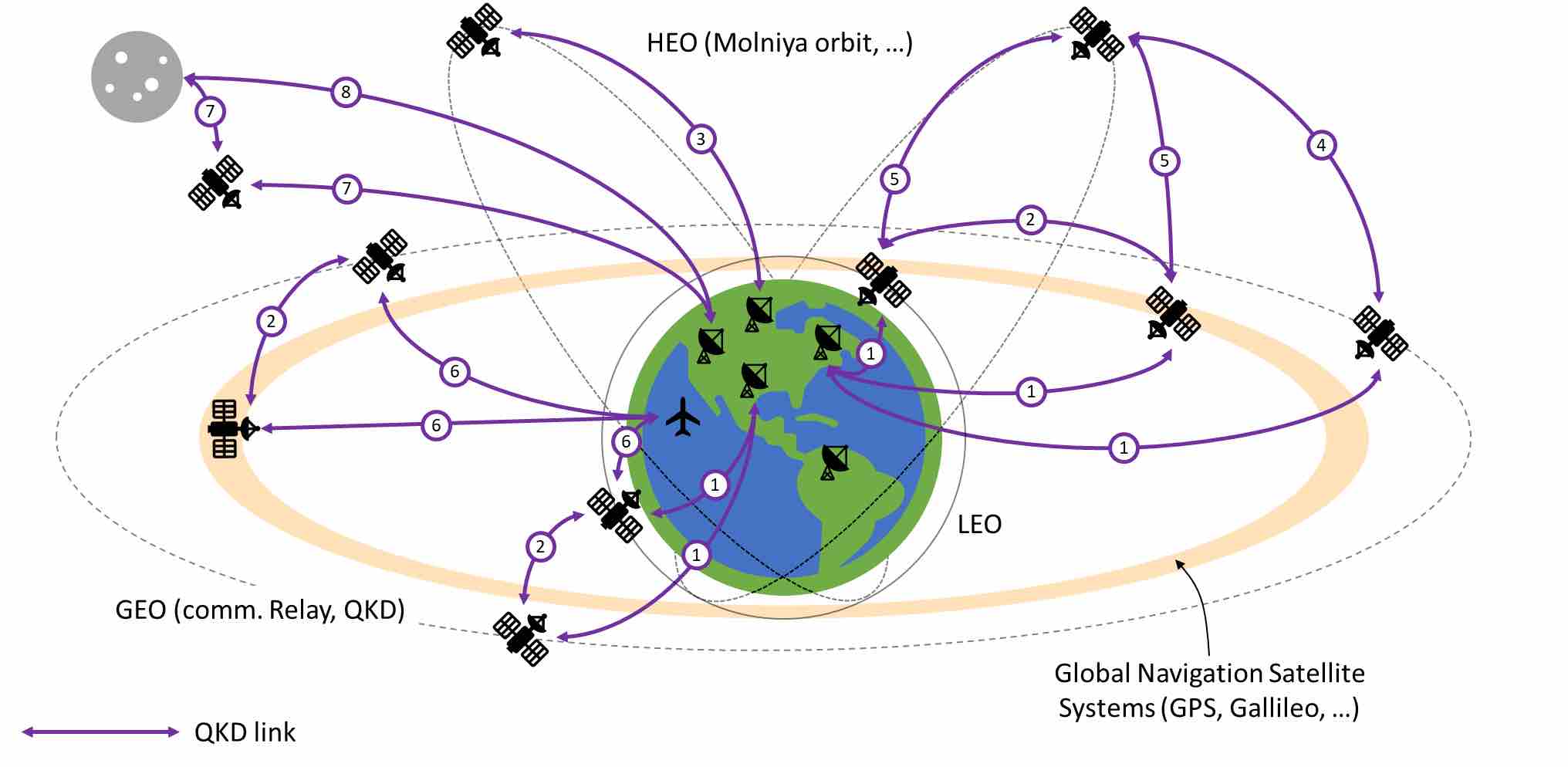}
\caption{Rough outline of the possible link scenarios for entanglement distribution} 
\label{fig:LinkScen}
\end{figure*}

\begin{table} [h!]
\centering
    \begin{tabular}{|c|c|c|} \hline
 \textbf{Link Num} & \textbf{Distances / \textbf{Orbits}} & \textbf{Description} \\ 
\textbf{(Fig. \ref{fig:LinkScen})} & \: & \: \\ [0.5ex] \hline
1 & LEO/MEO/GEO $\leftrightarrow$ Earth & Single-link, trusted-node QKD, \\ 
\: & \: & like QEYSSat 1.0 in LEO \\ \hline
2 & LEO/MEO/GEO $\leftrightarrow$ & Intersatellite links \\ 
\: & LEO/MEO/GEO & \: \\ \hline
3 & HEO $\leftrightarrow$ Earth & Useful orbit for \\ 
\: & \: & Canadian QGS locations \\ \hline
4 & HEO $\leftrightarrow$ GEO & Intersatellite links\\ \hline
5 & HEO $\leftrightarrow$ LEO/MEO & Intersatellite links\\ \hline
6 & LEO/MEO/GEO $\leftrightarrow$ HAP & Link avoids atmospheric loss \\ \hline
7 & SAT $\leftrightarrow$ Moon $\leftrightarrow$ Earth & Quantum repeater nodes on SAT \\ \hline
8 & Moon $\leftrightarrow$ Earth & Lunar Gateway \\ \hline
\end{tabular}
\caption{Quantum satellite link scenarios. LEO: Low Earth Orbit. MEO: Medium Earth Orbit. GEO: Geostationary Orbit. HEO: Highly Elliptical orbit. HAP: High Altitude Platform. SAT: Interplanetary Satellite. QGS: Quantum Ground Station.} \label{Tab:Links}
\end{table}

\begin{itemize}
\item Ground to Satellite Link (up-link)
\item Satellite to Ground Link (downlink)
    \item Intersatellite Link
    \item Constellations between Satellites
    \item Satellite to High Altitude Platforms (HAP)  
    \item Simultanous Double-links for entanglement distribution, including double photon and single photon protocols 
\end{itemize}

Table~\ref{Tab:Links} lists the different orbits we have considered in Figure~\ref{fig:LinkScen}.

\begin{table}
\centering
    \begin{tabular}{|p{4cm}|c|p{2.4cm}|p{4cm}|} \hline
 \textbf{Scenario} & \textbf{Link Num} & \textbf{Distances/} & \textbf{Comments} \\ 
 \: & \textbf{(Fig \ref{fig:LinkScen})} & \textbf{Orbits} & \: \\ [0.5ex] 
 \hline
 Quantum Key Distribution; Bell tests & 3 & GEO / MEO - Earth & $>10^8$ pairs per second, Narrow-band filtering (satellite is in daylight) \\
\hline
Quantum Clock Synchronisation  & 1,2,3,4,5,6 & LEO – MEO - GEO – Earth& Very narrow coherence time of the photons pairs.\\
\hline
Quantum Teleportation, (Entanglement Swapping), Network Ground-Satellite or satellite constellation& 1,2,3,4 & LEO (MEO) – Earth; GEO - Earth & $>10^9$ photon pairs, pulsed, variable delay (maybe); must be ‘narrow band’ in order to have suitable wavepacket timing. Scheme becomes much more feasible with quantum memory systems (e.g. atom-photon entanglement)\\
\hline
Relativistic scenario of the observation of entangled photons  & 2 & LEO-LEO/ or MEO-MEO  & $>10^7$ photon pairs, coherence time of the photons very short; requires ultra-fast detectors\\
 \hline
QKD using the Moon as a hub (8), e.g. quantum computing infrastructure on the moon  & 7,8 & Moon – Earth & $>10^9$ photon pairs per second \\
\hline
\end{tabular}
\caption{Overview of  some Entangled Photon Links: Application description and source requirements} \label{Tab:Scenarios}
\end{table}

\subsection{Orbits and Constellations}
For instance, the straightforward approach to achieve a pan-Canadian quantum network is to place an EPS on board of a satellite, and utilise a double-downlink to address the two different ground sites. However, a simplistic geometric analysis shows that for a ground distance of around 4,500 km (see Figure~\ref{fig:ground_distance}), and a minimum elevation angle of 45 degrees above the horizon, a single satellite EPS node must at least have an altitude of 4,200 km, which will require a MEO orbit (ca. 20,000~km). Alternatively, a constellation of LEO platforms or a single LEO satellite carrying a quantum memory with a storage time of several hours could be used to enable such long-range communications.


Furthermore, to assess the feasibility of the scenarios listed in Table~\ref{Tab:Scenarios}, a link model has been developed to estimate the average link loss for a $90^\circ$ ($\pm45^\circ$ zenith angle or the maximum angle allowed) satellite pass over a ground station or a lower-orbit satellite. This model assumes that the satellites are orbiting the Earth at the same inclination angle (passing the zenith of the ground station), and considers ideal apertures size for the receiver and the transmitter to obtain the corresponding flyby duration and link attenuation. (see Appendix~\ref{Appendix:link_analysis}). Figure~\ref{fig:time_required_for_QKD} (a) and (b) indicate the time required for a successful QKD transmission using 785nm over a (a) single link WCP and (b) double link EPS, based on recording a certain number of events. These figures declare the practicability of the scenarios regarding today's quantum technology and suggest the potential improvements required to enhance the link budget in high-loss scenarios.

\begin{figure}[ht]
    \centering
\begin{subfigure}{0.8\textwidth}
\centering
\includegraphics[width=\textwidth]{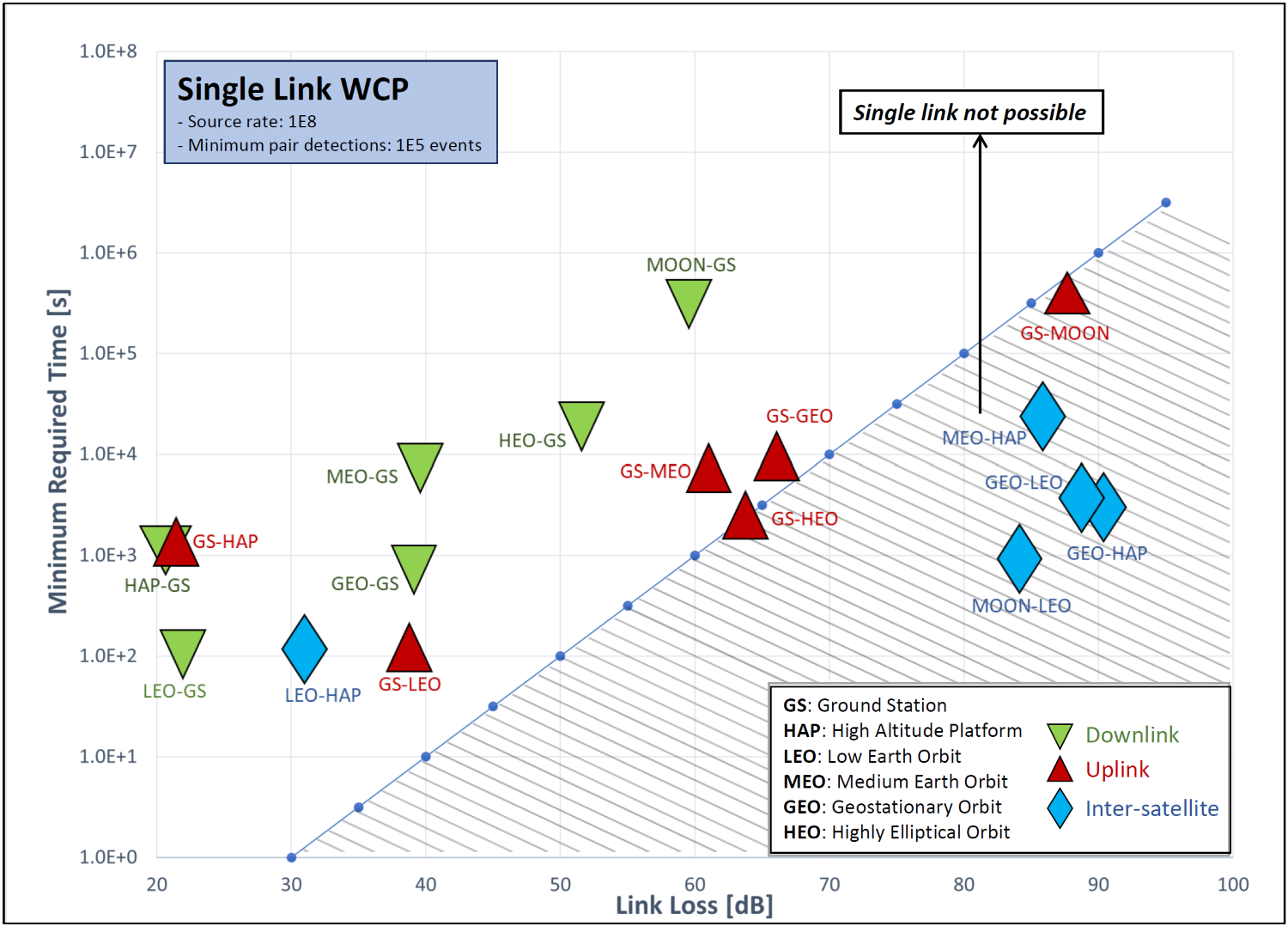} 
\caption{}
\end{subfigure}
\begin{subfigure}{0.8\textwidth}
\centering
\includegraphics[width=\textwidth]{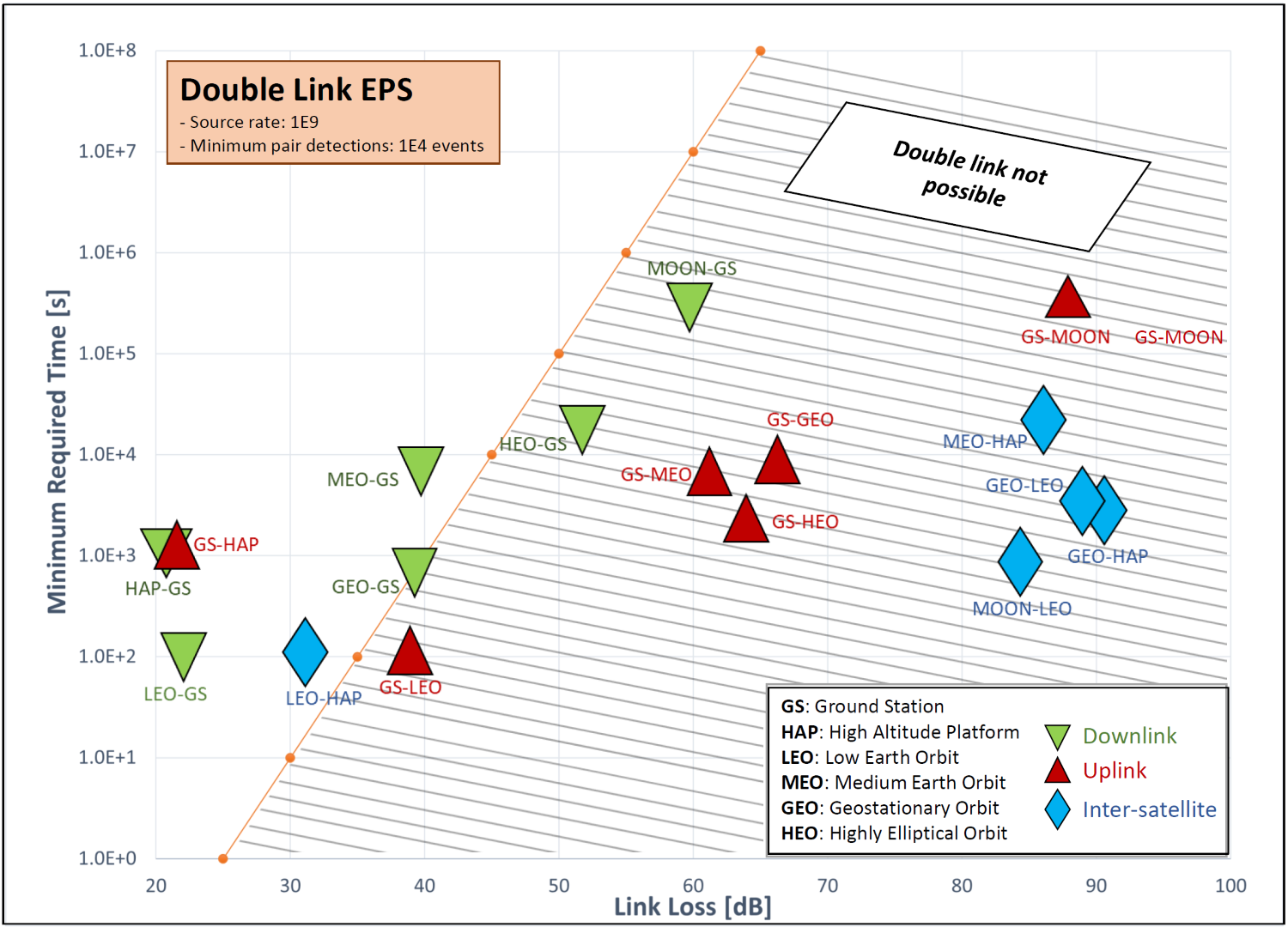} 
\caption{}
\end{subfigure}
    \caption{This chart outlines the time needed to complete a QKD transmission using 785nm over a high-loss link. The shaded areas indicate where a QKD single-link, and probabilistic double-link (EPS) are not possible. The assumption is that once a certain amount of events are recorded, the link is successful. Note that a memory-assisted double-link has the potential to perform as well as a single-link, depending on memory lifetime, due to the repeater enhancements. Each icon on the plot shows a free-space link with a $-45^\circ$ to $+45^\circ$ zenith angle (or the maximum angle allowed) coverage and the corresponding time and link attenuation with its ideal aperture sizes as listed in Table~\ref{fig:Link_Analysis_Dynamic}.}
    \label{fig:time_required_for_QKD}
\end{figure}

\subsubsection{Link Availability and Constraints}
\begin{figure*}[ht]
\centering\includegraphics[width=0.8\textwidth]{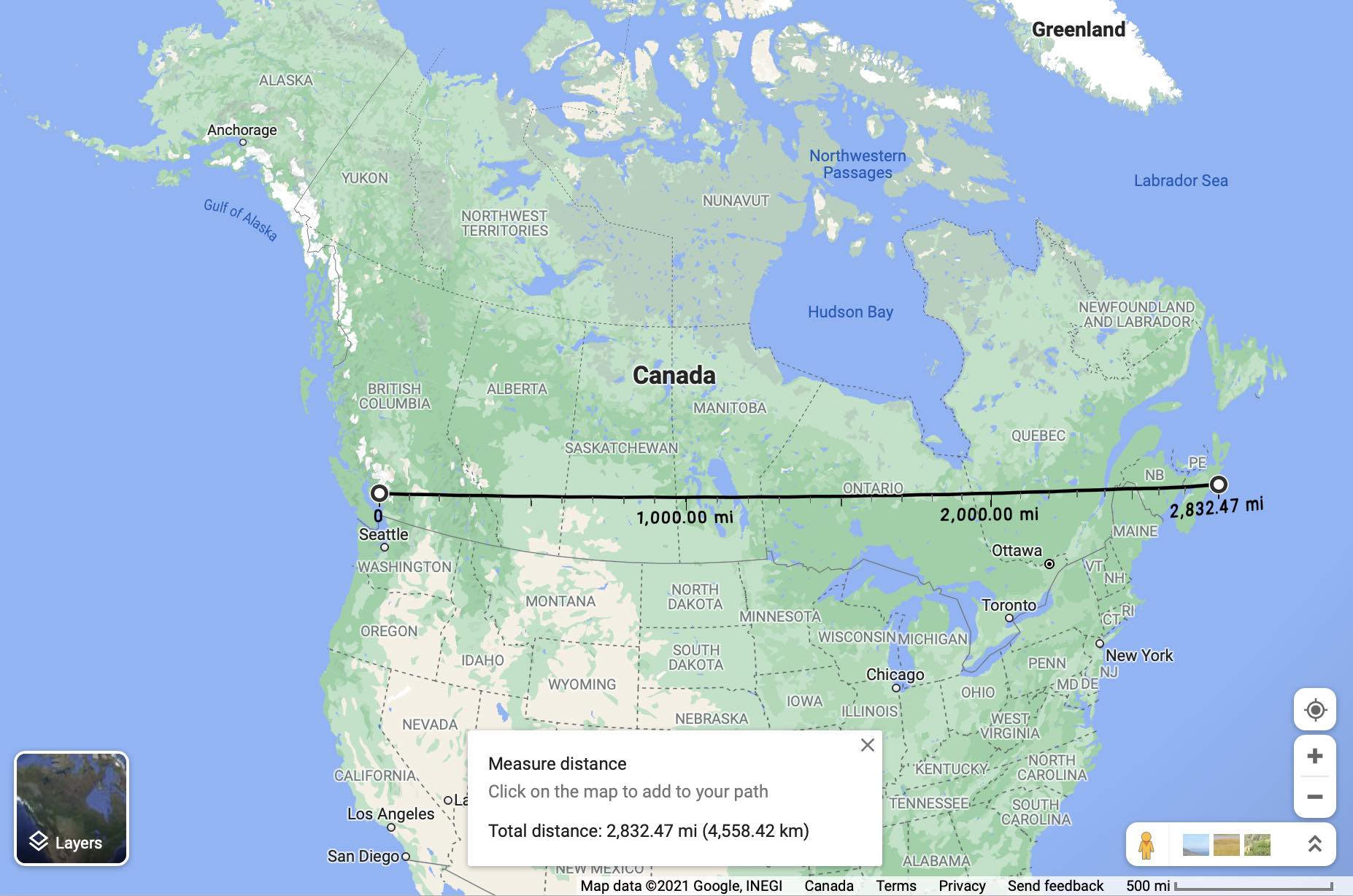}
\caption{Estimated ground distance for a pan-Canadian link from East to West coast.} \label{fig:ground_distance}
\end{figure*}

First off, the aim to achieve Canada-wide networking from a trusted and coherent space platform poses some interesting challenges given the size of Canada, as shown in Figure~\ref{fig:ground_distance}. One approach is to place a communication node at high-altitude orbits in space, to be accessible from two (or more) ground sites at the same time. GEO orbit is a good example that is used in telecommunication networks. However, given the higher latitudes of Canadian regions, other options might be more useful, such as highly-elliptical orbits, as shown in Figure~\ref{fig:LinkScen}.

Having a constellation of LEO or MEO satellites that are interlinked with each other, as well as the ground, is another approach. However, the coordination of multiple LEO and MEO orbits is certainly a formidable challenge, but is already addressed by large-scale classical networks such as Starlink or OneWEb.

The third approach, which may ultimately be the most scalable option, is to have a fully functional quantum repeater node on board the satellite, such as in LEO orbit. This system is very appealing from a conceptional perspective, but requires highly performing quantum memories, and also long-term storage (e.g. error correction) of quantum information, such that the coherence can be maintained over at least one orbit (ca. 90\:min for LEO).

\subsubsection{Light Pollution and Daylight Operation}

Solar background, light pollution and atmospheric turbulence can reduce the optical link efficiency due to the increased background noise, beam distortion and false detections. Atmosphere structure, $C_n^2(h)$ changes over different altitudes and weather conditions that cause fluctuations in the refractive index of the link path. Thermal stability during the day and night is another factor that impacts the strengths of these fluctuations. At low altitudes, the atmosphere has a thermal exchange with the Earth surface which is called the atmospheric boundary layer (ABL), and extends to 1-2 kilometres \cite{Andrews:2004aa}. In this region, the temperature gradient varies day and night. During the daytime, thermal plumes occur which guide warm and less dense air to rise. However, at night the Earth is surrounded by cold air that is more stable and less likely to rise. Thus, the thermally neutral condition happens near sunset and sunrise. Atmospheric structure parameter, $C_n^2(h) = C_n^2(h_0) (h/h_0)^{-b}$ changes with altitude by the power of 4/3 for unstable conditions (day) and 2/3 for neutral or stable conditions (night). 
As a result, a plot of path-averaged values of $C_n^2$ over 24 hours, near the ground (1.5 m above the Earth), shows a diurnal cycle that reaches its maximum in the mid-day, is almost constant during the night and has its minimum near the sunset and sunrise. Therefore, near the sunset, which is thermally stable and has less sun reflection than the sunrise has been suggested as the ideal time for free-space optical communication \cite{LaserBeamPropagation, PassiveInfraredDetection}.

Furthermore, the concepts of quantum LIDAR and quantum illumination to circumvent light pollution and daylight operation seem promising. Coincidence detection from entangled photonic pairs provides a scheme for a system more robust to background and even jamming signals \cite{zhao2021quantum}. Quantum LIDAR is enabled by refined coincidence counting electronics, such as those developed at the University of Waterloo \cite{UQDevices}.

\subsubsection{Wavelength Considerations}
Wavelength selection should be based on different aspects:

\begin{itemize}
\item \emph{Technological constraints:} Some wavelengths can be easily produced while others need to be generated through up/down conversion, OPO, OPA or other non-linear processes. The energy requirements for equivalent optical power can widely vary, and a global network would require readily-available high-efficiency sources, or the development of such sources.

\item \emph{Flexibility on the wavelength:} Different meteorological and atmospheric conditions lead to a variety of relative atmospheric transmittance through the spectrum. Quantum converters to make use of a different wavelength for propagation in different conditions could lead to the use of a single, wavelength-tuned emitter/receiver. For instance, sources/receivers in the communication bands, with converters to and from LWIR or THz bands for propagation in degraded environments could enable more general usage. However, using such long wavelengths for quantum communication is likely to be challenging because of blackbody radiation.
\end{itemize}

An optical link from the ground to a satellite with light at a wavelength between 700\:nm and 1600\:nm has challenges. For daytime operation, there is the problem of daylight filtering, which is discussed in Section~\ref{sec:DayFilter}. More significant for a reliable operational link is operation in adverse weather conditions. The geographic location of ground stations will have a significant impact. Any atmospheric phenomenon that reduces visibility at the operation wavelength will reduce the secure key rate, if not prevent it altogether. Haze, fog, cloud cover, smog, rain, and snow are all potential link problems, as is turbulence if no adaptive optics (AO) is implemented, or if the AO does not have the capacity to correct all distortions. Depending on the geographic coordinates of a ground station, different types of atmospheric particulates will generate the link budget loss. Maritime fog is not the same as fogs in the interior lands, and arctic fogs are different from mid-latitude fogs. It will be critical to choose a location where these atmospheric phenomena are at a minimum.

\begin{figure}[ht]
    \centering
    \includegraphics[width=\textwidth]{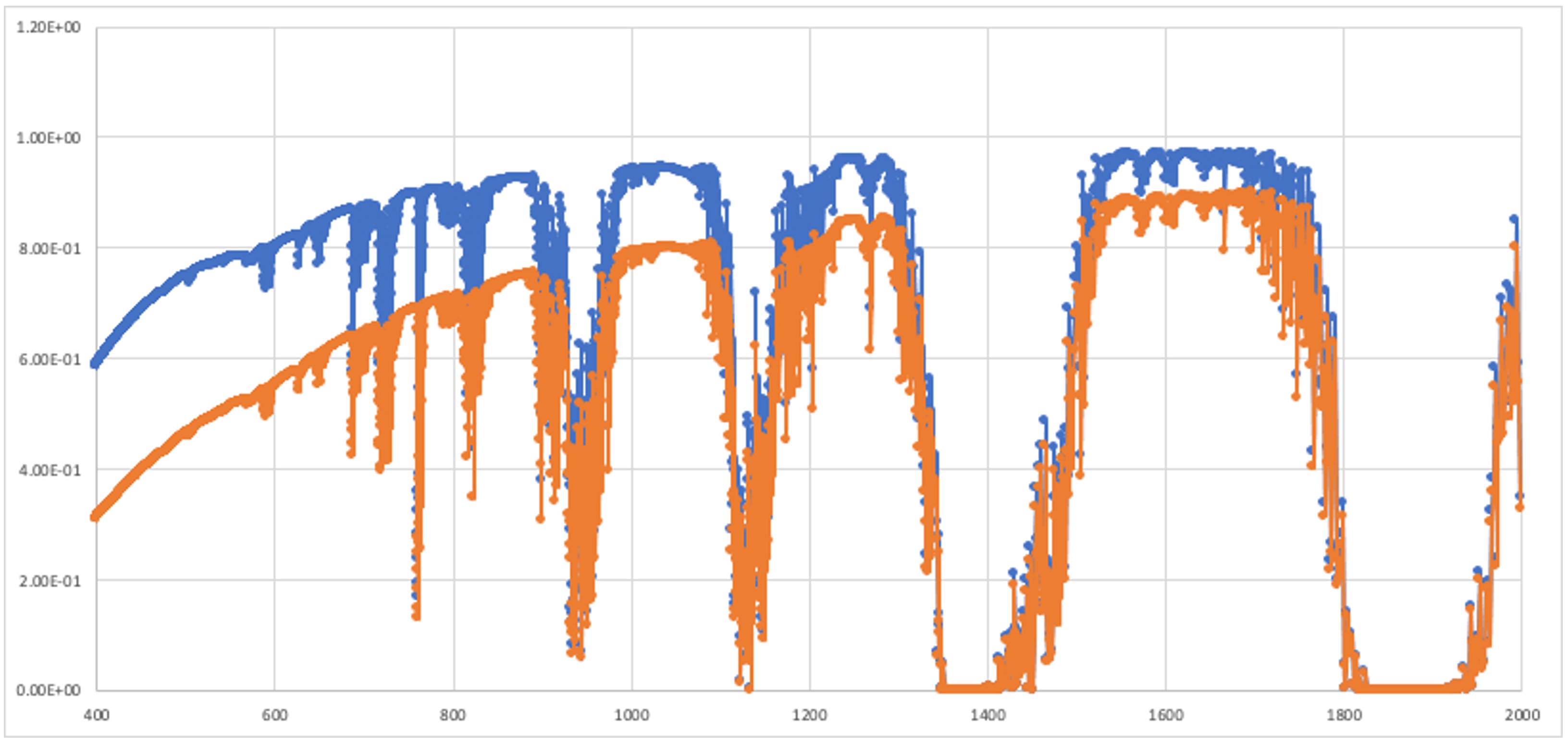}
    \caption{Calculated atmospheric transmission for a vertical pass from ground to space, where the aerosol compositions are rural (blue) and urban (orange). Here the x-axis shows the wavelength in nm. For slanted angles at Zenith angle $z$, this attenuation will be scaled according to $\sec{z}$ of the air mass (see  Appendix~\ref{Appendix:link_analysis}.)}
    \label{fig:Atmospheric_Transmittance}
\end{figure}

There are no significant differences when it comes to transmission losses for wavelengths from 700\:nm to 1600\:nm in the various fogs. However, there could be significant differences regarding backscattering radiation. There could be requirements in the choice of the wavelength for the uplink versus the downlink because of backscatter. Also, smaller wavelengths ($<$400\:nm in the UV for example) can be made to diverge less than longer wavelengths for the same emitter diameter, and thus potentially have less link loss due to beam size at the receiver. However, the very low wavelengths have more Rayleigh scattering losses, as shown in Figure~\ref{fig:Atmospheric_Transmittance}. 

The optical quality of surfaces is  more of an issue at shorter wavelengths. These wavelengths will not transmit through clouds, except very tenuous clouds. A high-altitude ground station (or high-altitude platform) could thus be a very interesting choice, if possible. The fibre network or another quantum link would then need to transfer the quantum information between the HAP and the ground. Even though the differences are not huge, in general, a more detailed analysis of specific fogs or clouds would be necessary to optimize the wavelength selection. Although it is widely believed that 1550\:nm light travels with less loss than 800\:nm light in fogs, this greatly depends on droplet size distribution, and in most real-life cases is not true. Very small droplet distributions are not so common. And 800\:nm light goes through water more easily than 1550\:nm light. For high transmission through fogs and clouds, microwave transmission down to the mm-scale would be preferable, but obviously there are critical technological challenges with operating at these wavelengths. 

Quantum sources are readily commercially available at wavelengths of 780\:nm, 810\:nm, 840\:nm, 1360\:nm and 1550\:nm. There are several commercially available entangled photon pair sources available,  including one provided by Oz Optics, in Ottawa. However, for daylight operation, a strongly filtered receiver and a custom-built source will most probably be necessary. For low spectral linewidth optical sources, molecular absorption is a concern. Absorption lines of water and CO$_2$, in some cases, are to be avoided.

\subsection{Network Design}
A worldwide quantum internet might be built using satellites and ground-based quantum communication channels. In general, whether we need to use quantum satellites alone or in combination with quantum repeaters depends on different factors such as the type of satellites, i.e., Low-earth orbit or geostationary, number of available satellites, technologies that satellites are equipped with (e.g., quantum memories), and potential applications.

\subsubsection{Satellites equipped with Entangled Photon Sources }
In the simplest case, a low-earth orbit satellite equipped with a photon pair source can be used to send entangled photons to the two distant ground stations where quantum non-demolition (QND) detectors allow for heralded loading of quantum memories. Entanglement can then be distributed between the neighboring ground-based nodes using entanglement swapping \cite{boone2015entanglement}. In this architecture, the satellite should be visible from both ground stations. Hence, the distance between the stations is limited to a few thousand-kilometer. 

\subsubsection{Satellites equipped with Quantum Memories}
To overcome the mentioned limitation, satellites should be equipped with quantum memories. Hence, in a more advanced architecture, a quantum memory and an entangled photon pair source are placed inside the satellite. One photon will be stored in the memory, and the other will be sent down over the first ground station. The satellite then flies to the proximity of the second station, where the stored photon will be sent down. For this architecture, the distance between the two ground stations is mainly limited by the storage time of the memory.

Using satellite links, the ultimate goal could be to entangle satellites using free-space optical repeater links. In this architecture, the middle satellite is equipped with a pair of memories, receivers, and QND measurement devices (or any other setup that herald a successful transmission of photons to a memory). In contrast, the first and last satellites are equipped with EPS. To distribute entanglement, entanglement swapping can be performed between the retrieved photons from the quantum memories of each satellite. As a result, only two of the channels are atmospheric, i.e., the first and last one, and the rest are space-based channels (i.e., inter-satellite links) established between satellites \cite{gundougan2020space}. Hence, the loss rate would be lower when satellites are equipped with quantum memories, and the weather condition would be important only in the first and last atmospheric links. 

Note that, it is also possible to avoid using QND by different means. For instance, one can use an additional BSM for each memory in the middle station \cite{panayi2014memory}. In this case, a successful loading of the memory can be heralded by performing the BSM between the incoming photon and a photon that is already entangled with the memory. Directly performing the BSM between the emitted photons is another way of eliminating the need for QND. In this situation, the first and last satellites are equipped with quantum memories and EPS. One of the photons from each source will be directed to the middle satellite for the BSM, while the other one will be stored in the relevant memory. These techniques can be used in up-links geometry as well \cite{gundougan2020space, boone2015entanglement}. Needless to say that carrying out a BSM between arrival photons can be challenging, as it requires real-time synchronisation of photon arrival times. Correcting for the turbulence-induced distortion of spatial photon modes (particularly in the uplink scenario) is also an important task (see Sections \ref{Technical-Feasibility} \& \ref{sec:TechBottlenecks} for more information).



\clearpage

\section{Critical Technologies}
In this section, we present an overview of what quantum technologies exist now, and identify technologies we deem necessary for future missions.

\subsection{Quantum Sources}
\label{sec:Sources}
The development of scalable non-classical light sources is critical for realizing quantum networks. Many quantum communication schemes are based on the following emitters.

\subsubsection{Weak Coherent Pulse Sources}
\label{sec:WCP}
At this time, most QKD solutions utilised photon emitters based on attenuated laser, usually called WCP due to the simplicity and low cost of the solution. 

However, this is contrary to the requirements of the original protocol published in 1984 by Charles Bennett and Gilles Brassard (called BB84) \cite{Bennett84a}, as its security is based on true single photon pulses transmitted by Alice. But because single photon sources are difficult to implement, WCP provides a good alternative - as long as a special protocol is used that will properly account for the  multi-photon emission, this is called the `Decoy-state protocol' \cite{PhysRevLett.94.230504}. 

The decoy-state protocol uses multiple different energy level for the emitted photon pulses, usually the signal level is an average photon number of $\mu = 0.5$, while the two decoy state levels are $\nu_1 = 0.1$ and $\nu_2 = 0.0$. The protocol will compare the received photon counts rates for the different levels and can then extract a secure, single-photon contribution from the received key in processing. 

The decoy-state-WCP protocol for QKD is therefore, less efficient than single-photon and also less noise resilient, however these disadvantages are often outweighed by the simplicity and low cost of a laser source.

However, in a space-ground QKD application, true single-photon emitters may have an advantage again, in particular, because a satellite link has limited time duration, high losses and potentially high background noise.

Several small-scale WCP sources have been demonstrated, most notably the UBristol's handheld QKD source \cite{Zhou2019} with 4 small LEDs which could be a good candidate for space integration, and indeed, the quantum downlink module for the Reference-Frame Independent (RFI) quantum communication for satellite-based networks (ReFQ) project is based on a UBristol design. Other examples of very high-rate chip-scale solutions such as the Giga-Hertz clocked system from Toshiba 
operating at 1550~nm, could also point towards satellite deployment. 

\subsubsection{Entangled Photon Sources}
\label{sec:EPS}
A perfect EPS should generate highly efficient, maximally entangled, and indistinguishable photons on demand. So far, spontaneous parametric down-conversion (SPDC) in nonlinear crystals is the most extensively used method for generating pairs of entangled photons \cite{kwiat1995new, kwiat1999ultrabright,wang2016experimental}. SPDC process occurs when one photon from a pumping laser passes through a second-order $\chi^{(2)}$ material and converts into two photons of lower energy. The same process is known as spontaneous four-wave mixing when materials with third-order $\chi^{(3)}$ are used \cite{caspani2017integrated}. SPDC-based sources produce photons in a probabilistic process following the Poissonian statistics. Hence, there is a non-zero probability for multiple photon pairs being generated during a single excitation cycle. On the other hand, these sources generate photons with a very high degree of entanglement fidelity and photon indistinguishability.

Alternative photon sources are those based on the cascaded radiative decay mechanism in semiconductor quantum dots \cite{benson2000regulated,akopian2006entangled}. These sources are guaranteed to only produce single pairs with very high efficiency. This advantage is especially important for some quantum repeater protocols \cite{sangouard2011quantum}. Over the last few years, there has been great progress in addressing some of the challenges of semiconductor sources of entangled photons. In particular, improving entanglement fidelity by eliminating the fine structure splitting of the neutral exciton state \cite{huo2013ultra, chung2016selective}, and increasing the extraction efficiency by incorporating quantum dots in a micropillar cavity \cite{dousse2010ultrabright} have been demonstrated. As a result, these sources can produce high fidelity and efficiency entangled photon pairs that are yet indistinguishable \cite{wang2019demand}. Recent demonstrations for practical fibre-coupling of such devices lead the way to stable, high-brightness sources \cite{northeast2021optical}. 

\textbf{Consideration of source wavelength} 
The selection of the wavelength depends on several considerations, the most important being: atmospheric transmission windows, beam diffraction and single photon detection sensitivity. We will identify the baseline for the EPS through trade-offs between
\begin{enumerate}
    \item Atmospheric absorption (given transmission windows)
    \item Diffraction loss (improves with shorter wavelengths)
    \item Single-Photon detectors (depends on detector technology) 
    \item Available wavelength depends on the particular source process
\end{enumerate}

\subsection{Single Photon Detectors}
\label{sec:detectors}
The ability to accurately detect single photons is crucial for several applications, including quantum communication, sensing and computation \cite{bouwmeester2000physics,bennett2000quantum, kimble2008quantum, zhong2020quantum}. For example, the security of QKD is based on encoding quantum information in single photons so both the generation and detection are important. Single photon detector technology for VIS-1000 nm is well established (Si technology), while single photon detectors for telecom wavelength (1000~nm - 1600~nm) have been continuously improved over the past decades, achieving higher efficiencies at faster detection rates, with lower probability of dark (false) counts and after pulses.

\begin{table}[h]
    \centering
    \begin{tabular}{|c|c|}
    \hline
Parameter & Value (typical, or range)  \\
\hline
Photon Detection Efficiency (PDE) & $> 50$~\% \\
\hline
Timing resolution  & $< 500$~ps \\
\hline
Dark Count Rate (DCR) & $< 1000 $~cps \\
\hline
After pulse probability & 5~\% \\
\hline
Saturating  count rate & $10^{6}$~cps to $10^{8}$~cps\\
\hline
    \end{tabular}
    \caption{Overview of typical photon detector parameters for devices used in long distance QKD.}
    \label{tab:my_label}
\end{table}

The most common detector technologies for quantum communications are Single-photon Avalanche Diodes (including Geiger-mode APD and SPAD), both in Silicon and Indium-Gallium-Arsenide (InGaAs) technology. The better performing Silicon devices have good efficiency, relatively low dark noise, and are free-running. These have already been used in Space multiple times \cite{abshire2005geoscience, SPEQTRE}, and are also well characterised for radiation damage \cite{Anisimova2017,Lim2017,DSouza2021}.
However, they only operate in the spectral range of visible to 1000~nm. The much preferred telecom wavelengths, including 1310~nm and 1550~nm, however, are usually observed using InGaAs APDs, which have drastically inferior performance, and to our knowledge have not been utilised in space yet. An important drawback for a space-to-ground channel is that the InGaAs devices would require a gated operation, which would be demanding for a fluctuating free-space channel. 

An important new technology in particular for detecting telecom wavelength photons is superconducting nanowire detectors (SNSPD), which have far superior detection efficiency, speed and signal-noise than any of the APD devices. While such systems are widely available from several vendors, they  require cryogenic operations (2.5 K), and space operation is therefore a challenge. Furthermore, these devices are typically coupled via a single-mode fibre, which makes the coupling of a free-space beam a challenge, in particular for a ground based receivers.

Another rapidly emerging single-photon detector technology is array devices, which have the advantage that they could be used to extract wave-front and tracking information from a quantum signal, while detecting the actual quantum information \cite{Sajeed:2021aa, Donaldson:2021aa}. However, their overall efficiency to detect a single photon is currently very limited due to the pixel fill-factor, as well as the shallow depth of the absorptive material. But such systems are under development, including with QEYSSat Science Team collaborator at the University of Sherbrooke. Array detectors based on SNSPD have also been developed \cite{Wollman:2019aa}.

Photon number resolving (PNR) photon counters are of general interest in quantum optics, and are helpful for some quantum applications. In quantum communication and quantum information processing, the photon-number-resolving function could be beneficial for some protocols including quantum repeaters and linear optics quantum computing. Their use for quantum communication could be heralding of single photons using an EPS, however the benefit for long range quantum communications is not clear unless the detection efficiency is greater than 90\%.   PNR detectors can be implemented using APD in a linear photon counting mode \cite{Thomas:2010aa}, and essentially measuring waveforms at few photons light levels. One of the best technologies are cryo-cooled bolometric detectors using transition-edge sensing (TES) \cite{cabrera1998detection} has been demonstrated with discrimination up to 20 photons in the same spatiotemporal mode \cite{harder2016single} with efficiency reaching 95\% in the telecom wavelength range \cite{lita2008counting} and up to 98\% at 850nm \cite{fukuda2011titanium}. TESs can also be optimized to any wavelength in the visible and IR ranges. A significant shortcoming of TESs remains their slow operation, currently at least two orders of magnitude slower than what would be considered practical in photonics. Optical TESs do not appear to be commercially available at this time, and given their operating constraints and also speed of operations their use for space-to-ground quantum communications is not clear at this time.


\subsubsection{Photon Multiplier Tubes}
Traditionally,  single photons were detected using photo multiplier tubes (PMT). It is still the case in a large number of applications. PMTs are very low noise detectors. But they are electronic tubes and they require high voltages. Their high gain also makes them susceptible to cosmic rays. Lastly, they are not very efficient in the 700 – 900~nm window. Best in class PMTs have less than 1\% efficiency at 900~nm, have $<5$~\% at 700~nm (but up to 40\% at 400~nm). Therefore, typically  PMT would only be considered for quantum communication applications at shorter wavelengths (400 to 600~nm), such as considered for the IQC proposed NanoQEY concept \cite{Jennewein:2014aa}. Advantages that speak for the use PMT are their large active area (typically a diameter of 5~mm), as well as a very good radiation tolerance, where an IQC-led radiation test showed no observable change with PMT dark counts \cite{Anisimova2017}.

\subsubsection{Avalanche Photo Diodes - APD and SPAD}
The most common single photon detectors for the 700 – 1100~nm spectral window and the near-infrared (up to 1600~nm) are avalanche photodiodes (APD) operated in Geiger mode (a breakdown condition that generates easily measured high currents).  They are based on semiconducting materials such as Si, Ge, GaAs and InGaAs \cite{migdall2013single}. These Geiger-mode detectors can only detect one photon at a time, and they must be quenched to detect another photon. There are thus dead times and a maximum photon count rate due to quenching. Biasing over the photodiode breakdown voltage to get to the Geiger mode generates a large number of dark counts, photon detection like events that are the greatest noise contributors, and are deleterious to effective quantum communications, InGaAs (1500 – 1600~nm) much more so than Si (700 – 900nm). These detectors show detection efficiencies in the 10 to 65\% range depending on the amount of bias over the breakdown voltage. The higher the overbias, the larger the amounts of dark counts. Si APDs covers the range from 300 to 1100~nm, InGaAs typically from 800 to 1700~nm. Extended range of InGaAs can reach up to 2500~nm. SI and InGaAs APDs do not need to be cryocooled for operation, although dark counts are significantly reduced at temperatures of ~90°K. In order to have the possibility of counting more than one photon per time bin, arrays of small Geiger mode APDs are built. While one APD is being quenched, the others in the array can detect a photon. These are called Si photomultipliers (SI-PMT) or Multi-Pixel Photon Counters (MPPC). There are Canadian manufacturers of semiconductor detectors, such as Excelitas (in Vaudreuil Qc), as well as teams at the University of Sherbrooke working on custom SPAD-Array ASIC with built-in quantum analysis. Another advantage of APD's is that unlike PMTs, there are multiple developers and manufacturers of solid-state, semiconductor detectors.

More recently, a new material composition has been of interest, mercury-cadmium-telluride (MCT). APDs made of MCT can be tailored to respond to the wavelengths of interest, although they have been developed essentially for the infrared. MCT APDs can have very high gains, and can be used for photon counting, without being in Geiger mode. They are used for linear mode photon counting, as are the PMTs. They thus generate much fewer dark counts. On the other hand, they are necessarily cryocooled. Both NASA and ESA have MCT APD programs for LIDAR applications in space \cite{krainak2016novel}. MCT APD was used for GHz detection scheme within the ESA program \cite{pes2021reaching}.

Finally, a remaining limitation of APD photon counting technologies is the lack of photon-number-resolving (PNR) capability \cite{slussarenko2019photonic}. Some APD operating circuits have also demonstrated photon-number resolution\cite{Thomas:2010aa} using linear readout.

\subsubsection{Superconducting Nanowire Single-Photon Detectors}
Superconducting nanowires single photon detectors (SNSPD) have recently emerged as the state of the art in single photon detection technology \cite{esmaeil2021superconducting, marsili2013detecting, reddy2019exceeding, chang2021detecting} surpassing APDs in terms of detection efficiency (reaching close to 99\% efficiency) \cite{chang2021detecting}, speed and dark count rates. They are however more complicated to operate than APDs, as they require cryogenic temperatures (typically 0.8~K to 3~K depending on the superconducting material). SNSPDs can be fabricated for different wavelengths by embedding the appropriate superconducting material in a suitably designed optical cavity. SNSPDs can be designed to be PNR detectors. Some SNSPD schemes have been investigated \cite{mattioli2015photon}.

SNSPDs are better suited for applications where high detection efficiency, low dark count rates, and low timing jitter are required by a protocol, and where good APD's are not available, such as for telecom wavelengths, 1550~nm. SNSPDs have enable several noteworthy quantum information experiments \cite{esmaeil2021superconducting}, such as several record-breaking quantum communication demonstrations \cite{boaron2018secure, yu2020entanglement, chen2020sending}, loophole-free test of local realism based on Bell experiment \cite{shalm2015strong}, and large scale boson sampling \cite{zhong2020quantum}.

There has been a lot of effort in improving the manufacturing process to make better detectors. In particular, three independent groups have reported recently on $>98\%$ system detection efficiency based on different material systems: MoSi with distributed Bragg reflectors \cite{reddy2020superconducting}, dual-layer NbN meanders \cite{hu2020detecting}, and NbTiN with a membrane cavity \cite{chang2021detecting}. 

Black-body radiation can be a major noise source creating false detection events. Every dark count is adding noise to the measurement, which affects the maximum achievable secure QKD rate or degrades the quantum state being teleported. However, the trade-off of using SNSPDs is mainly around the engineering challenges, such as operating ultra-cold detectors (mK) in space. The lowest dark count rate reported so far is $10^{-4}$ per second \cite{shibata2015ultimate, hochberg2019detecting}. 

High time resolution is another noticeable advantage that SNSPDs offer over SPADs due to the nature of their operation. The timing jitter of the system can be significantly reduced by engineering the length, width and meander pattern in specific ways. For example, the best reported time jitter has been reported for short straight nanowires made of NbN ($<3$\:ps) \cite{korzh2020demonstration} and WSi (4.6\:ps) \cite{korzh2018wsi}.

SNSPDs have been demonstrated to operate at a large range of wavelengths from x-ray to mid-infrared, offering more flexibility in their detection range compared to SPADs. SNSPDs are powerful tools in astronomy and deep-space applications such as exoplanet transit spectroscopy, deep-space optical communication, and aid in the search for dark matter. These detectors may be suitable to measure sub-GeV particles on Earth from dark-matter particles from the halo of the milky way \cite{hochberg2019detecting}. The stringent power, size and weight requirements on satellites make it difficult to achieve deep-space links. Better detectors like SNSPDs have the potential to increase the optical communication bandwidth to satellites and improve link quality \cite{robinson2006781, hemmati2006deep, grein2011design, messerschmitt2020challenges, robinson2011overview}. However, a major challenge with using these detectors in quantum space applications is relying on large amounts of data in and out of cryostats operating at mK temperatures as classical approaches like coaxial lines will limit the bandwidth. 

Waveguide-integrated SNSPDs are another technology currently being developed, with on-chip quantum optics experiments being demonstrated \cite{khasminskaya2016fully, elsinger2019integration}, including on-chip two photon quantum interference \cite{schuck2016quantum}, and on-chip secure quantum communication \cite{beutel2021detector}.

Polarisation-based QKD requires the detection of the polarisation state of incoming photons. By themselves, APDs are not sensitive to polarisation. However, SNSPDs have recently been proposed for polarisation resolved single photon detection \cite{guo2015single}. They are limited to linear polarisation detection and do not appear to be commercially available at this time. polarisation resolved single photon detectors sensitive to circular polarisation states would expand the possibilities of polarisation-based detection from linearly polarised single photon detection potentially provided by SNSPDs.

Another important new direction for SNSPD is multi-mode coupled devices, or even arrays of thousands of pixels\cite{Wollman:2019aa}. This is important as the coupling of free-space optical beams to the typical single-mode fibre can be very lossy. NASA/JPL is developing even free-space coupled SNSPD as a detector in their Deep-Space Optical  Communication (DSOC) receiver using 5-meter telescope\cite{JPL_DSOC:2021aa}.

\subsection{Space use vs. Ground use of Photon Detectors}
In space-to-ground quantum communication, it is important to consider if the detectors will be used in the space platform or on the ground.

In summary, for {\bf space use},  single-photon avalanche diodes (APDs), offer low-power operation that is well suited for space use, as well as ground or airborne operation, and can operate with moderate cooling (-20 to -100 C). APDs are readily used in applications like sensing and QKD, however, for applications using telecom wavelengths, APD's, unfortunately, have some limitations due to higher dark count rates and lower detector efficiency, and high afterpulse probability as compared to SNSPDs). SNSPD will be difficult to use in Space given the cryogenic operation requirement.

For {\bf ground use}, the size-power-mass requirements are obviously relaxed, but other considerations must be considered. While SNSPD clearly provides the best possible performance across all wavelength ranges, a specific challenge is that the coupling to the single-mode fibre input of the SNSPD will cause additional losses due to atmospheric turbulence, in particular, if a larger aperture ground station is used ($> 0.5$~m ). Possible approaches are to use multi-mode fibre coupled SNSPD, or even free-space coupled arrays of SNSPD, which are under development. Also, InGaAs APD is relatively small ($25 \mu$m ), and even free-space coupling could be lossy. For adaptive optics techniques, see Section~\ref{sec:adaptive_optics}.

\subsection{Quantum Memories}
\label{sec:Qmemories}
A quantum memory allows us to store quantum information of a light field in the internal states of materials and recall it on demand. A current research target is to combine a long memory lifetime with high efficiency in a single device. Over the last decade, several quantum memory protocols such as those based on electromagnetically induced transparency (EIT), Autler-Townes splitting (ATS), and atomic frequency comb (AFC) have been demonstrated in different platforms \cite{lvovsky2009optical,saglamyurek2018coherent, afzelius2009multimode}, including rare-earth ions (REIs), defects in diamonds, Bose-Einstein condensates (BEC) or warm atoms. In the following, we elaborate on some of the most promising ensemble-based and single-spin memory platforms.

\subsubsection{Rare-earth Ions doped solids}
Solid-state systems doped with rare-earth ions stand out as one of the most promising candidates for quantum memories at low temperatures. Long coherence times and narrow optical transitions are the most important properties of many RE ions. In particular, using a clock transition and dynamical decoupling, a hyperfine coherence time of several hours has been reported \cite{zhong2015optically}, which can result in long-term quantum state storage and successful use in quantum memories. Even for Kramers ions such as erbium that are more subject to interact with the environment, the hyperfine coherence time of 1.3~s has been demonstrated \cite{ranvcic2018coherence}, although in the presence of a large magnetic field. Large multimode capacity in all degrees of freedom, i.e., temporal, spectral and spatial, is another advantage of certain rare-earth ions \cite{yang2018multiplexed, afzelius2009multimode,sinclair2014spectral}. So far, a low-noise memory with an efficiency of up to $69\%$ \cite{hedges2010efficient}, storage bandwidth of $5$ GHz \cite{saglamyurek2011broadband}, and conditional fidelity of around $99.9\%$ \cite{zhou2012realization} have been reported for rare-earth-based memories.

 \bigskip
 \emph{Wavelength range ---}
 Within the rare-earth ions, erbium (Er), europium (Eu), praseodymium (Pr), ytterbium (Yb), neodymium (Nd), and thulium (Tm) are of great interest. Erbium ions offer photon emission at telecom wavelengths 1536~nm (ITU optical grids in C-band), where absorption losses in silica fibres are minimum. Europium, for which the coherence time of six hours has been reported \cite{zhong2015optically}, can emit optical photons at 580~nm. For praseodymium, the memory can be carried out using the 605.98~nm optical transition. Amongst rare-earth ions with non-zero nuclear spins, ytterbium has the most simplest level structure and emits photons at 980~nm. And finally, optical transitions of thulium (neodymium) ions can be used to emit photons at a wavelength of 795.5~nm (880~nm).
  \bigskip
  
  \emph{Storage time ---}
  The spin dynamics of rare-earth ions are mainly governed by the flip-flop mechanisms and spin-lattice relaxations. By controlling these mechanisms, for instance, through the application of an external magnetic field, lowering the temperature, and performing dynamical decoupling, very long coherence times are achievable. Storage time of up to 25\:$\mu$s in Pr doped into Yttrium orthosilicate (YSO) with 62 temporal modes using AFC memory \cite{lago2021telecom} has been demonstrated, as well as storage of up to 1250 (100) temporal modes using the Yb:YSO (Nd:YVO) \cite{businger2022non} (\cite{tang2015storage}). To date, an hour-long memory storage time has been demonstrated in Eu:YSO crystal using the atomic frequency comb (AFC) protocol \cite{ma2021one}. This is the longest measured storage time of all optical memories.

\subsubsection{Bose-Einstein Condensates}
As one of the first proposed platforms, BEC played a pivotal role in developing quantum memories. Large atomic density, as required by most of the high-efficiency quantum memory protocols, is probably the most important advantage of BEC-based memories compared to other systems. In addition, the ability to inhibit thermal diffusion at ultra-low temperatures can result in a relatively long coherence time in BEC systems. So far, storage of a polarisation qubit (20\:ns long pulses) of a single photon with a write-read efficiency of $53\%$ using ATS memory in a rubidium BEC has also been demonstrated \cite{riedl2012bose}.

Recently, a BEC of rubidium atoms has been produced in the Cold Atom Laboratory (BECCAL) of the International Space Station (ISS). In light of this technological progress, the initial demonstration of BEC-based memories in space is not far \cite{frye2021bose,aveline2020observation}.

\bigskip
\emph{Wavelength range ---}
To date, different alkali atoms such as Rubidium (Rb) and Sodium (Na) have been studied in the context of BEC-based memories. For the Rubidium atoms, the optical transition wavelengths are 794.7~nm ($D_1$) and 780.24~nm ($D_2$). For Sodium, the emission wavelengths are 589.6~nm ($D_1$) and 589.0~nm ($D_2$).
\bigskip

\emph{Storage time ---}
The spin-wave coherence time of a BEC is dominated by thermal diffusion, magnetic dephasing, collision loss, and recoil motion.  So far, a storage time of more than 1\:s has been demonstrated for a classical light using EIT memory in Na atoms at ultralow temperatures \cite{zhang2009creation}. Generally, by operating at ultra-low temperatures, using relatively weak trapping frequencies, isolating the system from different magnetic field noise sources, and proper selection of ground states, one can increase the coherence time of the system.
Furthermore, the microgravity condition of space can prevent the need for gravity compensating trapping potentials, and therefore, an improvement in the coherence time is expected. Considering these treatments, a few minute-long coherence times are imaginable for the BEC systems.
\bigskip

\subsubsection{Warm vapours}
Of the different platforms, warm vapour systems are attractive as they do not need optical trappings or do not need to operate at cryogenic temperatures. Ensemble-based memories at room temperature have been studied using different memory protocols \cite{guo2019high,  reim2012multipulse, dou2018broadband}. For instance, employing the iteration-based optimization strategy, a high memory efficiency of more than $82\%$ has been demonstrated at $\text{T}=78.5\, ^o\text{C}$ using an Rb atomic vapour \cite{guo2019high}.

Rare isotopes of noble gases such as $^{3}\text{He}$ are optically inaccessible (unless using an optical UV addressing); however, their nuclear spins are shielded from the environment by the full electronic shells. As a result, in noble gases, the nuclear spin coherence times are in the order of hours. Recently, the coherent coupling of alkali-metal vapour and noble gas spins has been demonstrated \cite{shaham2022strong}. This can open up the opportunity for designing long-lived memories at room temperatures.

In general, four-wave mixing (FWM) noise is an important impediment to employing hot vapour for quantum networks. Therefore, different solutions have been suggested to suppress this noise \cite{zhang2014suppression, nunn2017theory, thomas2019raman}. Among them, cavity engineering is shown to be very effective in FWM noise suppression \cite{saunders2016cavity}. Recently, motivated by this demonstration and also the spin-exchange interaction between noble-gas spins and alkali atoms \cite{shaham2022strong}, the use of hybrid alkali-noble gases has been suggested for a non-cryogenic repeater protocol \cite{ji2022proposal}.

\bigskip
\emph{Wavelength range ---}
Cesium and rubidium atoms have been studied intensively for designing room-temperature memories. Cesium D-line transitions have wavelengths of 894.0~nm ($D_1$) and 852.0~nm ($D_2$). 
\bigskip

\emph{Storage time ---}
Atomic collisions and motions mainly limit the coherence times of alkali vapours at room temperatures. However, using the spin-exchange relaxation-free mechanism at low magnetic fields, a memory storage time of 1\:s has been achieved in a room-temperature cesium vapour \cite{katz2018light}. Besides, by transferring the spin-wave coherence of alkali vapours to the ultra-stable nuclear spins of noble gases, hour-long memory lifetimes even at room temperatures are expected.

\subsubsection{T Centres in Silicon}
A recently emerging and promising technology for quantum memories are individual spin defects in silicon lattices, the most important of which is the T centre~\cite{Simmons.2020, Simmons.20208p8}. 

Given the scalability of silicon integrated photonics, thousands or even millions of high-fidelity quantum memories are conceivable on a single chip. These T centres have been shown to be individually addressable via commercially standard silicon integrated photonics~\cite{higginbottom2022optical}, and show promise not only as memories but for QND measurements and distributed quantum information processing. A significant benefit of operation in silicon is the immediate applicability of standard silicon processing techniques and foundries. The T centre contains one optically-addressable electron spin and one or more long-lived nuclear spins, allowing both excellent spin-photon interfacing and qubit storage. They are substantially brighter than rare-earth atoms in solids.  

One challenge is the cooling requirement of $\sim2$~K for high-quality photon emission, which, while technologically feasible, can be a challenge in a space environment. However, long storage times and large bandwidth could make such a trade-off worthwhile. Given the advantages, the T centre has been recently proposed as an effective memory platform that can function well in both free-space and in the presence of a cavity \cite{higginbottom2022memory}.

\bigskip
\emph{Wavelength range ---}
The T centre’s photon emission is at 1326~nm, in the telecom-
munications O-band. Therefore, it is compatible with existing telecommunications
systems and fibres for terrestrial transmission. This wavelength is acceptable for atmospheric free-space propagation, albeit close to a large water absorption region. Further colour centres in silicon can reach other relevant wavelength bands (e.g. M center at 1630~nm \cite{Lightowlers.1997}, for very low-loss atmospheric transmission), though these are less studied.
\bigskip

\bigskip
\emph{Storage time ---}
Coherence times for the electron spin of 2.1~ms and for the nuclear
spin of 1.1~s have been demonstrated in isotopically purified $^{28}$Si \cite{katz2018light}. Nuclear storage times of minutes to hours are expected via dynamic decoupling techniques \cite{Thewalt.2013}.
\bigskip

\subsubsection{NV Centres in Diamond}
The nitrogen-vacancy (NV) defect centre in diamond is another promising candidate for solid-state quantum memories. So far, NV centers have been used to show entanglement generation between two independently controlled qubits \cite{humphreys2018deterministic, bernien2013heralded}, and between the electronic spin of a single NV$^-$ center and optical photons \cite{togan2010quantum}. In addition, a multi-qubit register can be formed using the NV electron spins that manipulate neighbouring nuclear spins \cite{dutt2007quantum, reiserer2016robust}. 

In Ref \cite{heshami2014raman}, using an ensemble of NV$^-$ centers at low temperatures and based on the off-resonant Raman coupling (to circumvent optical inhomogeneous broadening), an optical quantum memory has been proposed. In this scheme, to accomplish the appropriate optical polarisation selection criteria, a large external static electric field and a low magnetic field are required. As a result, a total efficiency of 81$\%$, which can be further improved using a higher control field strength, is achievable. The main advantage of using NV centers-based quantum memories is a relatively narrow linewidth of the zero-phonon line even at temperatures of up to 50~K \cite{fu2009observation}. This can, to some extent, obviate the need for expensive and complex cryocoolers that operate at few kelvin temperatures.

\bigskip
\emph{Wavelength range ---}
The most commonly used optical transition in the NV center has a wavelength of 637 nm. It is also possible to use the singlet transition with a zero-phonon line at 1042 nm. 

\bigskip

\emph{Storage time ---}
Even at room temperatures, nitrogen-vacancy centers offer a millisecond-long electron spin coherence time \cite{bar2013solid, maurer2012room}, making them ideal candidates for use as quantum memories in quantum networks. Using nuclear spins at temperatures of around 50~K, 1~s storage time should be achievable.

\subsection{Quantum Non-Demolition Measurement}
Quantum non-demolition (QND) measurement is the capability to detect photonic qubits without absorbing the photon or unfolding their qubit encoding or state in a non-destructive manner. This ability is especially important as we need the photonic qubit to remain usable for quantum information processing applications after the non-destructive detection.
There are several approaches to the non-destructive measurement of photons. So far, nonlinear interactions have resulted in significant progress, such as large cross-phase modulations using the AC Stark shift \cite{beck2016large} and single-photon phase shifts using Rydberg atoms \cite{saffman2010quantum, tiarks2016optical}. Therefore, one approach in the non-destructive measurement of photons is by using nonlinear interactions in atomic systems between the photon to be non-destructively detected and a probe pulse \cite{sinclair2016proposal, hosseini2016partially}. Another approach includes reflecting the photon off the atom-cavity system. This approach has been discussed in both trapped atoms \cite{reiserer2013nondestructive} and rare-earth ions \cite{o2016nondestructive}. Another strategy is to design a probabilistic QND using linear optics and single-photon detection that works similar to the BSM used for quantum teleportation, which can operate with a maximum efficiency of 50$\%$ \cite{meyer2013entanglement, curty2011heralded}.

QND has many applications in quantum information processing and quantum communication since in most networking scenarios it is essential to know when the photon has arrived without destructively detecting it. This is especially important when interacting with satellites as most of the time the photon will be lost (see Appendix. \ref{sec:additional_teleport_schemes} ). Hence, one needs to employ a QND and an on-demand quantum memory, which together form a heralding memory. In this regard, heralded transfer of polarisation qubits into a trapped atom-cavity system has been demonstrated \cite{kalb2015heralded}. Alternatively, heralded quantum memories can be employed using emitted spin-entangled photons, and performing an optical BSM with the incoming photonic qubit to be stored \cite{barrett2005efficient}.

\subsection{Quantum Transducers}
In a quantum network, heterogeneous qubit types might be used for different quantum information science applications. Transducers can connect quantum systems of different nature with possibly very distant electromagnetic frequencies. In particular, quantum frequency translators can be used to shift the wavelength of telecom photons, which are well suited to transfer information over optical fibres, to microwave photons that interact with superconducting qubits, and vice versa. 

Quantum transduction can be carried out through atomic ensembles with optically and microwave-addressable transitions such as defects in diamond \cite{lekavicius2017transfer}, silicon \cite{higginbottom2022memory} or rare-earth ions doped solids \cite{o2014interfacing, asadi2022proposal}. In particular, the use of rare-earth ions at zero external magnetic field has been proposed for transduction between microwave and optical photons at telecom wavelength \cite{asadi2022proposal}. In addition, an experimental transduction efficiency of $82\%$ has been reported using Rydberg atoms \cite{tu2022high}. Another transduction approach employs mechanical motions to couple microwave-optical systems and is called electro-optomechanical conversion \cite{han2020cavity}. Conversion at the few-photon level with $47\%$ efficiency has been demonstrated using opto-mechanical systems \cite{higginbotham2018harnessing}, although with a low conversion rate. It is also possible to use an electro-optics that utilises a $\chi^{(2)}$  or $\chi^{(3)}$ nonlinear material to transfer energy between the two modes \cite{javerzac2016chip,fan2018superconducting}. 

A transduction protocol needs to have a long coherence time, large signal-to-noise ratio and conversion bandwidth. The latter is particularly crucial in frequency and time multiplexing and is mainly limited by the relatively weak microwave coupling. To date, magneto-optical conversion with bandwidth on the order of a few MHz \cite{hisatomi2016bidirectional} has been realized. Microwave-to-optical conversion using Rydberg atoms with a bandwidth of 15\:MHz has been recently demonstrated \cite{vogt2019efficient}. However, mature quantum transductions are not yet available and need to be developed.

\subsection{Quantum Optical Links}
\label{sec:quantum_link}

\subsubsection{Link Requirement}
The required link performance must clearly be handled on a case-case basis depending on the unique aspects of the protocol and the implementation. While a detailed analysis is subject to future studies, for the purpose of this WP we will use a QKD link as a baseline to determine a  link requirement.

The assumptions for this estimate are as follows.
\begin{enumerate}
    \item For a successful  key distribution using quantum signals, a certain number of photon detection events must be accumulated (\cite{scarani2009security}). 
    \item The number of successful detect ions required depends on the protocol. Two main cases that we will consider are that for a decoy state protocol (faint laser based), about 100,000 detections are required, while for an EPS source or a single-photon emitter, only about 10,000 events are needed.
    \item The link duration is limited by the satellite orbits as follows: for LEO, we assume 120 seconds link, for MEO we assume 20 minutes link, and for GEO (and similar) we assume 1 hour link duration.
    \item For the source Rate we will assume a clock rate of 1 GHz.
    \item We will consider that the photon rate might further be enhanced using multiplexing, and we will assume a 100 fold use of the channels.
    \item In this simple estimations, no noise will be considered, as we are interested in the best-possible case. Therefore, we assume the intrinsic dark counts, background signals are all set to zero, which is the best possible case.
\end{enumerate}
 
To highlight our approach, these assumptions on source rates allow us to draw a relation between link losses (in dB) and the measurement time as shown in Figure~\ref{fig:time_required_for_QKD} This relation will be utilised to assess different scenario parameters such as aperture sizes, orbits etc.


\subsubsection{Link Loss Estimation}
While the actual quantum optical link is demanding from a technical point of view, the technology is already under development for optical (classical) communications using laser systems.  

We implemented a simple link model for the purpose of this whitepaper in order to assist the selection of wavelength, and aperture sizes, based on rough-order-of-magnitude quantum link performances. (See Appendix~\ref{Appendix:link_analysis}).

The first factor for the link attenuation is the vacuum beam diffraction based on the wavelength, and the aperture. The second factor is the atmospheric absorption assuming a standard atmosphere, wave-front distortion caused by atmospheric turbulence, and finally tracking errors and coupling efficiencies. See Appendix~\ref{Appendix:link_analysis} for more details.

Based on this model several different scenarios (satellite altitude/orbit, apertures, wavelength are compared and the physical requirements are defined. See Appendix~\ref{Appendix:link_analysis} for a an overview. What is obvious is for MEO or GEO orbits, the quantum link losses are dramatic, and will easily reach 60~dB or even 80~dB using realistic technologies. In particular the main limitations are the aperture sizes for space telescopes (preferably less than 25 cm diameter), and photon emission rates of typically 100 MHz (currently). While the telescope aperture remains fixed, there is a lot of potentials to advance the photon rates through better sources and detectors, as well as employ channel multiplexing.

\subsection{Ground-based Systems}

\subsubsection{Quantum Ground  Stations}
Several optical ground stations have been built and demonstrated for laser communications and quantum communications. Various telescope systems with apertures of 1m to 1.5 m, have successfully used their tracking capability for optical links with classical and quantum communications  satellite terminals with very good quality.

\begin{figure}[ht]
    \centering
    \includegraphics[width=0.9\textwidth]{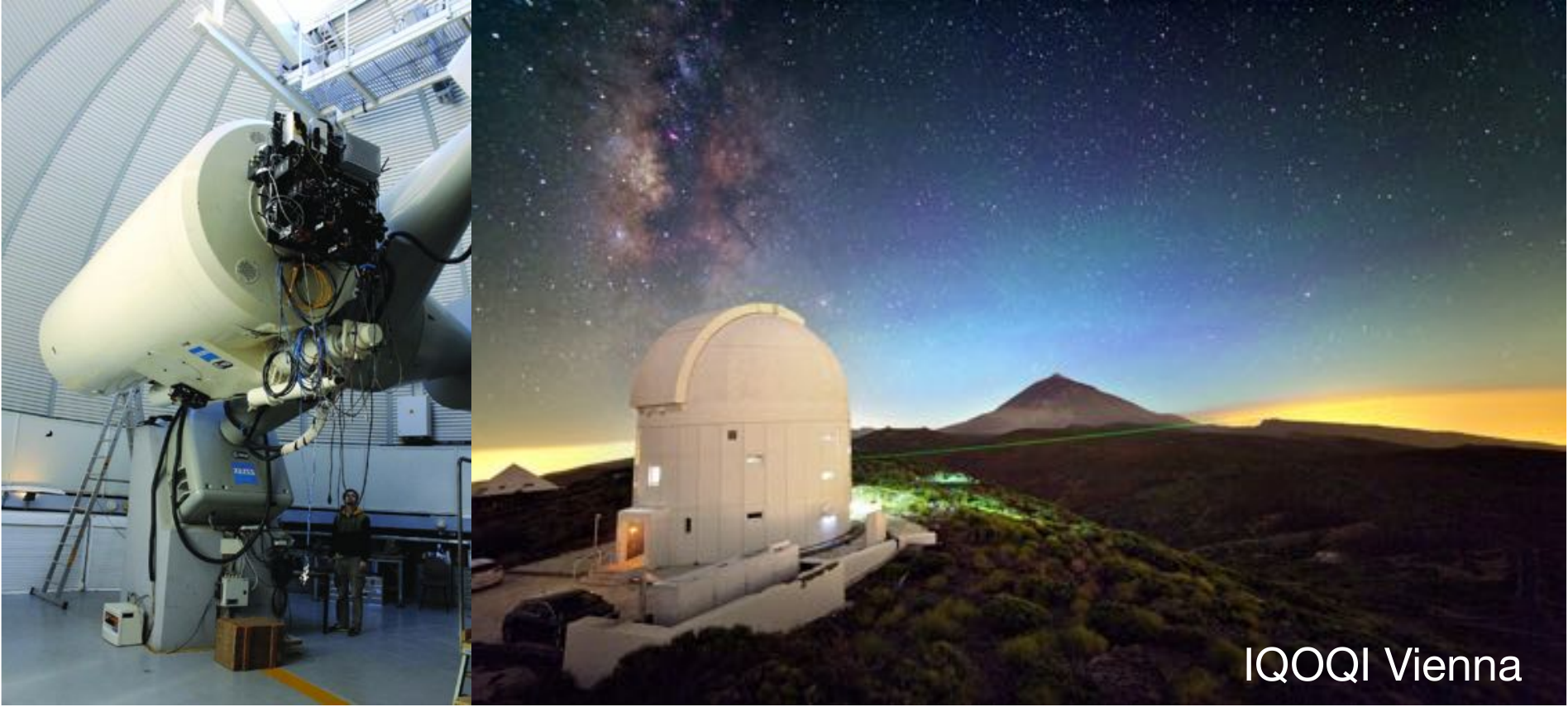}
    \caption{Pictures of ESA's Optical Ground Station on Tenerife, configures as a quantum communication transceiver. (Images by IQOQI Wien, retrieved from $https://www.esa.int/ESAMultimedia/Images$)}
    \label{fig:OGS_ESA}
\end{figure}

Here are a few notable examples: Japanese ground station,  ESA's OGS on Teneriffe (see Figure~\ref{fig:OGS_ESA}), China (multiple ground stations), Italy (Matera), just to name a few. In Canada, CSA's and University of Waterloo's optical ground stations are under development to be the primary and secondary OGSes of the QEYSSat. In addition, University of Calgary owns the tertiary ground station which has demonstrated successful satellite trackings for optical communication purposes (Figure~\ref{fig:CanadaST}).

Quantum ground stations are becoming commercial, and multiple telescope vendors are preparing turn-key systems, including Astro Systems in Austria (ASA), PlaneWave in USA, QuantumCTek in China, and Mylnarik Germany.

\subsubsection{Quantum Airborne Stations}
A link between Satellite - HAP - Ground is of particular interest. An airborne platform, in particular a slow moving HAP, can operate from a much thinner and quieter atmosphere, and therefore benefit from the reduced impact of scattering, turbulence and weather.  The link models estimated in Section \ref{sec:quantum_link} show that the availability and attenuation in particular for LEO satellites to HAP is very favourable.

Quantum and optical links from airplanes and drones to the ground have been demonstrated multiple times, specifically including the DLR demo (LMU 2014) and our Canadian demo (IQC 2016). 


\subsubsection{Underwater stations}
Underwater communication has long been distinct from the rest of the global communication due to its reliance on acoustic signals while the rest of the world is using radio signals which cannot be transmitted more than a few centimeters in water. However, as we move to optical wavelengths when exploring QKD channels between ground and satellite, we open the door to incorporating underwater channels where blue-green wavelengths can be transmitted for hundreds of meters which is enough to establish a channel between submarines while they remain submerged. There have already been many experimental explorations into the feasibility of such underwater quantum communication at optical wavelengths. Using the above-described adaptive optics system, one can pre-correct for the aberrations introduced by waves on the air-water boundary. Therefore, enabling a submarine to satellite QKD channel.

\subsection{Multiplexing}

\subsubsection{Multiplexed Quantum Channels}
One major benefit of a free-space quantum channel is that there is a little amount of wavelength dispersion, which makes the channel very suitable to implement extensive multiplexing methods. In order to achieve high rate operations, a multiplexing factor of 100 or even 1000 times is preferred.

Most importantly this would include wavelength-division multiplexing (WDM) of the quantum sources, and the photon detectors. 
For instance, gratings or filter systems could enable using 100 wavelength channels across a ground-to-space link.
An interesting version to consider is asymmetric WDM, 
using SPDC-based entanglement, where the long satellite link receives a single band, while the other receiver utilised WDM to fan-out the different wavelength channels. Another direction is to use very wide-band EPS such as the fibre-based source by UofT. WDM has also been combined with polarization to demonstrate a multiplexed state suitable for QKD applications\cite{Woo2020}.

The next proposition is the implementation of temporal modes to perform high-dimensional communication. The advantages of high-dimensional quantum communication include increased key rates and security in high-loss or noisy channels; however, there are many photonic degrees of freedom which can be used to implement this. The OAM of photons is one such approach which has some advantages in measurement and detection such as compact liquid crystal devices. Nevertheless, a large propagation distance, such as a ground to a satellite link, results in significant crosstalk between OAM modes due to turbulence. This turbulence is particularly destructive to OAM modes as transverse spatial distortions result in crosstalk between the modes, not just losses in the channel. However, prior knowledge of atmospheric turbulence through channel probing can lead to some criteria for clean OAM modes \cite{amhoud2019oam}. Effect of weather, fog/dust and other scatterers on cross-talk and demultiplexing efforts were studied for relatively long ranges $\sim$800m of communication channels \cite{singh2022performance}. It was also reported that the use of a pilot tone led to turbulent-resilient multiplexing of a 2-OAM channel \cite{zhang2020experimental} Temporal modes, e.g., in the Hermite-Gaussian basis, have their information encoded along the propagation direction, as opposed to in the transverse plane. Therefore, turbulent channels will have little effect on the modes encoding the information, which is primarily destructive to the transverse wavefront of the beam. This could allow for a Hong-Ou-Mandel measurement, or ‘super sorter’ based approach for the implementation of a QKD channel using temporal modes.

Hybrid multiplexing of OAM, polarisation and spectral was also investigated over turbulent channels for ranges over 1 to 10 km \cite{zhao2019effects}.

Higher-dimensional quantum encoding is another approach for enhancing the channel capacity. The main methods to achieve this are using multiple time-bins,  frequency encoded photons, and OAM. Some recent developments have shown that time-bin is indeed viable for high-loss, turbulent channels. While OAM is an interesting direction, the viability for high-loss channels remains to be determined.

\subsubsection{Multiplexed Memories}
It is possible to enhance the performance of quantum memories using multiplexed schemes. There are different degrees of freedom that can be used to design a multiplexed memory. In temporal multiplexing, multiple entanglement generation attempts per the communication time can be performed by sending a train of photons to the memory and beam splitter. Temporal multiplexing can significantly improve the entanglement generation rate. In some solid-state systems, inhomogeneously broadened atomic ensembles can be used to create temporal multiplexed memories. So far, storage of around 1250 temporal modes using the Yb:YSO in an atomic frequency comb memory has been demonstrated \cite{businger2022non}.

In a spatial multiplexing architecture for quantum memories, several sub-ensembles are required such that each of them needs to be addressed individually.  Photons emitted from these independent spatial modes will be collected and detected at will. In this regard, a multiplexed memory with 225 spatial modes has been demonstrated experimentally \cite{pu2017experimental}. Spatial multiplexing can also be accomplished by routing photons to many addressable colour centres, e.g. via integrated optics and fast switching. This type of multiplexing can significantly reduce the required memory storage time as entanglement swapping can be performed between non-adjacent memories.

Another type of multiplexing is angular multiplexing. Using angular multiplexing, one can store multiple overlapping excitations with different spin-wave vectors. In Ref \cite{chrapkiewicz2017high}, an angularly multiplexed holographic memory with up to 60 atomic spin-wave modes has been demonstrated. However, the operation is not in a fully quantum regime. 

In the above ensemble-based multiplexed schemes, the transition frequencies of the atoms in the ensemble were assumed to be the same. It is also possible to address individual atoms with different transition frequencies within the ensemble, or different defect centres with individually tuned emission frequencies. Spectral multiplexing uses different spectral modes to store multiple photons with different frequencies. To date, spectral multiplexing with up to 26 spectral modes has been demonstrated \cite{sinclair2014spectral}.

Multiplexed storage for quantum repeaters is achievable in Er-doped fibres as it has been very recently reported with up to 1650 single-photon modes. This is of particular interest as space-based classical telecommunications already exploit doped fibres. Interconnectivity and hybrid processes could be envisioned \cite{wei2021multiplexed}.

OAM also permits multiplexing with higher order than polarisation states. While polarisers were developed for both emission and reception in an earth-satellite comm link for Qeyssat 1.0, QKD protocols based on OAM could favour higher transmission rates.

\subsection{Synchronisation and Phase stabilisation}
There are two main levels of challenges to measure and resolve short pulses by the quantum receiver, one is to stabilise down to the arrival time of the photons, while the much harder challenge (but also more helpful challenge) is to stabilise even the phase of the channel. 
Currently, the channels will  synchronise the source and detector to the level of 1~ns, but the level of  100~ps is desired to achieve a high-speed gating resolution, as this would open up the channels for faster photon rates, but making more extensive use of temporal multiplexing.

For ground based systems, it is common to use optical fibres or even electrical cables to establish common time and phase references. Note that on a ground based, installed fibre, the  optical signals can be transferred over relatively long distances.

The quantum repeater schemes using BSMs or quantum memories will  have their own specific challenges. In order to establish long-distance entanglement using quantum repeater, one can either use single-photon or two-photon detection-based schemes. The straight-froward approach to entanglement generation protocols based on two-photon BSMs required that two photons reach the central node, each experiencing $\sqrt{\eta}$, which means the link rate scales as $\eta$, i.e. the square of the individual link efficiency \cite{barrett2005efficient}. On the other hand, the new single photon-based schemes scale only with $\sqrt{\eta}$, which is a huge improvement for the photon rates (maybe up to 20 dB!). However, these schemes are inherently more sensitive to phase fluctuations and are less robust against the photon and detector losses. 

The sensitivity of the scheme to the phase fluctuations can make designing scalable quantum networks challenging. Hence, a set of strategies should be employed to ensure the stability of fibre links. Otherwise, the remaining instability will be amplified upon each BSM. Phase stabilisation usually requires synchronisation between remote nodes to make sure that the wave packets overlap. In general, optical path stabilisation strategies include the use of self-compensating Sagnac-type setups \cite{minavr2008phase, childress2005fault}, or active feedback protocols \cite{chou2005measurement}.

Time transfer using quantum technologies depends in part on coincidence measurements, which themselves depend on photon detectors and detection electronics. High-speed photon detectors with low timing jitter are thus required. Commercial timing electronics can now tag a pulse with respect to a reference pulse to within $\sim$5\:ps to 10\:ps (FWHM) or better with time bins in the few hundreds of femtoseconds (fs). Using multiple coincidences (long measurement time) and frequency entanglement it has been demonstrated that time transfer to within $\sim$60\:fs is possible over 20\:km of optical fibre \cite{hou2019fiber}. These transfers require entangled photon pair sources and fast, stable photon detectors (SPADs in this case). These results can also be obtained using SNSPDs \cite{esmaeil2020efficient}. This, like other quantum technologies, could benefit from integrated photonics and integrated photonics with electronics.

Time transfer is also vulnerable to attack. Quantum technologies could render time transfers more secure \cite{dai2020towards}. The Micius satellite allowed for testing this quantum secure time transfer (QSTT) using hardware for QKD to within 30\:ps.

Long-range phase stabilisation could require interferometric sensors. Ultra-long coherence and advanced phase-retrieval algorithms are required for phase-sensitive processes as synthetic aperture [Patent US10564268B2]. Such technologies could limit losses if coupled with adaptive optics.

\subsection{Advanced Daylight Filtering}
\label{sec:DayFilter}

Current systems, including QEYSSat 1.0 mission, are designed for night-time only operation. Even under night-time operation, the susceptibility of the quantum channel to artificial light sources is an important issue, and has to be overcome for the operation of an OGS from a city. Another issue could be the sunlight scattered from the spacecraft itself.
Ultimately, the system should allow for daytime operations. The main contributors to background noise during daytime is obviously the sunlight  that is scattered by the atmosphere into the quantum channel. 

Operating the quantum link under high background signals is a very crucial technology, and will require filtering the signals in all three available degrees of freedom, including the photons
\begin{enumerate}
    \item spectral,
    \item temporal, and
    \item spatial modes.
\end{enumerate}
Several demonstrations on  ground systems have achieved this with the assistance of single-mode fibre coupling at a receiver. For instance, the first tests at around 800~nm signals including the first demonstration in 2002 by the LANL group \cite{Hughes:2002aa}, and a more recent demonstration which also incorporated adaptive optics \cite{Gruneisen:2021aa} to improve the fibre coupling. Another very interesting demonstration was at 1550~nm telecom wavelength \cite{Avesani:2021aa} in daylight, which made use of the  lower scattering cross section of 1550nm vs. 800 nm, due to  the longer wavelength. A demonstration of daytime entangled photon transfer in free-space use careful arrangements of baffles and blackout materials to suppress the daylight \cite{Peloso:2009aa}.
Another direction is to explore the absorption lines in the solar spectrum. The NIST group demonstrated a QKD system operating within the   656~nm H-Alpha line, thereby observing a notable reduction of solar background light \cite{Rogers:2006aa}.

Therefore, advanced optical filtering techniques such as etalons, atomic line filters (ALF), and mode filtering such as single-mode fibre coupling shall need to be explored.

To operate in daylight conditions, the effects of daylight must be reduced. Daylight reduces SNR and ultimately secure key rate if not addressed. There are multiple technologies to filter out unwanted light and reduce the amount of daylight falling on the detector. The most common is thin film interference filter (TFIF) technology. It is possible, today, to manufacture a filter with a full width at half maximum (FWHM) bandpass as small as 0.5~nm. In many applications this is sufficient. One drawback of thin film interference filters is that it is very sensitive to the incidence angle, both in terms of central pass wavelength and in bandpass. In other words, the wavelengths passed depend on the angle of incidence. The smaller the bandpass, the more acute the drawback. In an optical design, the TFIF must be in a path in which the light is well collimated. And the TFIF must be in a mechanically stable holder with respect to the incoming optical beam.  

The bandpass is not the only parameter of interest, the blocking optical density (OD) is almost as important. The single photon detectors are often sensitive over wide wavelength ranges, especially the semiconductor detectors. The filter must efficiently block light over the entire range of wavelengths over which the detector is sensitive. It is often not possible to do this with a single filter. A stack of filters is then necessary, since each filter has a transmission that will not be 100\%. Filter throughput needs to be incorporated into the transmission budget. The wavelength that is passed also somewhat depends on the temperature, although not a huge factor in most applications. There are multiple commercial vendors of thin film interference filters, although it could be necessary to have one developed if the laser wavelength is selected to be at a non-commercial wavelength. One important TFIF supplier is Iridian Spectral Technologies, based in Ottawa.

The angle of incidence, other than 0°, will be different for different polarisations, a parameter to consider in the quantum protocol. In addition, using well controlled optical filters has an impact on laser wavelength stability. In a quantum link operated in daylight with aggressive optical filtering, the laser wavelength will need to be stabilised to keep transmission losses to a minimum.   

If a TFIF is deemed insufficient, other technologies exist. Optical etalons are another type of filtering that can be used. Optical etalons are optical cavities, usually with two flat mirrors face to face (but other geometries are possible), on a single piece of glass/substrate or on separate substrates but with a well-controlled spacer. Again, there will be dielectric coatings with a well determined reflection coefficient. Multiple etalons would be required to significantly reduce the bandpass and cover a large blocking wavelength range. In fact, a mix of TFIF and etalons would be required. The same drawbacks are encountered here as for the TFIF. The ultimate bandpass can be very small, much less than 0.1~nm. This means even more stringent requirements on laser wavelength stability. An interesting etalon supplier, in particular of custom devices, is Light Machinery, based in Nepean, Ontario.



Finally, for filters that are very stable in wavelength, there are atomic line filters. Atomic line filters are sometimes used in LIDAR \cite{fricke2002daylight, gong2015lidar}. Atomic line filters are atomic vapour cells placed between crossed polarisers and subjected to a magnetic field. The magnetic field causes the rotation of light polarisation and transmission of the light resonant on an atomic transition to passthrough the second polariser \cite{keaveney2018optimized, menders1991ultranarrow}. Filters can be built around different atomic species, including Cs, Rb and Na. \cite{dick1991ultrahigh, chen1993sodium,kiefer2014faraday}. Atomic line filters will probably be an important part of a secure quantum link to space. This limits the available working wavelengths.

A very well established method for spectral narrow filters is atomic optical line filters. The benefit is that in addition to a very narrow line defined by the atoms, it also is tolerant against mode distortions. It is therefore an interesting technology, and could be applied to quantum channels. The spectral filtering for day-light quantum operation should be feasible and was recently shown by AFRL group \cite{Gruneisen:2021aa}. However, the compatibility with WDM and handling the gravitational and motion induced frequency shifts are of concern, and therefore it is not clear if such filters are useful.

Another established method is cascades of etalons made from stable materials, getting stable small line-widths (spectroscopy done with etalons in past). However, how their performance suffers from multi-modal beams is yet to be determined. Moreover, the multimode nature of light degraded by atmospheric distortions could also affect  narrowband filtering. Where single mode operation is needed for using etalons or other filters, there may be a  need for adaptive optics for single mode fibre coupling. Note that for an uplink, the spatial mode at the receiver aperture will be 'locally' very pure given the long propagation distance from the top layer of the atmosphere to the satellite.

In general, a big issue with narrow-band filters is that the line widths may be too narrow for the photon rates that are considered (Giga-hertz quantum sources may not be compatible with some of the ALF).


Another important aspect of background light suppression is to filter the spatial mode, and this can be accomplished by coupling the received signals into single mode fibres. Recent demonstrations of adaptive optics systems have demonstrated a day-light operating receiver \cite{Gruneisen:2021aa}. 


\subsection{Adaptive Optics}
\label{sec:adaptive_optics}
The implementation of an adaptive optics system in a ground-satellite link has at least two major benefits, decreased losses and increased mode fidelity. Adaptive optics systems, particularly using the pre-correcting approach, reduce channel losses thus increasing the key rates in quantum channels. The pre-correction approach involves probing the channel to receive the turbulence information and then making the phase correction before the signal is sent. Such an approach is also the most convenient for a ground-to-satellite approach as bulky adaptive optics equipment can be on the ground as opposed to taking up space on the satellite. 

Atmospheric turbulence causes tip/tilt fluctuations, resulting in beam misalignment, as well as higher-order beam distortions which in extreme cases results in the breakup of the beam, called scintillation. Indeed, the atmospheric distortions impede an optical link through random turbulent media, and also generally for any moving systems. The distortions impact the spreading of the beam (as is the case for uplink from ground to space), and reduce the coupling of the signals to filters or photon detectors (as is the case for a ground based receiver). In a ground-to-satellite channel, these distortions introduced in the lower atmosphere are exacerbated by the long distance travelled to the satellite. Any improvements to the quantum link through adaptive optics is therefore a very important technology. The wave-front distortions and pointing errors caused by atmosphere, as well as by tracking errors, will impact the uplink and downlink channels potentially differently.

Therefore, a pre-correcting adaptive optics system at the ground sender has the potential to greatly decrease losses from misalignment. The use of high-dimensional quantum states is often highly impacted by turbulent channels, particularly orbital angular momentum (OAM) states, but also those states which require some interference on a beam splitter such as temporal modes and coherent states. A system which can recover a high-fidelity mode after a turbulent channel will allow for the exploration of many protocols including those from the continuous variable world.

Adaptive optics have been a game changer in wavefront correction in space imaging. Closed-loop adaptive optics could be used for phase correction while maintaining quantum properties. Recent observations showed that spatial filtering through adaptive optics may better serve daylight filtering than spectral, interferometric methods \cite{Gruneisen:2021aa}. 

The quantum signal will experience atmospheric turbulence as it passes through the atmosphere, caused by changes in the refractive index due to small temperature fluctuations along the beam path. This effect induces temporal intensity fluctuations (scintillation), beam wander, and beam broadening \cite{smith1993infrared}.

For the purpose of quantum transmissions, the temporal intensity fluctuations can be ignored as the average intensity over time will remain unchanged and only the total number of received photons is relevant to the quantum protocols.

The beam wander and beam broadening, which together can be modelled as an average total beam broadening will result in an increase in the average loss over the link. This effect is minimal in the case of a satellite-to-ground transmission because it happens only near the very end of the link, when the beam is already large and has little distance left to travel. However, atmospheric turbulence is one of the dominating sources of loss in transmissions from the ground to a satellite, to the extent that this effect can dominate over intrinsic diffraction broadening.

Improvement can be attained using adaptive optics techniques. Adaptive optics (AO) involves using a deformable mirror (or lens)  to correct (or in some applications induce) wavefront errors in a propagating beam \cite{tyson2022principles}. It is often used in the astronomy field to correct distortions in a received image. High fluctuations in the signal are not relevant for our purpose. Here, we only require a high total signal intensity to be received. This makes our need somewhat different from most adaptive optics applications. A detailed preliminary study of the possible improvements that can be acquired through adaptive optics was performed for QEYSSat~1.0.

The results from this preliminary study were that there can be a gain of between 5 and 10~B by using adaptive optics when using a ground station at sea level (for a satellite on a 600~km orbit) when compared to tilt error correction only (done with the beacon beam). This improvement is when the link is in the best satellite pass (small radial speeds). There are “easier” improvements by using a large ground emitter aperture diameter ($>50$~cm) to reduce beam diffraction. This is also true when using shorter optical wavelengths (thus the 700-900~nm window is better than the 1500 - 1600~nm window). The feedback loop correction bandwidth is also of prime importance, the faster the better. There is also a significant gain in having a ground station at higher elevations, to reduce initial atmospheric turbulence in beam propagation.

There has been some work in Canada on adaptive optics for astronomy telescopes (such as the work done for the Mont Megantic telescope \cite{martin2015ino}). Work sponsored by the CSA on deformable optics and wavefront sensors. An adaptive optics add-on usually requires a guide star and a wavefront sensor along with deformable optics. Technology has evolved since the preliminary work on QEYSSat 1.0. Deformable optics are more readily commercially available and can be incorporated in optics design (for example see alpao.com). Some form of customization will be necessary because of the moving nature of the satellite and to adapt the AO system to the application.

\subsubsection{Uplink}
For the {\bf uplink} from ground to space, the issue is the wave-front distortion that the optical beam incurs after exiting the transmitter aperture, which directly impacts the beam shape and pointing at the far-field. Therefore, the channel  will  be impacted due to the coherence length of the atmosphere. Previous studies, including the joint work done by UW and INO  (\cite{ISI:000507254300010}), showed that the AO correction of the uplink to a satellite is rather difficult to implement, given the anisoplantic error, and could only really achieve its full potential if an artificial guide-star is involved, which is created at the correct point-ahead angle. Future technologies could further study the use of manipulated downlink laser beacons, and attempt to resolve the point-ahead angle that way.

\subsubsection{Downlink}
For the link {\bf downlink} from space to ground, the main issue of wave-front distortion is that the received signal in the focal plane of the receiver telescope will fluctuate and wander a lot, and therefore be difficult to couple into the optical fibres or very narrow-band filters, without beam stabilisation and mode correction. However, the advantage of the downlink scenario is that the co-propagation beacon laser can act as a bright, very good guide star, and therefore a beam correction in real time should be very efficient. For instance, the AFRL team demonstrated daylight quantum communications using a horizontal link of 1.6~km that was enhanced by an Adaptive-Optics system \cite{Gruneisen:2021aa} (and so have others at NASA and Onera).

Interesting alternatives to stabilise a quantum channel are that the quantum entanglement could enable direct solutions where the quantum states themselves can be used to stabilise the quantum signal intensity. 
This would correspond to 'adaptive optics in post processing'.

\subsection{Compact and On-Chip Polarimeter}
Recent advances in silicon-based chip polarimeters could enable polarisation QKD protocols as defined in QEYSSat. 
Losses, polarisation extinction ratio and FOV might suffer in comparison to a bulk polariser, but adaptation and optimisation could lead to a significant submodule in a quantum network.

\clearpage

 \section{QEYSSat~2.0 Workshop Summary}
 \label{chap:Workshop}
\begin{figure}[ht]
    \centering
    \includegraphics[width=1\textwidth]{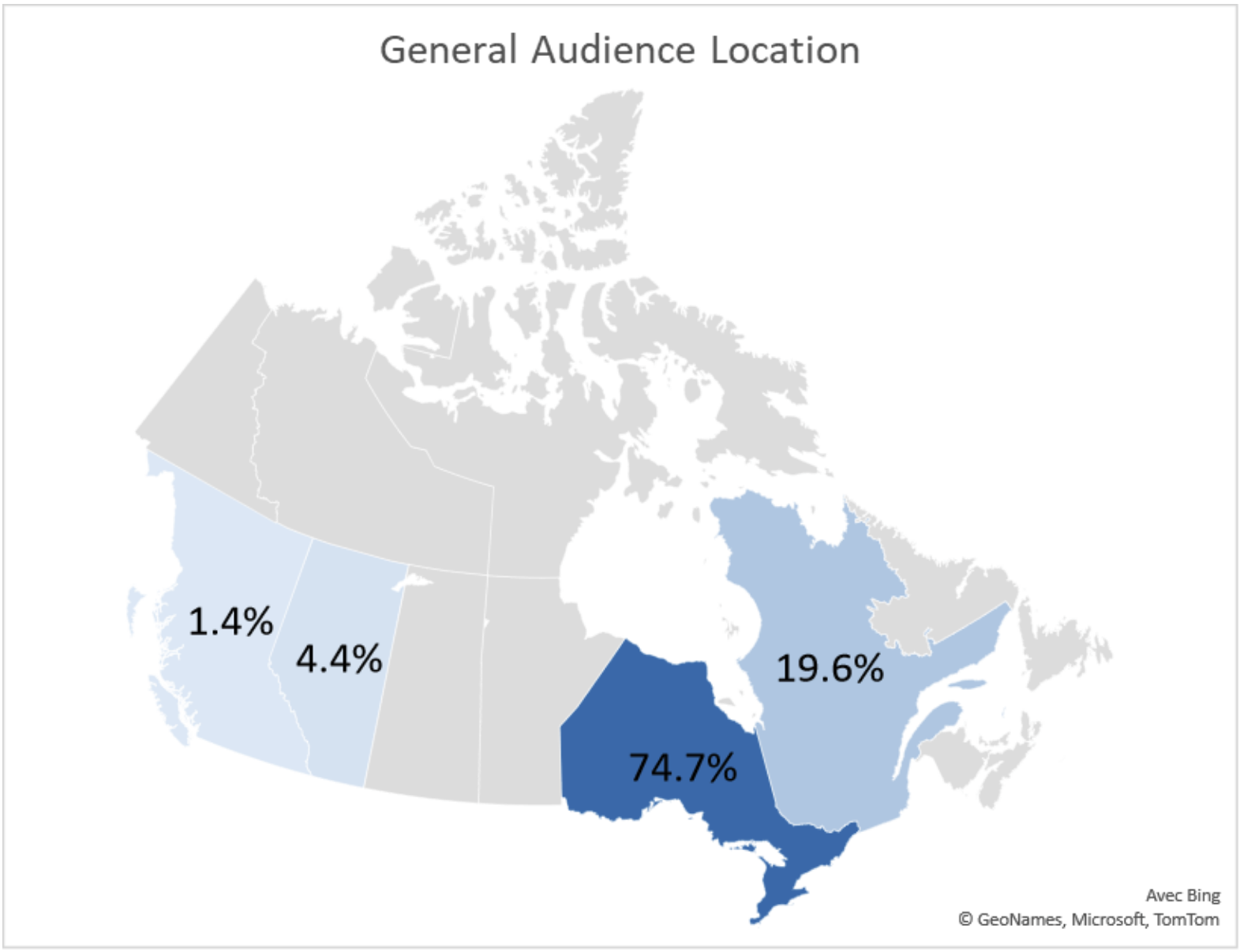}
    \caption{The geographical distribution of the participants that virtually attended the QEYSSat 2.0 Workshop.}
    \label{fig:workshop_audience}
\end{figure}

\begin{figure}[ht]
    \centering
    \includegraphics[width=1\textwidth]{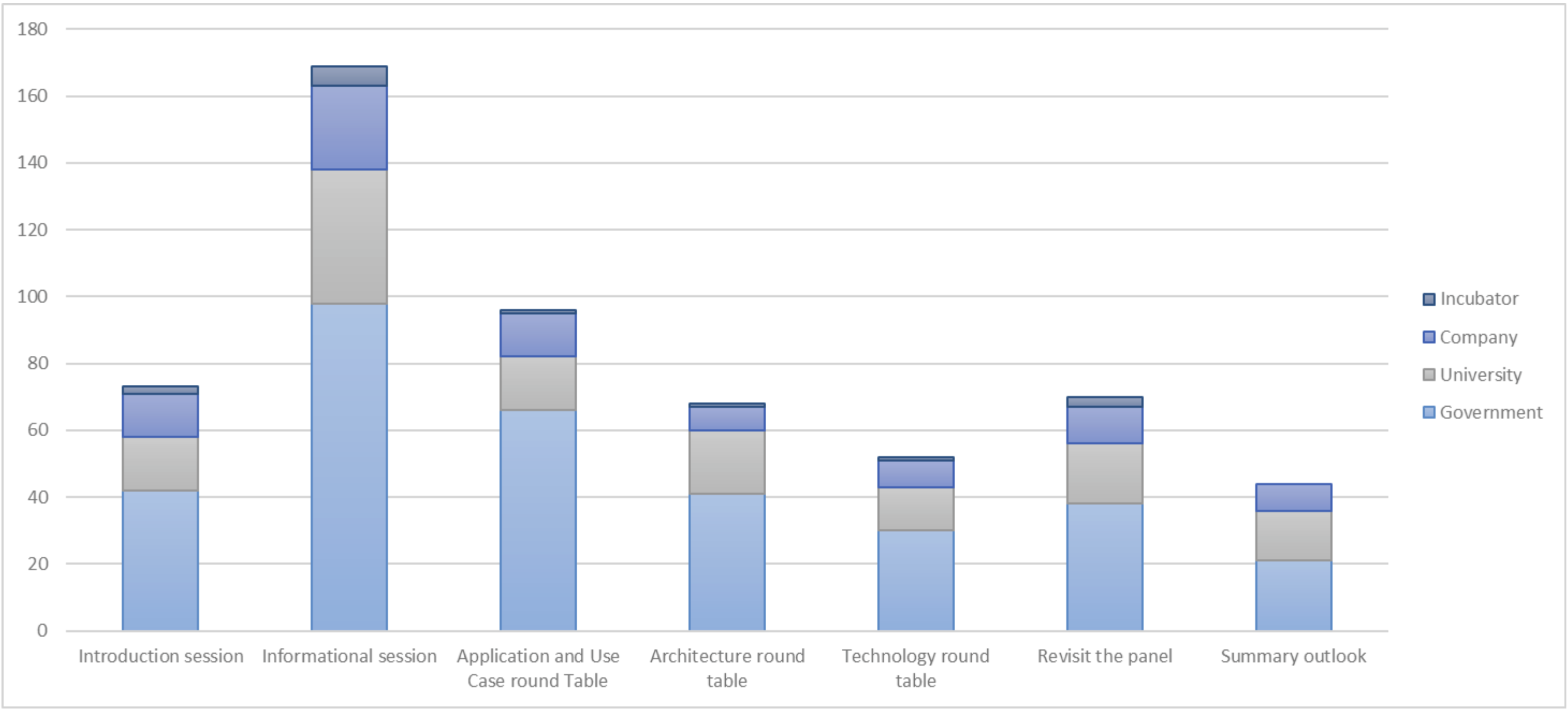}
    \caption{Attendance breakdown per session at the QEYSSat 2.0 Workshop in the categories University, Government, Company and Incubator.} 
    \label{fig:workshop_attendence}
\end{figure}

The objectives of the QEYSSat~2.0 workshop that was held virtually on 15 February 2022 were to:

\begin{itemize}
    \item Identify directions for what Canada should pursue for future quantum communication satellite missions. \\
    \item Educate workshop attendees about status and opportunities that exist now/near future. \\
    \item Gather input and consensus on technologies of priority (Bottlenecks), and when and how to pursue them (Road-map).
\end{itemize}
We had a total of 91 attendees of the 109 invited participants, with 71 colleagues participating in the Applications Round Table, 61 colleagues in the Architectures Round Table, and 54 colleagues in the Technologies Round Tables. The geographical distribution of the audience is shown in Figure~\ref{fig:workshop_audience}, and the attendance breakdown per workshop session is shown in Figure~\ref{fig:workshop_attendence}. 
 
\subsection{Quantum Applications and Use-Cases Round Table Questions}

We asked the workshop attendees the following discussion questions during the Quantum Applications and Use-Cases round table:

\begin{enumerate}
    \item What is the most exciting application of quantum communications technology? \\
    A) Quantum secure communications (i.e. QKD), \\
    B) Distributed quantum computing, \\
    C) Quantum sensing, \\
    D) Other \\
    \item What concerns do you have about the Quantum Internet? Choose all that apply: \\
    A) Difficulty updating infrastructures, \\
    B) Costs, \\ 
    C) Post-quantum algorithms will replace the need for QKD, \\ 
    D) Quantum communication will not be useful for mainstream use, \\
    E) Useful quantum computers are too far in the future, \\
    F) National security concerns, \\
    G) Negative social/economic impacts from new quantum technologies on society \\
    \item What do you feel are Canada’s strengths, weaknesses, opportunities and threats in quantum communication applications? \\
    \item Ethical aspects of Quantum Internet: Do you think the Quantum Internet will be employed by everyone? If niche then are there ethical considerations if some people cannot access Quantum Internet? People with quantum computers could hack people without that technology. \\
    \item Which applications require what rates? \\
    \item How many access points could the future Quantum Internet have? (e.g. will every mobile phone have a quantum access point?) \\
    \item How important is an untrusted satellite vs trusted satellite? (e.g. QKD offers a different kind of security. Entanglement distribution is needed for the Quantum Internet, not just for untrusted-node QKD.) \\
    \item Is latency and link availability a concern? (e.g. QKD allows for pre-storage of keys but sometimes the satellite link is not available.) \\
    \item What are the opportunity cost considerations of implementing the Quantum Internet now vs implementing in the near future vs never? \\
\end{enumerate}

 \subsection{Quantum Architectures Round Table Questions}

We asked the workshop attendees the following discussion questions during the Quantum Architectures round table:

\begin{enumerate}
    \item In your opinion, what architectures/technologies should be the focus of the next Canadian quantum satellite mission? Choose one: \\
    A) Satellite-to-ground quantum communication across Canada, \\
    B) Entanglement distribution across Canada, \\
    C) Quantum teleportation across Canada, \\
    D) Networking quantum devices across Canada, \\
    E) On-board quantum memories on satellites \\
    \item What do you feel are Canada’s strengths, weaknesses, opportunities and threats in quantum communication architectures? \\
    \item How do you see the possible convergence of the classical and quantum internets? \\
    A) Quantum Internet replaces classical internet, \\
    B) Complementary services on separate infrastructure, \\
    C) Future infrastructure will support both modes of communication, \\
    D) Other \\
    \item How many access points could the future Quantum Internet have? (e.g. will every mobile phone have a quantum access point?) \\
    \item How important is a coherent interface to the ground Quantum Internet? \\
    \item What to do about latency issues? Is the variable availability of the link a concern? Quantum links are sensitive to weather, daylight operation, etc. Quantum memories and HAP could help with link times. \\
    \item What are any software layer considerations specific to satellite links? \\
    \item Are there any considerations regarding Standards you wish to discuss that are specific to satellite architectures?
\end{enumerate}

\subsection{Quantum Technologies Round Table Questions}

We asked the workshop attendees the following discussion questions during the Quantum Technologies round table:

\begin{enumerate}
    \item What do you think is the most critical technology of the Quantum Internet? Choose all that apply: \\
    A) Frequency transducers, \\
    B) Quantum memories (ground/satellite), \\
    C) Dedicated fibre networks, \\
    D) Satellite links, \\
    E) High-rate single photon sources, \\
    F) High-rate single photon detectors, \\
    G) Daylight operation, \\
    H) Other \\
    \item Discuss/identify what you think are the main technology bottlenecks? \\
    \item What do you feel are Canada’s strengths, weaknesses, opportunities and threats in quantum communication technologies? \\
    \item Memories in space or on the ground? What are the potential benefits, drawbacks? \\
    \item How to deal with intermittent service disruptions in optical links from satellite passes? (e.g. Optical links are affected by weather whereas RF links provide an ‘on-demand’ service.) \\
    \item Importance / relevance to overcome daylight operation (or under light pollution) \\
    \item Potential to improve photon sources for on-demand, high rate operation (both single-photons and entangled photon pairs) \\
    \item Multiplexing and higher dimensionality of quantum encoding protocols \\
    \item Novel protocols and schemes for quantum communication implementations.
\end{enumerate}

\clearpage

 \section{Quantum Satellite Road-map}
 \label{chap:Roadmap}

\subsection{Phase 1: QEYSSat Extended Science Activities - Year 1 in-orbit}
\label{sec:QEYSSatPhase1}
\begin{figure*}[ht]
\centering\includegraphics[width=0.9\linewidth]{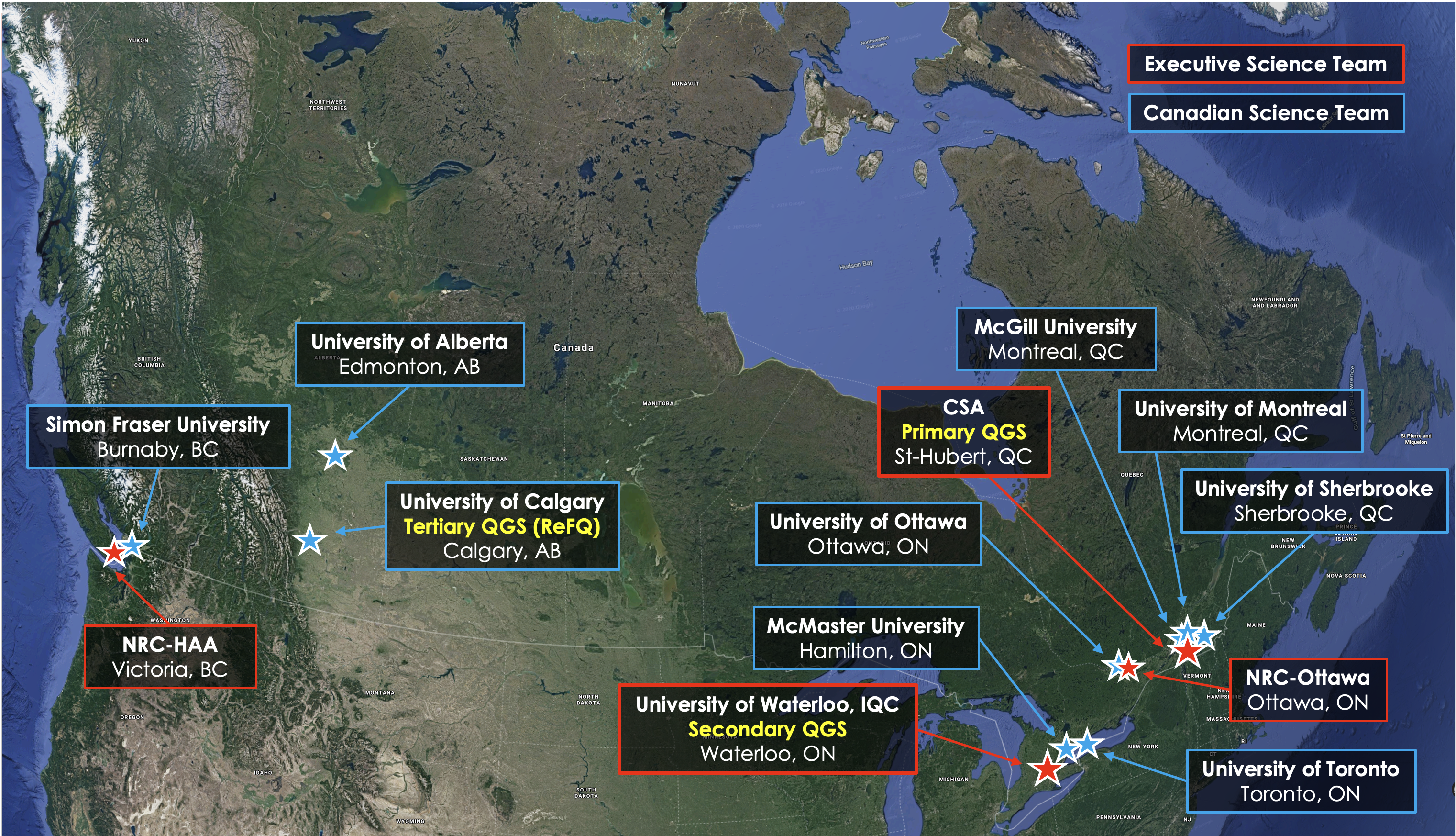}
\caption{Map of the Canadian Science Collaborator Team for the QEYSSat mission. The primary quantum ground station (QGS) will be located at the CSA, and the secondary QGS is planned for the Research Advancement Centre 1 (RAC1) building at IQC, University of Waterloo. A tertiary QGS is planned as part of the ReFQ project at the University of Calgary.} \label{fig:CanadaST}
\end{figure*}

The extended science activities of the QEYSSat mission will be proposed and executed by Canadian and International Science Teams with final approval of all mission activities by the CSA. These activities consist of various experiments and studies to maximise the scientific results of the QEYSSat mission, and to advance scientific knowledge. The Science Teams are composed of researchers across Canada, as well as international collaborators, and will help demonstrate quantum links from multiple geographically-dispersed locations. The Executive QEYSSat Science Team includes the CSA, NRC and the UW/IQC PI Team, as shown in Figure~\ref{fig:CanadaST}. The primary QGS is located at the CSA, the secondary QGS is at UW/IQC (operated by the PI Team), and the tertiary QGS is at the University of Calgary.

After the initial demonstration phase of QEYSSat, the mission will be available to the teams of collaborators to conduct several scientific experiments, as shown in the timeline Figure~\ref{fig:QEYSSatTimeline}. Figure~\ref{fig:CanadaST} provides an overview of the Canadian Science Team collaborators. This work will include:
\begin{itemize}
    \item Ground Stations
    \item ReFQ: Reference Frame Independent Quantum Communications using a Quantum Source on QEYSSat
    \item Quantum Link Science
    \item QGS Site Characterisation \& Light Pollution Studies
    \item Quantum Ground Networks
    \item Quantum Memories \& Interfaces
    \item Quantum Sources
    \item Novel Space Technologies
    \item Quantum Communication Theory
    
\end{itemize}

Detailed descriptions of QEYSSat Canadian Science Team mission activities were omitted from this white paper and will be subsequently published elsewhere.

An additional ground station, owned and operated by the University of Calgary (UC) team, will contribute several areas of expertise to the mission, including a QGS to link with QEYSSat, and the capability to receive quantum signals from the downlink source for the ReFQ project. The ReFQ project is funded by the joint UK-CAN program, called the `Reference-Frame Independent (RFI) quantum communication for satellite-based networks' (ReFQ), and aims to launch a WCP source on QEYSSat to study novel protocol space-based QKD demonstrations. The ReFQ project is very important for QEYSSat in multiple ways: It allows for the implementation of the goal of studying a QKD downlink, as well as studying novel protocols that could simplify the design and operation of a satellite QKD link. This project is a collaboration of Canadian partners UW, Honeywell-Canada, UC, and the CSA, as well as UK partners Craft Prospect Ltd., University of Strathclyde (UStrathclyde), and the University of Bristol (UBristol).

\subsection{Phase 2: QEYSSat Extended Science Activities - Year 2 in-orbit}
\label{sec:QEYSSatPhase2}
\begin{figure*}[ht]
\centering\includegraphics[width=\linewidth]{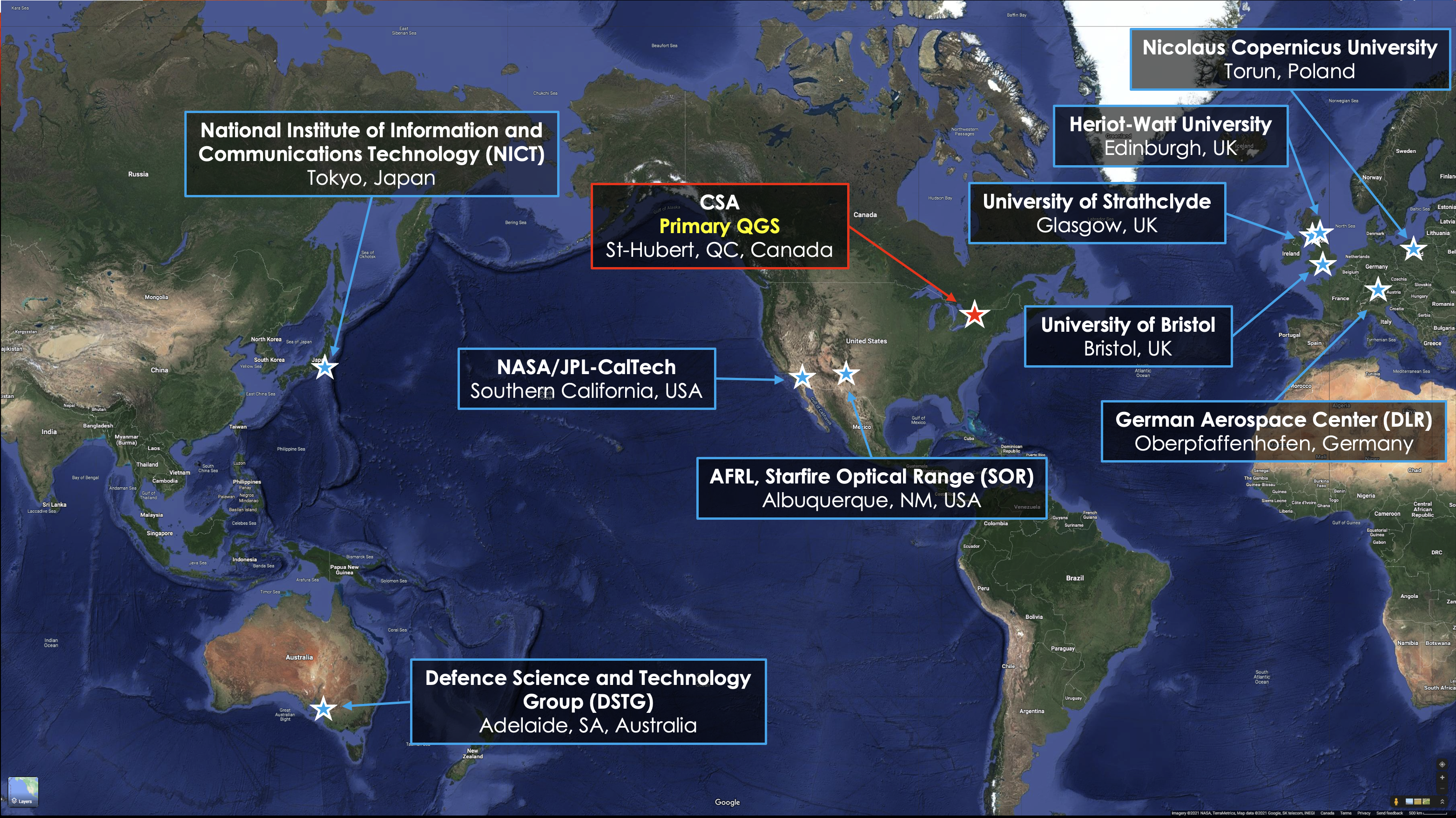}
\caption{Map of International Science Team collaborators for the QEYSSat mission (Updated Q2 2023).} \label{fig:InternationalST}
\end{figure*}

The International Science Team activities will commence after the initial demonstration phase of QEYSSat, and after the Canadian teams have had opportunities to link with QEYSSat, as shown in the timeline Figure~\ref{fig:QEYSSatTimeline}. An overview of the current International collaborators is shown in Figure~\ref{fig:InternationalST}. QEYSSat Year 2 in-orbit will also include the next stage of extended science activities with both the Canadian and International Science Teams. This work will include:
\begin{itemize}
    \item Quantum Ground Stations
    \item Quantum Sources
    \item QGS Site Characterisation \& Light Pollution Studies
    \item Quantum Memories \& Interfaces
\end{itemize}

Detailed descriptions of QEYSSat International Science Team mission activities were omitted from this white paper and will be subsequently published elsewhere.

\subsection{Phase 3: QEYSSat 2.0 Mission}
Recommended Mission Objective: {\bf Quantum Teleportation Across Canada.} \\

\begin{figure*}[ht]
\centering\includegraphics[width=0.9\linewidth]{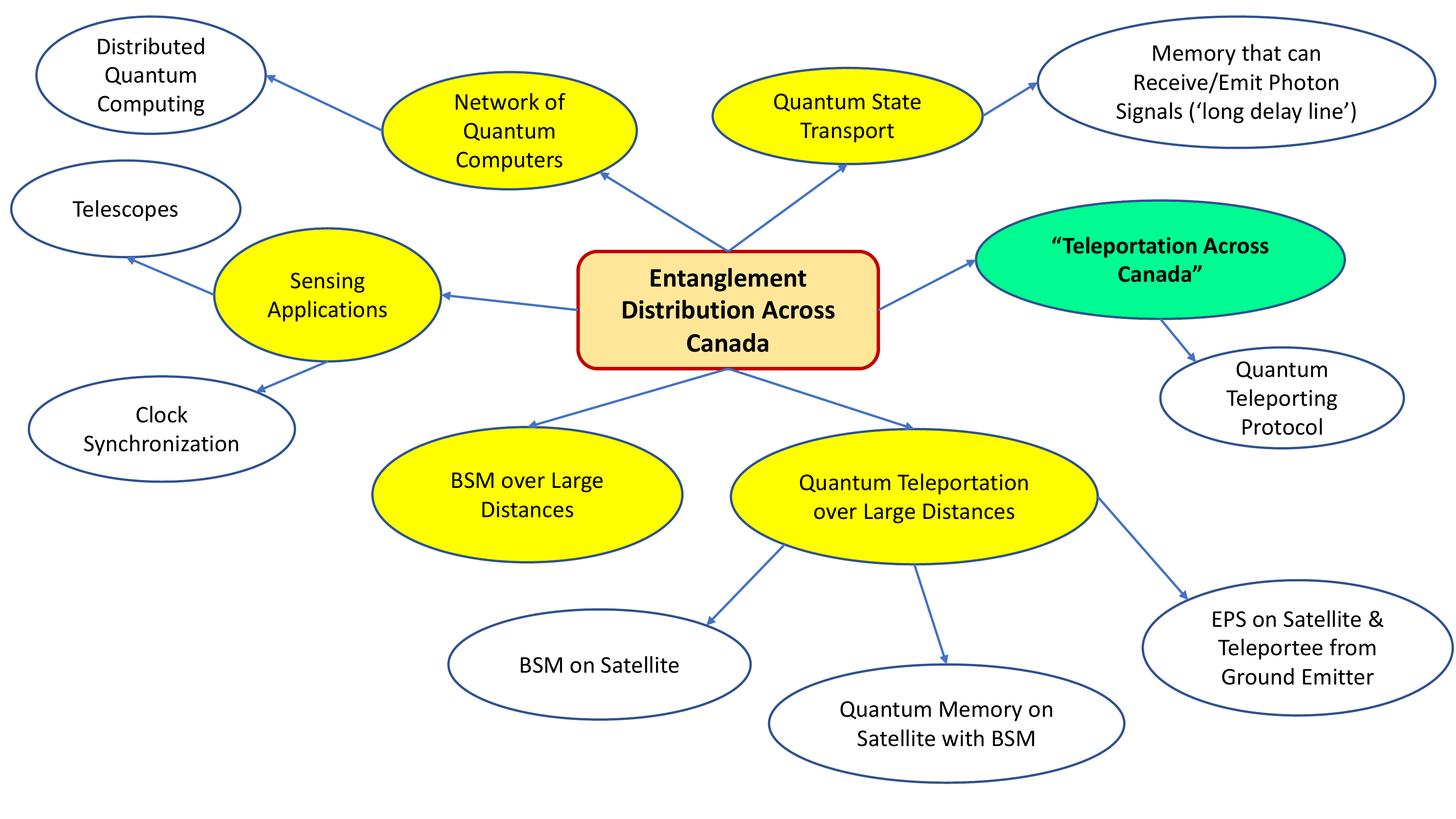}
\caption{Mind map of possible use cases for a Canadian entanglement distribution network.} \label{fig:CanadaEnt}
\end{figure*}

We recommend that future QEYSSat~2.0 missions should implement a network that distributes quantum entanglement that spans the entire country in order to make major steps towards a Canada-wide Quantum Internet. This entanglement resource can be used for various applications, with the ultimate objective to enable quantum teleportation across Canada (see Figure~\ref{fig:CanadaEnt}). This system will therefore  provide an important communications backbone that will enable some of the application scenarios.  Furthermore, this scenario was selected because solving its technical challenges will provide immense advance and boost to the quantum technology in this area, and  provide Canada with new world leading capabilities for quantum communications. 

It is therefore recommended that the primary QEYSSat~2.0 mission objective is {\bf Quantum Teleportation Across Canada.}

An important criterion for the future QEYSSat~2.0 mission is that it must enable quantum communication that can reach all regions of Canada in the North-South and East-West directions. Furthermore, interfaces to international / global locations is also desired, and use-cases could include to serve Canadian entities operating outside of the country. Another important aspect is that given space-QKD solutions are becoming commercially available, we recommend that the QEYSSat~2.0 mission and the developed technologies should represent a significant technological advance over the state-of-art, and enable Canada to take a clear leadership in this technology.

This technology could be suitable for the following applications we outline next.

\subsubsection{Canada-Wide Quantum Secure Communication}
This QEYSSat~2.0 mission concept should provide the means to establish secure quantum keys that, first of all, cover ALL of Canada. Any ground station site at Point-A to any Point-B shall be connectable, and the accessible area stretches from coast to coast to coast.

Specifically, the mission should involve:
\begin{enumerate}
	\item The technical capability for QKD between ground locations, as well as satellite to ground (e.g. securing satellite data/ satellite control)
    \item Should involve entanglement distribution to enable `untrusted' operation of the satellite.
    \item Achieve quantum channels under high background light such as daylight and light pollution, and also reject the light scattered off high-altitude satellites.
	\item Interface with stationary as well as moving ground sites (including aircraft, HAP, ships). 
	\item Be suitable to interface with ground networks at the ground station (for instance, an EPS located at a ground station is a good example scenario).
	\item Enable the relatively easy deployment of secure communication links (easy / rapid deployment) in remote areas
	\item This mission should also be suitable to interface with other missions and enable global QKD links to locations around the world. 
	\item Allow agile networks, which includes portable ground stations, but also could include movable platforms such as HAP, aircraft, UAV, ships and even under-water links.
\end{enumerate}

\begin{figure*}[ht]
\centering\includegraphics[width=0.9\linewidth]{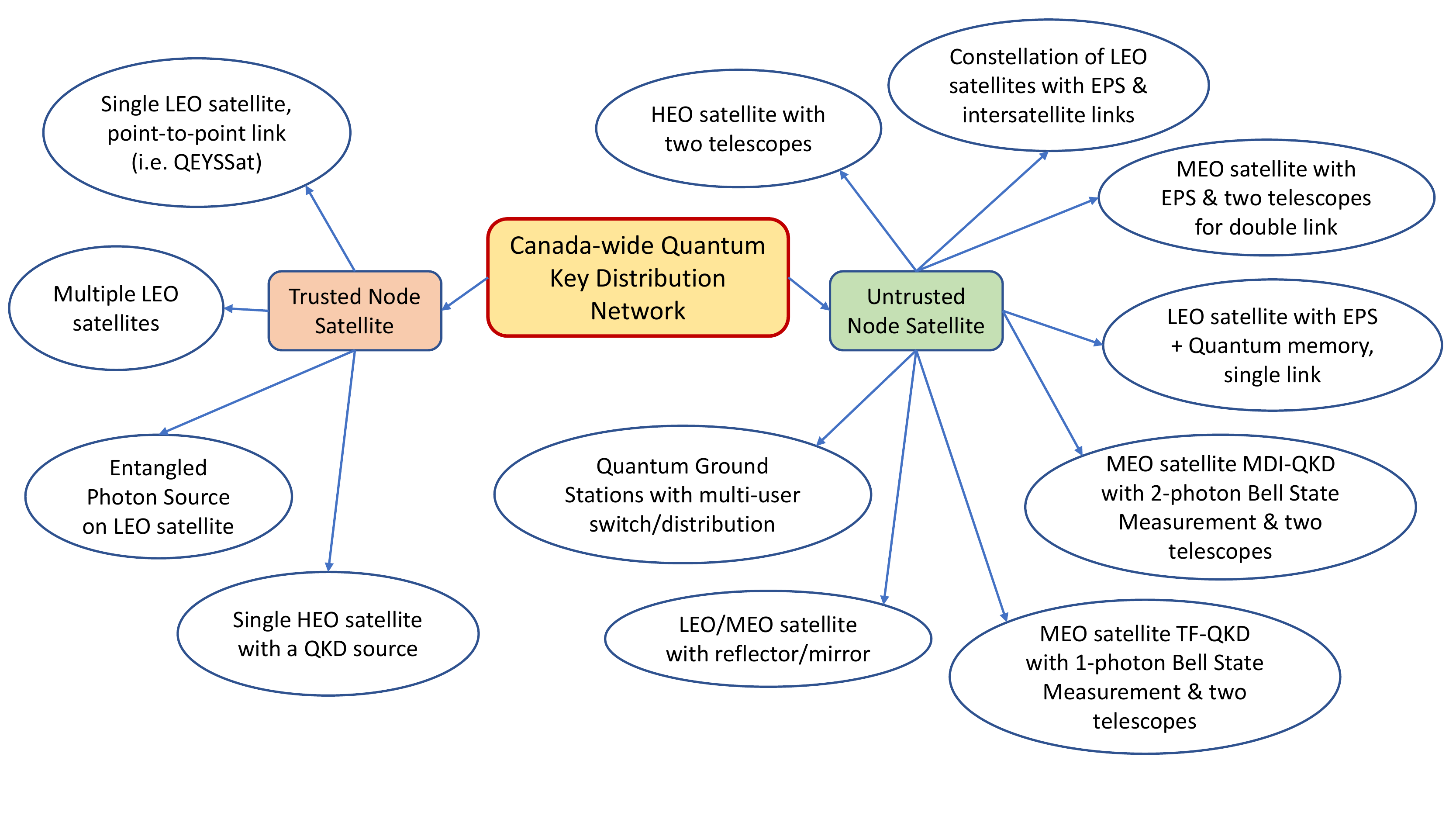}
\caption{Mind map of possible quantum satellite solutions to achieve both trusted and untrusted-node QKD across Canada.} \label{fig:CanadaQKD}
\end{figure*}

An overview of several of the possible technical solutions are shown in Figure~\ref{fig:CanadaQKD}, and the exact selection of technical solutions will be a matter of future studies. The most likely solution will be an entanglement source for operation on-board a MEO satellite, equipped with a double downlink.

Specifically, this use-case will require technology advances such as operation over very high channel losses in order for ground apertures to be manageable and portable. This will demand new development into better single photon sources with very high rates ($>1$\:GHz), multiplexing of optical channels, and photon detectors with improved timing/synchronisation. Furthermore, the newly developed quantum links should be suitable for very high link losses (increase tolerance up to 60\:dB, currently its 40\:dB!) to enhance the versatility and practicality of the system.

\subsubsection{Canada-wide Quantum Entanglement Network} 
The QEYSSat~2.0 mission should form the backbone for a Canada-wide entanglement distribution network as this is an important prerequisite for establishing a future quantum internet, and its applications include distributed quantum computing, secure communication, and quantum enhanced sensing.

\begin{figure*}[ht]
\centering\includegraphics[width=\linewidth]{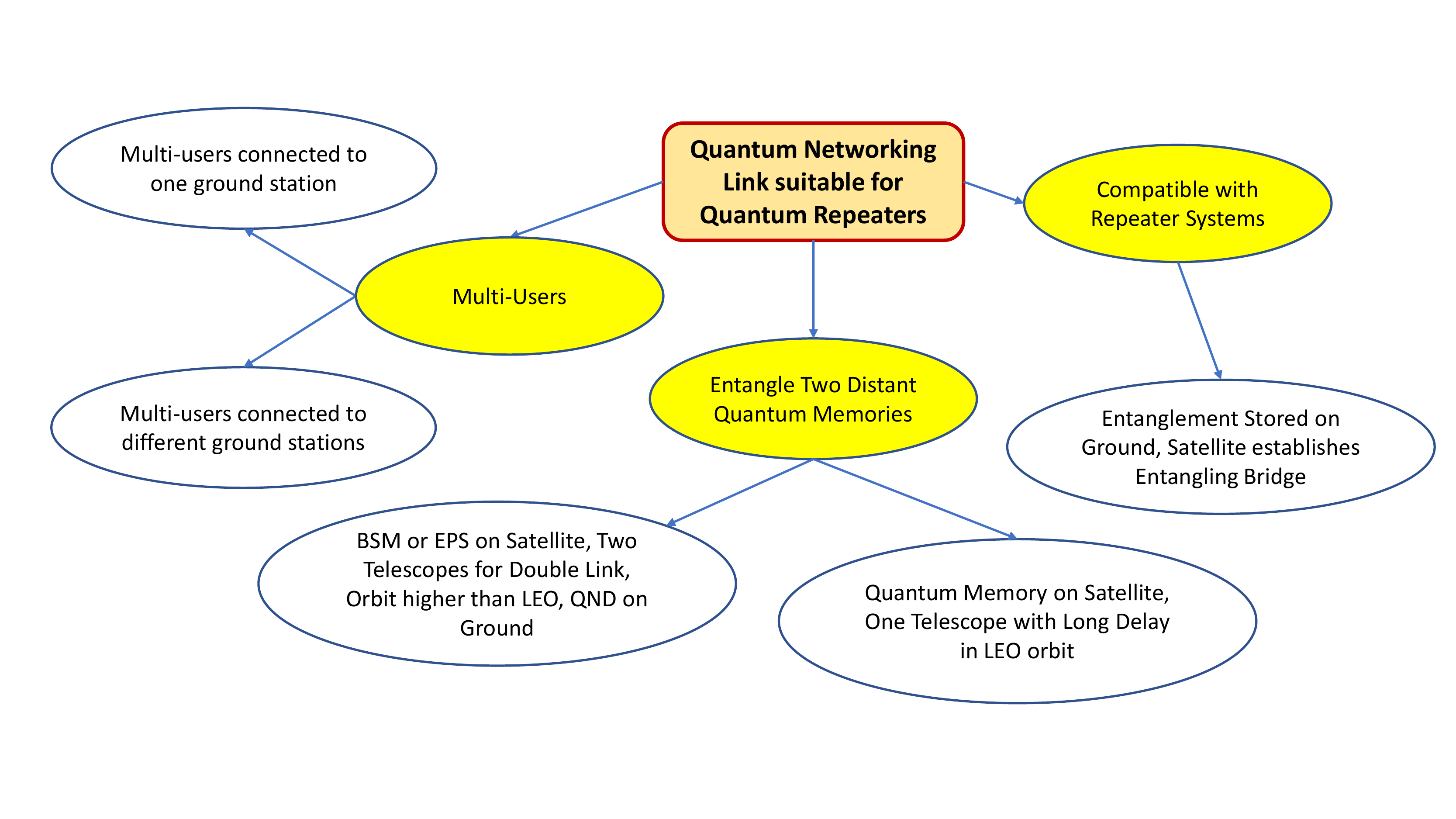}
\caption{Mind map of possible solutions for large scale quantum networks.} \label{fig:CanadaRepeat}
\end{figure*}

A Canada-wide quantum entanglement network would enable the following scenarios:

\begin{enumerate}
    \item Provide means to interface quantum processors / simulators between two ground quantum computing nodes, or interface a user with a quantum processor.
    \item Provide interfaces for quantum computing / processing systems in space that may be used for efficient pre-processing of satellite data prior to their ground communication. 
    \item Enable entanglement swapping and teleportation over large distances, and between different systems such as quantum memories or stored qubits.
    \item Backbone for secure communications.
    \item Assist in sensing applications by entangling sensors (e.g. magnetometers, gravimeters, or probing at one wavelength but detecting at another, detecting methane, other gases, etc).
    \item Enhance large-baseline telescopes.
\end{enumerate}

\subsubsection{Technology Advancement Considerations}
An important aspect of the technical solution will be to implement \textbf{deterministic sources of single and entangled photons.} There are several ways to accomplish this using either quantum-dot based systems or \textbf{quantum memories}, see Figure~\ref{fig:CanadaRepeat}. Another important aspect is to involve quantum memories in either heralding, synchronising or delaying photons to achieve the desired operation. Both scenarios with quantum memory systems on the ground and in space are to be considered.

A very interesting long-term perspective is to implement quantum memories, and ultimately quantum repeater nodes in space, with sufficient storage times at long enough time scales that the system can await a favourable link availability, such as using certain orbital positions or even for good weather. The space-based quantum computing or quantum network resources, including space quantum
 processors, could also deliberately make use of an environment that offers good pre-cooling, vacuum and micro-gravity.

The orbit of the satellite segment will most likely involve a higher altitude platform, such as MEO satellite, or constellations of LEO satellites.  Given the high latitudes of much of rural Canada, constraints on the allowable orbits will further limit the possible scenarios, and GEO may not be a preferred solution.

\subsubsection{Tests on the Foundations of physics} 
These missions allow us to study several fundamental scientific questions about quantum physics in relativistic settings, which are important to fully understand to operate a quantum network in a space environment. These explorations also help shed new light on the interplay of quantum mechanics and gravity theories. 

Specifically, the QEYSSat~2.0 tests and experiments could help explore the following:

\begin{itemize}
    \item Long-range entanglement and long-distance teleportation over distances (and velocities) not possible on ground.
    \item The stabilisation of quantum channels could improve mapping of gravitational potentials and structures.
    \item Entangled quantum memories in space offer unique opportunities into probing gravitational shifts that only massive quantum systems will experience.
\end{itemize}

\subsection{Technology Road-map}
This section aims at presenting the technology development road-map to achieve the QEYSSat~2.0 mission. It includes the presentation of an exemplary design concept of the possible QEYSSat~2.0 mission. This will be used to identify some of the technological bottlenecks in quantum and classical technologies. The technical readiness level (TRL) as well as their space suitability will be highlighted, and a proposal for a timeline of the technology development is given.

\subsubsection{Conceptual Design}
\begin{figure}[ht]
    \centering
    \includegraphics[width=0.8\textwidth]{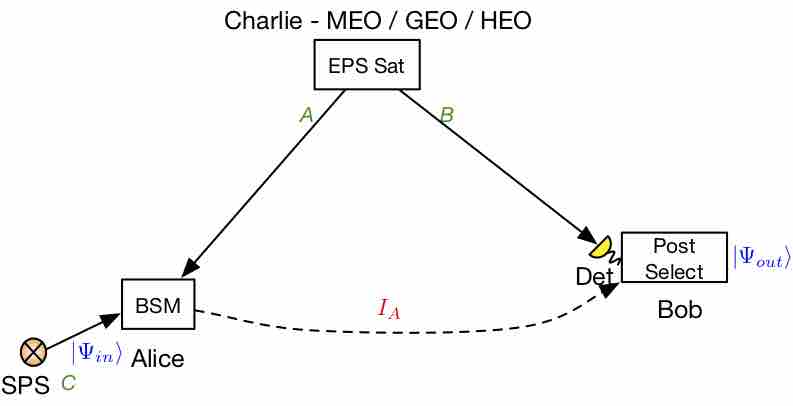}
    \caption{The generic teleportation protocol uses central node (`Charlie') for entanglement distribution, who releases the photons $A$ and $B$ to the stations `Alice' and `Bob'. Alice performs the BSM operation with the input photon $C$ in a state $\ket{\Psi_{in}}$, while Bob performs a corrective unitary operation, to obtain the output photon in a state $\ket{\Psi_{out}}$. The exchange of classical information needed to execute the protocol is shown as dashed line, $I_{A}$.}
    \label{fig:teleportationScheme_generic}
\end{figure}

A possible long distance quantum teleportation scheme is highlighted in Figure~\ref{fig:teleportationScheme_generic}. 
The satellite node (`Charlie') establishes entangled signals between the two ground sites (`Alice' and `Bob') using one of multiple possible protocols. The original (input) photon $C$ is emitted in a state $\ket{\Psi_{in}}$ by a single photon source (SPS), which is synchronised with the arrival of the photon $A$, which was sent from Charlie to Alice. Alice performs a BSM between photons $A$ and $C$, and after this measurement is complete, sends the classical result (dashed line, $I_{A}$)  to Bob. Bob, must ideally  perform the corrective Unitary $U_{B}$, in order that the state of photon $B$, $\ket{\Psi_{out}}$ becomes identical to $\ket{\Psi_{in}}$. For a demonstration experiment, Bob's unitary operation can also be done in post-selection (i.e., after the detection of Bob's photon) by sorting the measurement results. However, preferably Bob has a delay memory so he can hold on to the photon until the BSM result arrives from Alice.  Note that in the fully coherent protocol, Bob must employ a quantum memory to store the received quantum state until the result from Alice's BSM is available.

\subsubsection{Conceptual Design Feasibility}
We provide a high-level link performance estimation in order to get a general insight into the feasibility of the overall concept. It might be surprising that while Canada-wide quantum teleportation is challenging, the photon rates and duration for a successful demonstration are clearly feasible.

The simplest protocol is a so-called `memory-less' scheme, which operates entirely in post-selection (see Figure~\ref{fig:teleportationScheme_generic}). We make the following realistic assumptions for the baseline demonstration, using the well-proven method to implement teleportation since the first experiments in 1997 \cite{Bouwmeester1997}. `Charlie' is located on a GEO satellite (range of 36,000\:km), and sends entangled photons via transmit apertures of 0.5\:m at a wavelength of 810\:nm. The ground sites (`Alice' and `Bob') have receiver apertures of 2\:m, and the resulting link attenuation is approximately 40\:dB per free-space link. Charlie uses an EPS based on SPDC, with a 5\% probability to create a photon pair,  pumped at 1\:GHz rate. The input photon at Alice is created using another SPDC-based source with the same parameters. The rate for 4-fold photon detection events, which is also the rate of teleportation, is approximately 0.018 per second. At that rate, it will take about 16\:hrs of continuous data accumulation to perform about 1000 teleportation events, and thus conclusively demonstrate the protocol. However, all other things equal, if Alice where to use a \textit{deterministic} single-photon source, rather than SPDC which is probabilistic, to create the input photon, then the rate will go up by about a 10-fold, and only about 1.5\:hrs of continuous accumulation of data are needed to collect 1000 events. A more comprehensive overview of this and other experimental configurations, and their rates is given in  Appendix~\ref{sec:additional_teleport_schemes}. 

There are many technical challenges to overcome to accomplish long-range teleportation. However, it is very promising to see that even the least efficient setup from 1997 would still allow for reasonable transfer rates and a successful demonstration. The rates and success probabilities will only improve with technological advances as discussed below.

It is clear that other, much more efficient teleportation schemes are possible and should be pursued for QEYSSat~2.0.  In particular, these could involve quantum memories on ground or space to overcome post-selection, as well as deterministic photon sources.  A detailed analysis and trade of all the schemes can be undertaken in full detail in the future. Note that in order to perform a complete, or full teleportation protocol, a quantum memory system is required to ensure the received photon at Bob can be stored until the BSM result from Alice arrives via classical communication channel, as this is needed to set the unitary operation at Bob.

\subsubsection{Technological Bottlenecks}
\label{sec:TechBottlenecks}
\begin{figure}[ht]
\begin{center}
  \includegraphics[scale=0.65]{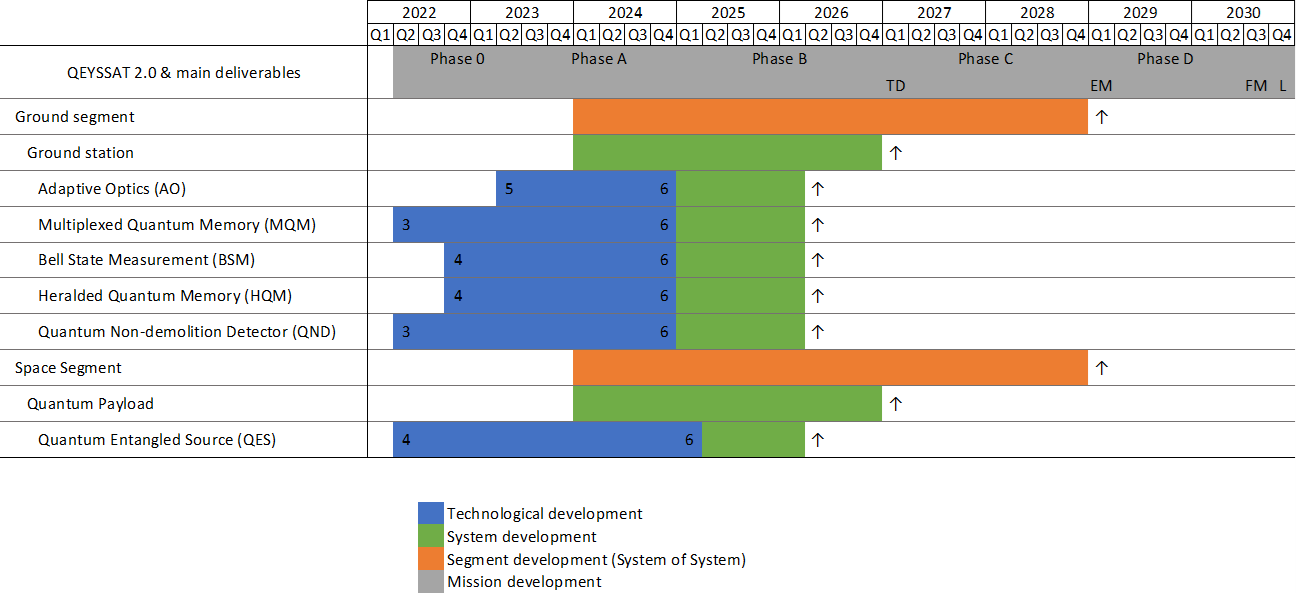}
  \caption{Proposed high level schedule for the quantum and ancillary technology development}
  \label{fig:GanttHighLevel}
\end{center}
\end{figure}

In order to achieve Canada-wide quantum teleportation, many technologies - quantum related or not - shall be used in a space environment, far from being as ideal as the one in a laboratory. To get a complete picture of the development to be done, a Technology Readiness Level (TRL) assessment is needed. The following Table~\ref{tab:technology_list_priority} identifies and presents the different critical technologies along with their descriptions and priority of development. Figure~\ref{fig:GanttHighLevel} presents their initial TRL and the level to obtain prior to integration into a system.


\newpage

\begin{longtable}{|p{2.2cm}|p{10cm}|c|}
\hline
Technology & Description & Priority \\
\hline
\hline
 Deterministic and high-rate quantum sources & Emitters of single and entangled photons with deterministic photon statistics, such as quantum dot or memory based systems. This could also involve SPDC-based sources with multiplexing features. To be efficient, an EPS with a repetition rate of $\geq1$\:GHz and an efficiency of $\geq50\%$ should be employed. & High \\
 \hline
  Multiplexed quantum memory & This is  needed to enhance the channel transfer rate, and could involve temporal, spatial or spectral multiplexing. & High \\
 \hline
  Heralded quantum memory and QND & Fully scalable  quantum communication  requires the capability of observing when a photon successfully passed the channel while leaving its quantum information intact. This requires either QND measurement with an efficiency of $\geq90\%$, or a heralded quantum memory. & High \\
\hline
BSM between photons sent from satellite to ground & Quantum teleportation inherently relies on a BSM operation. Substantial R\&D effort is required to overcome the  challenges imposed by the rapidly varying links between ground and  satellites.  Solutions needed to establish real-time  stabilisation and synchronisation of photon arrival times. The success rate of entanglement swapping based on linear optics is restricted to $50\%$. Auxiliary photons can be used to improve the swapping probability. & High \\
  \hline
 Adaptive Optics & Wave-front correction for the ground-based systems are required for various applications, including coupling received signals to a single mode system for improved coupling for two-photon interference, quantum memories and better detectors (e.g., SNSPD). AO could also be useful for improving the uplink beam pointing. & High \\
  \hline
 Background noise rejection & The operation of quantum links under conditions of daylight or severe light pollution will require very stable and precise filtering methods. Impact of Doppler and other drifts must be accounted for, as well as compatibility with multiplexed channels. & Medium \\
 \hline
 High-performance detectors & Specific development points are: InGaAS APD devices with higher efficiency, and fast gating for free-space channels, array-SPAD devices (any technology) with high efficiency (i.e., $>90\%$) and high rates, and suitability for multiplexing of channels. & Medium \\
 \hline
 Multiplexed channels & The bandwidth  of a typical optical  channel should allow for 100's of superimposed channels using WDM, as well as temporal and spatial multiplexing. The suitable multiplexed quantum emitters and quantum detectors must be developed. The channel itself and necessary filtering must also be developed. & Medium \\
 \hline
 Quantum memory in space & The quantum memories need to be robust and stable while also suitable for very high read / write rates. Storage times need to be long to account for the round trip times. The required storage time depends on the network design. However, to be efficient, memories with at least ms-range storage time and $\geq 50\%$ efficiency are required. & Medium \\
\hline
 Phase-stabilised channel & In addition to the BSM operation, phase stabilisation is desirable as the most efficient quantum communication protocols and applications demand phase stabilised quantum channels. Clearly this poses a huge challenge given that a satellite-to-ground link is inherently variable. & Medium \\
 \hline
 Ultra-narrow filters & Narrow filters such as atomic line filers should be employed to extract the faint quantum signals from strong background light such as daylight or light pollution. One major issue will be the Doppler shift caused by relative motion and gravitational shifts. & Medium \\
 \hline
  Classical Support System & DAQ such as time-tagging, processing, feedback and control & Low \\
  \hline
  \caption{Overview of the technologies identified as required for the QEYSSat~2.0 mission.}
    \label{tab:technology_list_priority}
\end{longtable}

\newpage

\subsubsection{Space Suitability of Systems, sub-systems and modules}
This section covers different aspects of the space suitability of the possible QEYSSat~2.0 mission, particularly radiation hardening of the payload and its critical electrical components, SWaP considerations, and cooling capabilities to allow for the use of superconducting cryogenic detectors and of memories (cryogenic or laser cooled).

\bigskip
\emph{Quantum sources in space ---} Typical ground-based quantum sources have power, size and mass specifications that are beyond space use. While some space-suitable systems have recently been demonstrated, it remains to be seen if these sources are suitable for QEYSSat~2.0.

WCP-based QKD source modules for space have been developed, and becoming commercially available, with systems from CRAFT Prospect (UK), as well as from others, with several under development for national space missions such as in Germany, Italy, and Japan. 

EPS for space have been developed by multiple academic groups,  and are also commercially available, including from SpeQTRAL (Singapore) and Fraunhofer (Germany). These devices include the pump laser system, temperature control and stable fibre coupling. However, to our knowledge, these commercial devices are not suitable for teleportation or quantum repeaters.

\bigskip
\emph{Quantum detectors in space ---} Space radiation is one the main issues when looking at performance for an optical payload. 
Radiation tends to degrade the performance of electronic parts over time. 

For any of the semiconductor based single-photon detector technologies (see Section~\ref{sec:detectors}), it is well established that the dark counts (a critical limiting factor in quantum experiments) and read-out noise will increase after exposure to radiation due to the ionizing dose and displacement damage. In the case of silicon-based SPAD, which are useful for wavelengths of visible light to 1000\:nm, multiple radiation tests have shown their susceptibility, and mitigation via annealing has been studied. \cite{Anisimova2017,N.Sultana,EPJQuantum}. Detectors for telecom wavelengths (including InGaAs APD or SPD) have not been demonstrated in space yet, however multiple studies are under way.

Radiation exposure may also induce chaotic behavior (single events) in electronics. There are known mitigation techniques that can be used to protect the detectors or other optical or electronic parts from degradation: 

\begin{itemize}
\item On one hand, a high performance shielding can be used, such as the use of high-Z material layers (mainly tungsten or tantalum) \cite{NASAShields-1}
\item On the other hand, active heating and cooling devices attached to the detector can be used to either decrease the temperature of the detector to increase its performance while in use, or to anneal the detector when dormant.
\end{itemize}

\bigskip
\emph{Radiation hardening of optical components ---} Radiation also degrades optical materials that includes phenomenon such as darkening of optics, hence inducing loss in optical transmission, as well as changes in birefringence and polarisability of optical components - even more when quantum encryption uses polarisation states of light. Indeed, radiation can create a densification of optical materials \cite{D.Doyle} inducing an undesirable and uncontrollable birefringence change of optical materials, hence interfering with the measurement of the polarisation state of the incoming photon flux. 

Mitigation for this issue should be considered carefully during the optical materials and coatings selection. Review of literature as well as de-risking campaigns should be completed to obtain expected performance at EOL. This mitigation applies to the aft-optics - since they will possibly impact the quantum state detection - as well as memory capabilities - if one uses optical fibres.

\bigskip
\emph{SWaP limitations ---} Size, weight and power are what drive the architecture and design of a satellite and its payloads. From a QEYSSat~2.0 mission perspective, this means a limited size for the optical system assembly (OSA) including the front-end optics, the aft-optics and the QKD modulator/demodulator, as well as for the electronics box (EB). 

As examples, current state of the art of commercial OSA sizes and weights range from being able to be on-board a CubeSat to being on-board large Telecommunication and Earth Observation satellites as shown below.

\begin{figure}[ht]
\begin{center}
  \includegraphics[scale=0.2]{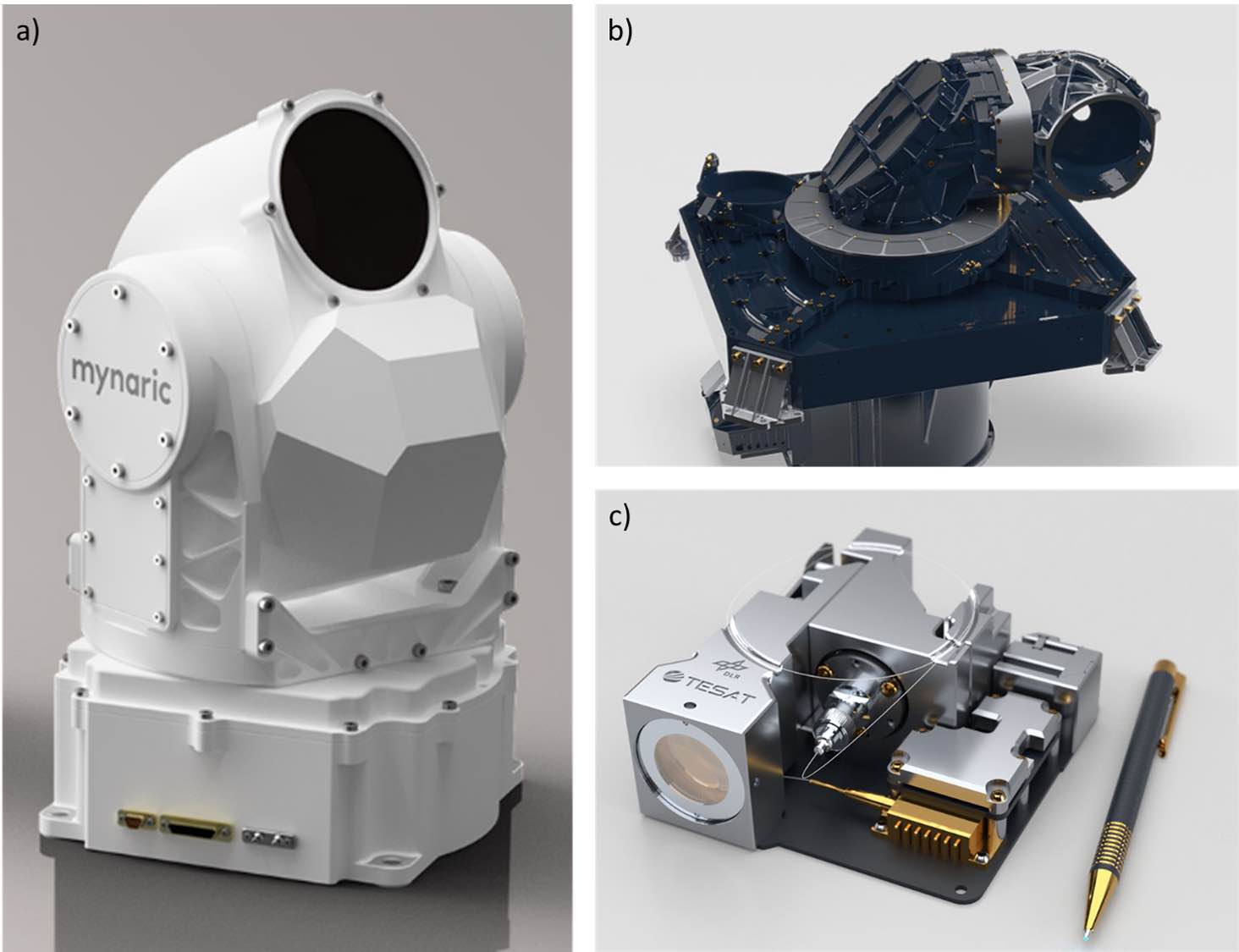}
  \caption{Examples of optical terminals for satellite-to-satellite and satellite-to-ground communication.}
  \label{fig:CommercialOpticalTerminal}
\end{center}
\end{figure}

\begin{figure}[ht]
    \centering
    \includegraphics[width=7cm]{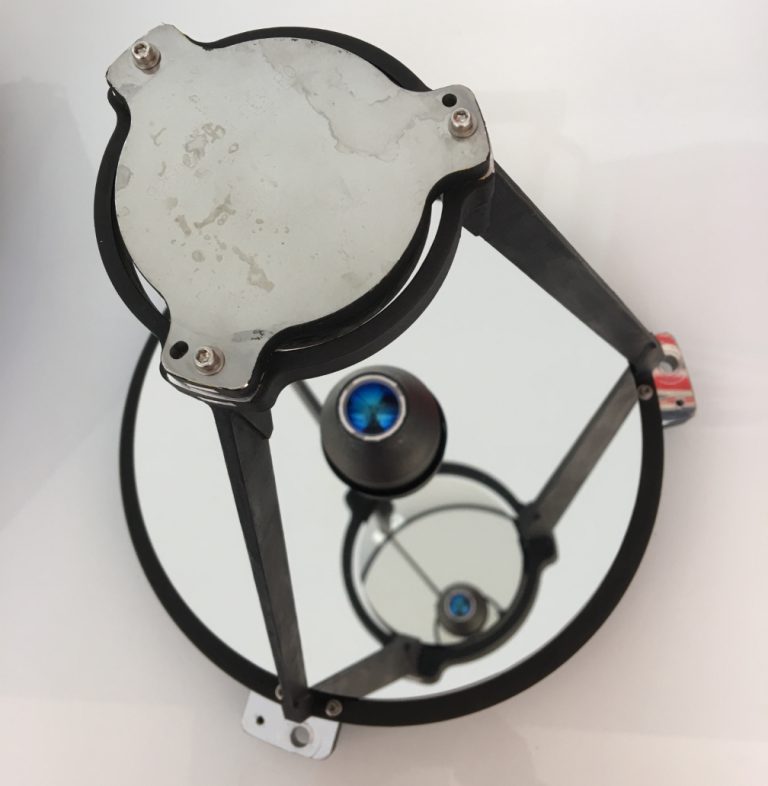}
    \caption{Honeywell-COMDEV On-axis all aluminum 25\:cm Telescope – Low cost precision mirror telescope target for image, Quantum Key Distribution and Geo feeder links. Credit: Honeywell.}
    \label{fig:HoneywellTele}
\end{figure}

\begin{figure}[ht]
\begin{center}
  \includegraphics[scale=1.25]{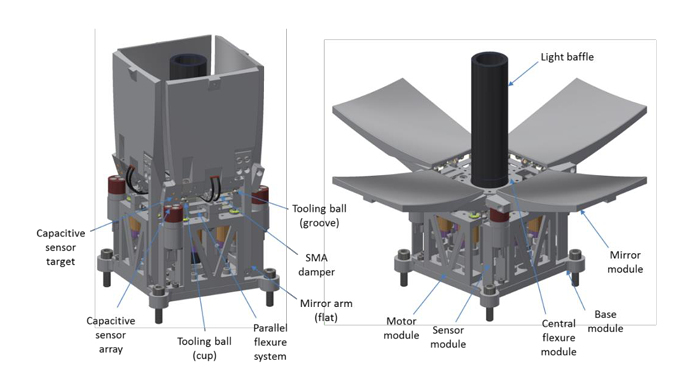}
  \caption{Concept of a deployable telescope with an effective aperture of 30\:cm while fitting in a 10 x 10 x 15 $cm^3$ volume \cite{schwartz2021active}.}
  \label{fig:DeployableTelescope}
\end{center}
\end{figure}

Some commercial examples of laser terminals and their parameters are listed in  Table~\ref{tab:optical_terminals} and illustrated in Figure~\ref{fig:CommercialOpticalTerminal} \cite{mynaric,tesat}.

\begin{table}[]
\caption{\label{tab:optical_terminals}
Optical Terminal Parameters.}
\begin{center}
\begin{tabular}{ |p{1.75cm}|p{2.75cm}|p{1.25cm}|p{1.25cm}|p{1.75cm}|p{3cm}| } 
 \hline
 Unit & Dimension ($cm^3$) & Weight ($Kg$) & Power ($W$) & Optical Aperture ($mm$) & Link Data Rate \\ 
 \hline
 Mynaric Condor Mk3 (a) & 35x21x17 (OSA)\newline16x34x26 (EB)
 & N/D & N/D & 80 & Up to 10Gbps @ 8000 km \\ 
 \hline
 TESAT LCT135 (b) & 60x60x70 & 53 & 150 & 150 & Up to 1.8Gbps @ 80000 km bi-directional \\ 
 \hline
 TESAT CubeLCT (c) & 9x9x3.5 & 0.4 & 10 & 25 & 100 Mbps LEO to GND \newline 1Mbps GND to LEO \\ 
 \hline
\end{tabular}
\end{center}
\end{table}

Larger optical apertures are also possible but usually require a custom design and build. For instance, 20-25\:cm aperture telescopes begin to be the standard for earth observation \cite{simera}, and it is possible to consider building a quantum payload based on an existing telescope platform to decrease the risks as well as the cost. Indeed, Canadian entities are working on telescope solutions with these apertures, such as Honeywell's COMDEV unit (Kanata), implementing a 25\:cm all-aluminum telescope (see Figure~\ref{fig:HoneywellTele}) that is the baseline for the QEYSSat mission. Other optical payload providers such as ABB and INO also work on optical missions in which they develop optical telescopes with size ranging from 25\:cm to 50\:cm.

Increasing the size of the telescope up to 50\:cm can be a bit more challenging to fit within a smallsat mission, since conventional telescopes will impose stringent design rules and larger size satellite bus, hence increase the complexity and the cost of the overall system. The use of a composite material as well as deployable optics \cite{schwartz2021active} are possible ways to increase the aperture of the communication payload while keeping the mission cost reasonable. Figure~\ref{fig:DeployableTelescope} presents the concept of the deployable telescope in the aforementioned reference.

\bigskip
\emph{Cryostat for quantum detector and memory ---}
Rare-earth systems work efficiently at a few Kelvin temperatures or even less. Hence, to extend the usage of RE-based quantum memories to satellites, a space-compatible cryostat needs to be developed. So far, for cooling far-infrared (atmospheric sub-millimeter emission) detectors, a 65.4\:kg weight cryocooler operating at 1.7\:K (4.5\:K) has been developed \cite{narasaki2006mechanical}. In addition, to cool SNSPDs, a space-compatible cryocooler with a minimum operating temperature of 2.7\:K and a total weight of 55\:kg has also been demonstrated \cite{you2018superconducting}.

We remind that in some memory technologies, the need for cryocoolers that operate at low temperatures (less than a few Kelvin) might be relaxed, e.g., by using NV centers, or even eliminated by techniques such as using hybrid alkali-noble atomic vapors (see Sec.\ref{sec:Qmemories} for more information).

\bigskip
\emph{Laser cooling and trapping techniques ---}
In general, standard laser cooling and trapping techniques are required to produce ultracold atoms. NASA’s Bose-Einstein condensate and cold atom laboratory offers different magnetic and optical trapping techniques and can produce single species of rubidium and potassium BECs \cite{elliott2018nasa, frye2021bose,aveline2020observation}.

\subsubsection{Timeline and dependencies}
While all these aforementioned technologies are related in the final application - quantum teleportation - they can mostly be developed independently. The proposed baseline is to develop these technologies over a time-span of about five years. 

The proposed development for each technologies passes through a two step sequence:
\begin{itemize}
\item A first step that consists in bringing each to a TRL 4 (i.e., a breadboard with the desired performances and a design effort in keeping the SWaP as low as possible foreseeing the future space application).
\item A second step will bring them to a TRL of 6 or higher by making them suitable for space by looking at elements of design to reduce furthermore the SWaP, and make them suitable for space qualification (vibration, thermal excursion and radiation).
\end{itemize}

\subsection{Critical Tests}
Some major experimental steps in the implementation of these technologies could involve the following critical tests:

\begin{itemize}
    \item Demonstrate a high-rate (1\:GHz) EPS source and single photon source for on-demand emission with a generation efficiency $> 50\%$.
    
    \item Free-space BSM (or two-photon interference) involving moving systems. For instance, one photon produced on the ground site, (stationary location) while the other photon is sent from a flying aircraft or moving vehicle (moving location) to the ground site. High-speed range variation up to 100\:m/s should be demonstrated.
    
\end{itemize}

\clearpage
\newpage

\section{Summary and Outlook}
This report tackled the interesting but daunting task of providing an overview of the current landscape of quantum internet technologies and use-cases for quantum communication and networking. It furthermore provided an extensive outline of the current state of the art for the main technologies. Finally, as a result of the study and of the consultation with many stake holders from government, academia and industry, the most promising direction for a future Canadian quantum communication satellite mission was determined to be \textbf{Quantum Teleportation across Canada.} Most importantly, this would be a big stepping stone towards a fully scalable Canadian quantum network (the `Quantum Internet'), where some of the required technologies already have a high TRL, however need customisation for the specific application, for instance adaptive optics or deterministic photon sources. Other technologies have lower TRL and will benefit from directed R\&D efforts, in particular quantum memories.

\subsection{QEYSSat 2.0 Mission Proposal}
 The QEYSSat~2.0 team  proposes the following mission scenario: 
 \begin{itemize}
     \item {\bf The Mission Objective: }Demonstrate quantum teleportation across Canada. The teleportation shall achieve a rate of 1 event per second, and the ground distance shall be greater than 4000\:km.
     \item {\bf Satellite node:} The most likely approach is to have a high-rate, deterministic entangled photon pair source on-board a high-altitude satellite (Medium-Earth-Orbit (MEO) or Highly-Elliptical-Orbit (HEO)). Our performance estimates (Appendix~\ref{sec:additional_teleport_schemes}) show that a satellite in Low-Earth-Orbit (LEO) orbit appears to be an interesting alternative, except that the large separation of ground stations requires that the telescope elevation angles are very low (close to horizontal), which is difficult to achieve on the ground. However, platforms in LEO, and possibly supported with high-altitude platforms (HAP), could be a potential alternative that requires further investigation.
     \item {\bf Ground node:} Each ground node will receive one of the entangled photons. At least one ground node should have the ability to perform a quantum information swapping operation (Bell-state measurement (BSM)) between one of the entangled photons and a ground-based input qubit, and therefore act as a teleport device. The other ground node should be able to perform a photon analysis and detection so it can act as a teleportation receiver. A slightly more ambitious architecture would allow both ground nodes to act as teleport devices and receivers, which would allow teleportation to be performed in both directions. 
     \item {\bf Main technological advances:} This QEYSSat~2.0 mission would showcase multiple core technologies including  high-rate, deterministic sources of entangled photons and single photons, advances in quantum memory and heralding technologies, and the synchronisation of photon sources between space and ground.
 \end{itemize}
  
Our finding is based on the extensive amount of information gathered during this study, as well as the consultations with all stakeholders. Furthermore, we undertook a preliminary performance assessment in order to determine the general feasibility of this proposal.

\begin{figure}[ht]
    \centering
    \includegraphics[width=0.95\textwidth]{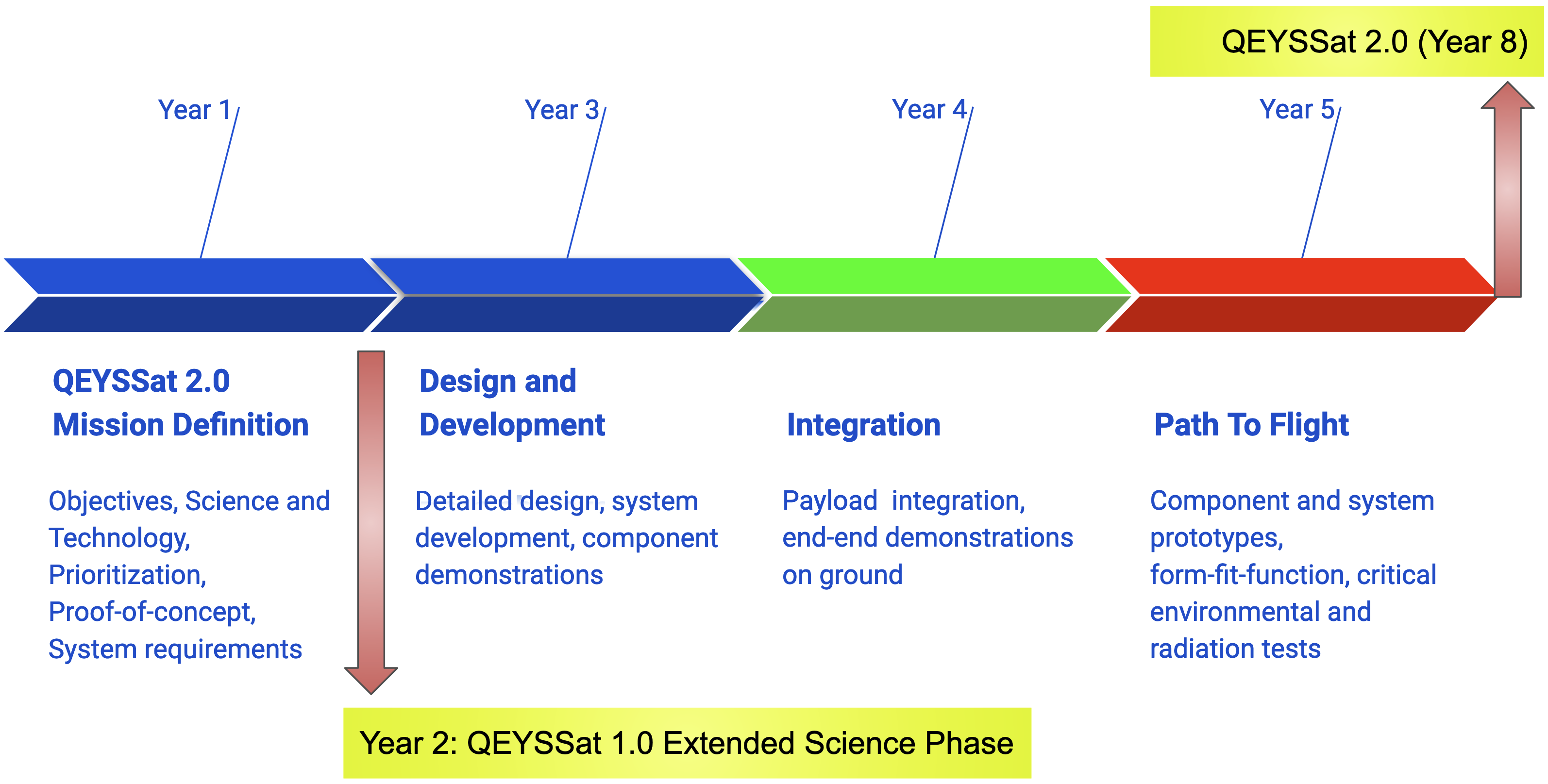}
    \caption{Tentative mission timeline of the QEYSSat~2.0 mission for Quantum Teleportation across Canada as proposed by the science team.}
    \label{fig:Qeyssat2.0timeline}
\end{figure}

This mission could be achieved with 8 years if dedicated and concentrated support of the project is allocated. A tentative time scale is provided in Figure\ref{fig:Qeyssat2.0timeline}.

The Canadian quantum information community is certainly ready and prepared to take on this challenge, which would ensure Canada is indeed taking a world-leadership in this area.

\subsection{Suggested Follow-on Studies}
Clearly, this QEYSSat~2.0 study is only the initial starting point of a QEYSSat~2.0 mission. Based on the technologies and  TRL levels, the following studies are suggested to define the full QEYSSat~2.0 mission:

\begin{itemize}
    \item \textbf{Extensive study of satellite link performance for different orbits and constellations.} This should also include a study of the relevant ground infrastructure and involve relevant Canadian industry with extensive satellite experience. 
    
    \item \textbf{Quantum memory development: } Conduct research to determine the leading / preferential technology for quantum memories, preferably with heralding or quantum-non-demolition (QND) measurement capability, and advance the technology to meet the required performance specifications for a given network architecture, such as bandwidth and storage time. Important considerations include multiplexing capacity to boost communication rates and (for deployment on-board satellites) the ability to function with limited cooling.
    
    \item \textbf{The development of new cryogenic technologies for space platforms for quantum applications.} This should address cryo-cooling systems that allow to go to temperatures lower than classical systems (i.e., 70\:K). The purpose of such systems is to enable the use of quantum devices, such as quantum memories, deterministic single photon sources and quantum detection modules (e.g., SNSPDs and high-ends SPADs). 
    
    \item \textbf{Photon sources and detectors: } Identify and select the preferred implementation of deterministic photon emitters and photon detectors suitable for these links, taking into account efficiencies, photon rates, multiplexing (frequency, time, polarisation), bandwidths and centre wavelengths.
    
    \item \textbf{Synchronisation of the BSM: } This measurement, also known as the 'Swap-node', must be suitable to handle the synchronisation between space and ground based photons down to the width of the wave-packet (ca. 100ps -- 1 ns). The primary challenge is to overcome the large and rapidly varying time-of-flight of a space to ground link, and synchronise it reliably with the ground based photons. This is a formidable challenge, and unique solutions must be developed and demonstrated on the ground.
    
    \item \textbf{Demonstrate leading approaches and technologies for adaptive optics suitable for quantum links} such as coupling beams that are collected in a telescope of 1 -- 2\:m aperture, into single-mode fibres. In particular at low elevation angles this will be a technical challenge.  This technology will be critical to achieve the high-quality spatial mode overlap between the  photons as they interfere in the BSM.
    
    \item \textbf{Investigate orbital solutions unique for Canada} such as the Highly-Elliptical Orbits to access northern latitudes, as well as LEO orbits with very low angles.
    
    \item  \textbf{Study the used of High-altitude platforms for ground to space links} as these could operate above the weather and allow for lower elevation angles operations to LEO satellites.

    \item \textbf{Study the use of low-cost space platforms} which are uniquely developed by Canadian entities, and could be utilized in LEO and MEO orbits.
    
    \item \textbf{Study the use of two/three LEO platforms allowing a quantum entanglement distribution over a long distance range} e.g., Canada to Europe or Canada to Asia. The satellites may be equipped with EPS, memories, and Bell-state analyzers depending on the architecture.
    
    \item \textbf{Study the architecture of a future constellation of satellites} enabling the achievement of a quantum internet by using all the outcomes of all the aforementioned studies.
    
\end{itemize}

\subsection{Opportunity for Canada}
The future applications of quantum networking are many-fold, including secure communications,  distributed quantum computing, and improved metrology and sensing. Small-scale quantum computers are already commercially available, with several systems under development. It is expected that this will further advance the push for quantum technologies. 

Internationally there have been astounding investments into quantum technologies.  Indeed, Canada has played a strong role, and is considered one of the top five nations in terms of quantum technology investments.  Leveraging our investments, Canadian entities are in a world-leading position in long-distance quantum communications, with the QEYSSat~1.0 mission well underway. In this study several research topics for utilisation of the QEYSSat~1.0 system have been presented.  

Canada has the opportunity to take further leadership in this area and start to work on next-generation systems for a follow-on mission such as QEYSSat~2.0. In this study the team identified that one of the most promising directions at this stage is to implement the world's first satellite enabled quantum teleportation bridge. Given the status of quantum technologies, this mission could be implemented within eight years. This would allow Canadian researchers and industry to maintain their world leadership in long-range quantum communication deployment.

Canadian researchers and industry are well positioned to develop the critical quantum technologies at a world leading performance level, and therefore strengthen and build industrial capacity and help ensure economic prosperity.  Canada already has extensive  expertise in quantum technologies, including quantum sources, quantum detectors, communication systems, and protocols but also in related classical technologies, such as spacecraft platforms, space optics and telescopes, as well as large satellite operators. This QEYSSat~2.0 program would be a pan-Canadian joint effort with partners in Government,  Industry, and Academia, as it will likely require involvement from many different directions, including technology development, basic research, commercial users, government procurement, and policy. Furthermore, several directions for basic research have been identified, which would be applicable for later mission scenarios, but should also be considered for basic research at this stage.

Canada is unique in land size and population density, which will require the involvement of satellites for a  quantum network across the country. In particular this will be important for connecting locations from northern latitudes to a quantum network, as terrestrial communication infrastructure is sparse.

The proposed mission scenario for QEYSSat~2.0 could realise the world's first teleportation across Canada, and would be an important stepping stone towards meeting the unique demands for a Canada-wide Quantum Internet.


\section{Acknowledgements}

The authors acknowledge funding from the National Research Council of Canada (NRC) and the Defence Research and Development Canada (DRDC) (Contract No. 969895). In particular we are grateful for valuable inputs to the draft version of the white paper by Étienne Boulais, Oleg Djazovski, Guillaume Faubert, Aimee Gunther, Khabat Heshami, Ebrahim Karimi, Alexander Koujelev, Lindsay LeBlanc, Charles Lemoine, Jeff Matthews, Lydia Philpott, Stephanie Simmons, Yoan St-Onge and Tricia Willink.  

The team further acknowledges  all participants to the QEYSSat 2.0 Workshop, as well as the many colleagues that engaged in discussions, and provided comments and suggestions on the study and the white paper. The UW and UC teams acknowledge support from the Natural Sciences and Engineering Research Council of Canada. The UW team is grateful for support from the Ontario Research Fund, the Canadian Space Agency and the Canada First Research Excellence Fund. INO acknowledges support from the Minist\`ere of \'Economie, Innovation and \'Energie from the province of Quebec.

The authors acknowledge that several papers on the topics discussed in this white paper have been published since the completion of this project and manuscript. The QEYSSat~2.0 study was undertaken between June 2021 and March 2022. Here is a non-exhaustive list of relevant recent papers that have been published since completion of this manuscript \cite{Belenchia2022,Spellmeyer2023}.

The authors declare no competing interests. This study did not produce any original data that needs to be shared or made available. All modelling/calculations are explained in the paper and were used to generate the graphs/data.

\appendix

\begin{appendix}

\section{List of Acronyms and Definitions}

\begin{tabular}{|c|p{11cm}|}
\hline
ABL & Atmospheric Boundary Layer \\
\hline
ACES & Atomic Clock Ensemble in Space\\
\hline
AFC & Atomic Frequency Comb\\
\hline
ALF & Atomic Line Filters\\
\hline
AO & Adaptive Optics \\
 \hline
 APD & Avalanche Photo Diode \\
\hline
ATS & Autler-Townes Splitting\\
\hline
BEC & Bose-Einstein Condensates\\
\hline
BSM & Bell-state Measurement\\
\hline
CDR & Critical Design Review \\
\hline
CSA & Canadian Space Agency \\
\hline
DCR & Dark Count Rate \\
\hline
EIT & Electromagnetically Induced  Transparency\\
\hline
 EPS & Entangled Photon Source \\
 \hline
 ESA & European Space Agency \\
\hline
FOCAL & The McMaster Free-Space Optical Communication Algorithms Laboratory\\
\hline
FWHM & Full Width Half Maximum\\
\hline
 GEO & Geostationary Earth Orbit \\
 \hline
 GPS & Global Positioning System\\
 \hline
 HAP & High Altitude Platform \\
\hline
InGaAS & Indium Gallium Arsenide \\
\hline
INO & Institut National d’Optique\\
\hline
 IQC & Institute for Quantum Computing \\
 \hline
  IQST & Institute for Quantum Science \& Technology \\
 \hline
 JPL & Jet Propulsion Laboratory\\
\hline
  LEO & Low Earth Orbit\\
 \hline
 LISA & Laser Interferometer Space Antenna\\
 \hline
 MCT & Mercury-Cadmium-Telluride\\
 \hline
  MEO  & Medium Earth Orbit\\
 \hline
 MPPC & Mutli-Pixel Photon Counters\\
 \hline
 NASA & National Aeronautics and Space Administration\\
\hline
 NRC & National Research Council (Canada)\\
 \hline
 NCR-HAA & NRC Herzberg Astronomy \& Astrophysics\\
 \hline
 OAM & Orbital Angular Momentum \\
 \hline
 OD & Optical Density\\
 \hline
OGS & Optical Ground Station \\
\hline
OSA & Optical System Assembly\\
\hline
PDE & Photon Detection Efficiency\\
\hline
PMT & Photo Multiplier Tube \\
\hline
PNR & Photon Number Resolving  detectors\\
\hline
 QEYSSat & Quantum Encryption and Science Satellite mission \\
 \hline
 QEYSSat~2.0 & Quantum Encryption and Science Satellite future mission(s) \\
 \hline
 QKD    & Quantum Key Distribution \\
 \hline
QM & Quantum Memory, coherent storage and retrieval of a photon  with its encoding. \\
\hline
\end{tabular}

\newpage

\begin{tabular}{|c|p{11cm}|}
\hline
QND & Quantum Non-Demolition measurement, detect the presence of a photon without disturbing its encoding. \\
\hline
QND+QM & QM that flags the successful storage of a photon, also known as a 'heralded quantum memory'. \\
\hline
Q(O)GS & Quantum (Optical) Ground Station \\
\hline
 Qubit & Quantum Bit, the unit of quantum information\\
 \hline
 ReFQ & Reference-Frame Independent QKD\\
 \hline
 REIs & Rare-Earth Ions \\
 \hline
 RFI & Reference-Frame Independent\\
 \hline
 SI-PMT & Si PhotoMulTipliers\\
 \hline
 SNR & Signal-to-Noise Ratio\\
 \hline
  SNSPD & Superconducting Nanowire Single Photon Detector \\
 \hline
 SPAD & Single Photon Avalanche Diode \\
\hline
 SPDC    & Spontaneous Parametric Down-conversion \\
 \hline
 SPS & Single Photon Source\\
 \hline
 TES & Transition-Edge Sensing\\
 \hline
 TFIF & Thin Film Interference Filter \\
 \hline
 UA & University of Alberta\\
 \hline
 UBristol & University of Bristol\\
 \hline
 UOttawa & University of Ottawa\\
 \hline
 UStrathclyde & University of Strathclyde\\
 \hline
 UT & University of Toronto\\
 \hline
 UW & University of Waterloo\\
 \hline
 WCP & Weak Coherent Pulse\\
 \hline
 WDM & Wavelength-Division Multiplexing\\
 \hline

\end{tabular}

\newpage

\section{Canadian Quantum Expertise}
\label{appendix:CanadaList}

 In the following table, we highlight important research groups in Canada in addition to their area of expertise. Note, this list is not meant to be exhaustive and is still being populated.

\begin{longtable}{|p{3cm}|p{3cm}|p{3cm}|p{5cm}|}
\hline
University/ Organization & Name & Position & Expertise \\
\hline
\hline
UWaterloo/IQC, Waterloo & Thomas Jennewein & Associate Professor, QEYSSat Science Team PI & Quantum communication, QKD, quantum optics, entanglement \\
\hline
UWaterloo/IQC, Waterloo & Norbert L\"{u}tkenhaus & Professor & Theory QKD, quantum communication \\
\hline
UWaterloo/IQC, Waterloo & Michele Mosca & Professor & Quantum algorithms, quantum security \\
\hline
\hline
University of Calgary/IQST, Calgary & Paul Barclay & Associate Professor & Quantum optics, quantum optomechanics, nanophotonics \\
\hline
University of Calgary/IQST, Calgary & Shabir Barzanjeh  & Assistant Professor & Quantum circuits, quantum optomechanics, quantum optics \\
\hline
University of Calgary/IQST, Calgary & Daniel Oblak & Assistant Professor & Quantum networks, quantum optics, quantum memory \\
\hline
University of Calgary/IQST, Calgary & Barry Sanders & Professor & Quantum information, quantum optics, quantum algorithms \\
\hline
University of Calgary/IQST, Calgary & Christoph Simon & Professor & Quantum optics, quantum networks, quantum memory, biophysics \\
\hline
\hline
Institut National d'Optique, Qu\`{e}bec & Andr\'{e} Foug\`{e}res & VP Innovation \& Technology & Quantum optics \\
\hline
\hline
University of Ottawa, Ottawa & Anne Broadbent & Associate Professor & Quantum computing, QKD, quantum information \\
\hline
University of Ottawa, Ottawa & Ebrahim Karimi & Associate Professor & QKD, quantum communication, quantum optics \\
\hline
\hline
National Research Council of Canada, Ottawa & Dan Dalacu & Research Officer (NRC) \& Adjunct Professor (UOttawa) & Quantum optics \\
\hline
National Research Council of Canada, Ottawa & Khabat Heshami & Research Officer (NRC) \& Adjunct Professor (UOttawa) & Quantum communication, quantum optics, quantum memory \\
\hline
National Research Council of Canada, Ottawa & Ben Sussman & Research Officer & Quantum optics, quantum communication, quantum sensing \\
\hline
\hline
Universit\`{e} de Montr\`{e}al, Montreal & Gilles Brassard & Professor & QKD, quantum information \\
\hline
\hline
University of Alberta, Edmonton & John Davis  & Professor & Quantum optomechanics, nanomechanics, superfluids \\
\hline
University of Alberta, Edmonton & Lindsay LeBlanc  & Associate Professor & Quantum memory, quantum optics \\
\hline
University of Toronto, Toronto & Hoi-Kwong Lo & Professor & Quantum information, QKD, quantum computing \\
\hline
University of Toronto, Toronto & Li Qian  & Professor & Quantum optics, QKD, nonlinear optics \\
\hline
\hline
Universite de Sherbrooke/ Institut Quantique, Sherbrooke & Alexandre Blais & Professor & Quantum information, quantum circuits, quantum optics \\
\hline
Universite de Sherbrooke/ Institut Quantique, Sherbrooke & Serge A. Charlebois & Professor & Nanofabrication, microelectronics \\
\hline
Universite de Sherbrooke/ Institut Quantique, Sherbrooke & Jean-Francois Pratte & Professor & Quantum optics \\
\hline
\hline
Simon Fraser University, Vancouver & Stephanie Simmons & Assistant Professor & Quantum memories, quantum optics \\
\hline
\hline
McGill University, Montreal & Lilian Childress & Associate Professor & Quantum optics \\
\hline
\caption{Overview of Canadian expertise in quantum technologies}
    \label{tab:CanadaList}
\end{longtable}

\newpage

\section{International Quantum Space Missions}
In the following table, we highlight key quantum space missions from numerous countries. Note, this list is not meant to be exhaustive as new missions are being frequently announced. 

\begin{longtable}{|p{4cm}|p{8cm}|}
\hline
Country & Mission \\
\hline
\hline
International consortium of six research entities & CubeSat quantum communications mission (CQuCoM) \cite{oi2017cubesat} \\
\hline
Canada & Quantum EncrYption and Science Satellite (QEYSSat) \cite{Qeyssat} \\
\hline
China & QUantum Experiments at Space Scale (QUESS), satellite name: Micius \cite{zhang2019quantum} \\
\hline
Japan & Space Optical Communications Research Advanced Technology micro-satellite (SOCRATES)  \cite{carrasco2017leo} \\
\hline
UK-Singapore & Space Photon Entanglement Quantum Technology Readiness Experiment (SpeQtre) CubeSat \cite{SPEQTRE} \\
\hline
Germany & CubeSat (CUBE) \cite{haber2018qube} \\
\hline
UK & Quantum Research CubeSat (QUARC) \cite{mazzarella2020quarc}, and national network of quantum technology hubs (UK NQT Hub) \cite{UKQTHub} \\
\hline
France-Austria & CubeSat (NanoBob) \cite{kerstel2018nanobob} \\
\hline
Austria & CubeSat (Q$^3$ sat) \cite{neumann2018q} \\
\hline
International & G7 quantum encryption satellite network \\
\hline
\caption{Overview of International space missions with quantum technologies}
    \label{tab:WorldList}
\end{longtable}

\newpage

\section{Teleportation Schemes}
\label{sec:additional_teleport_schemes}

\begin{figure}[ht]
    \centering
    \includegraphics[width=1\textwidth]{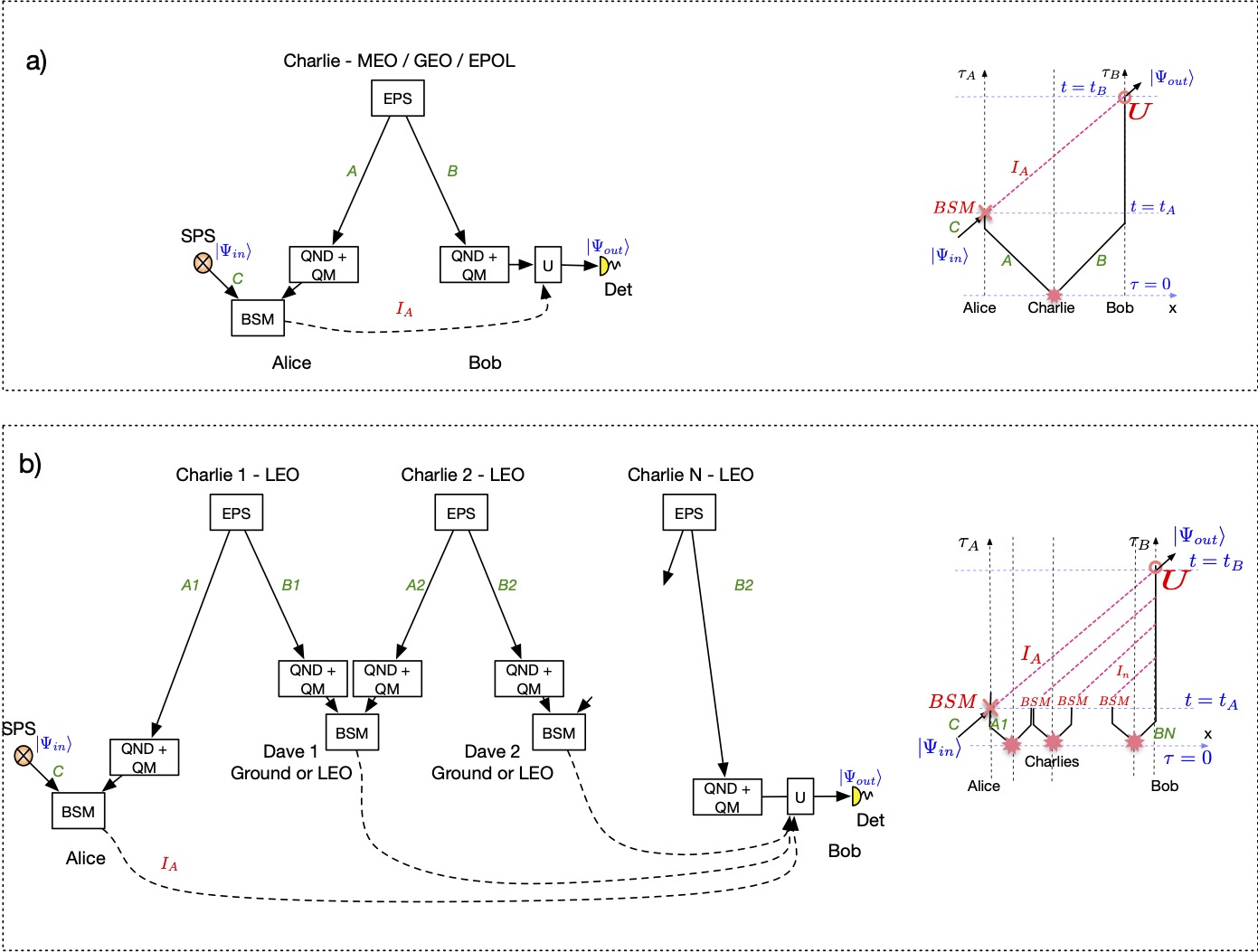}
    \caption{Teleportation schemes utilising quantum repeater infrastructure. Either one (a) or several (b) EPS on satellites, and quantum repeater nodes in between. If all the repeater nodes are on the ground, then this scheme does not require quantum memories on the satellite. However, scheme (b) could benefit from placing the repeater nodes on board satellites (LEO).}
    \label{fig:additional_telep_schemes_EPS}
\end{figure}

Figure~\ref{fig:additional_telep_schemes_EPS} shows further  approaches for implementing the  quantum teleportation using quantum repeater structures, with primarily the EPS on-board the satellites, and the memories on the ground. Such schemes could avoid the efforts of developing space suitable memory systems.
\clearpage

\begin{figure}[ht]
    \centering
    \includegraphics[width=0.9\textwidth]{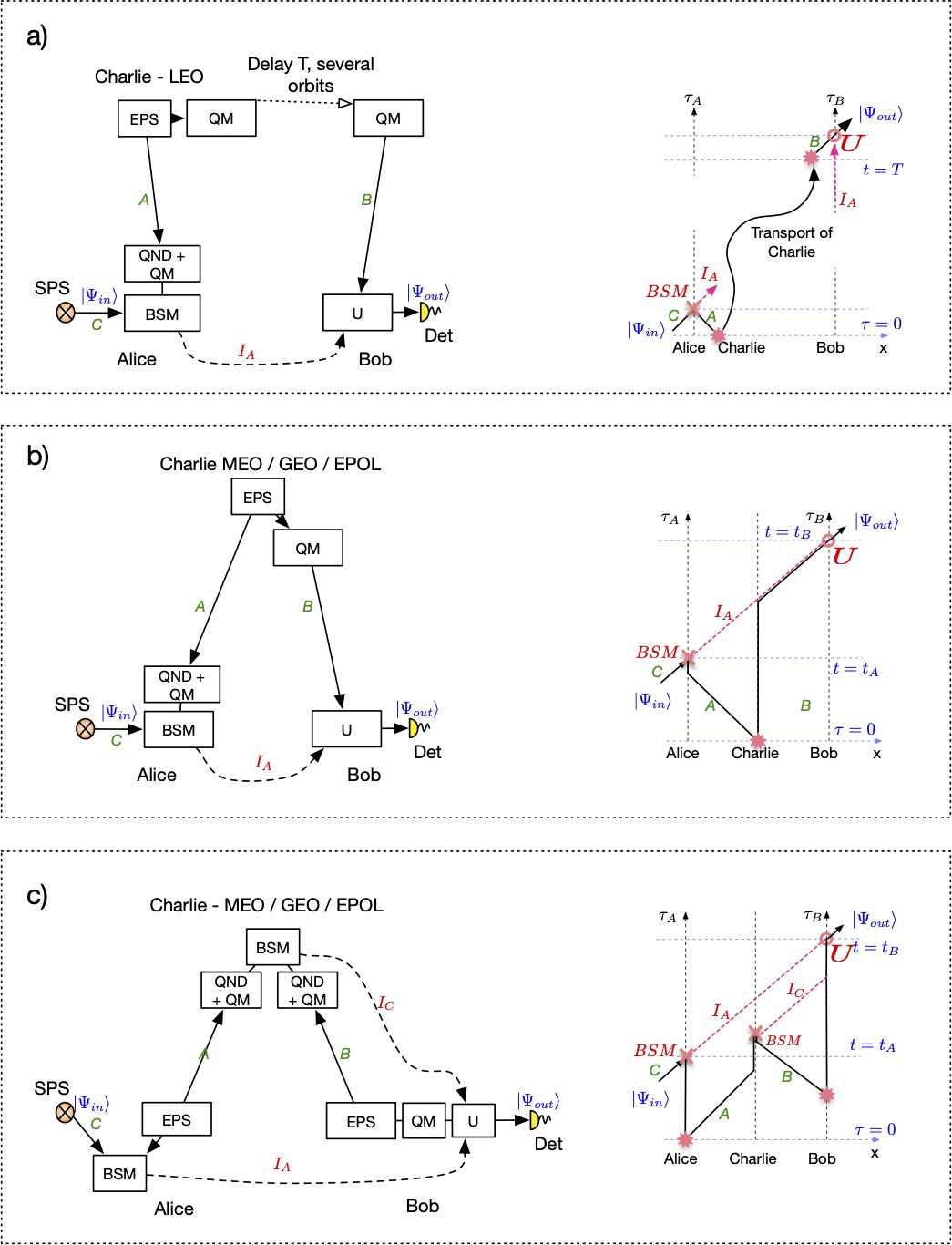}
    \caption{Overview of additional quantum teleportation schemes that use  quantum memories aboard the satellite, as well as on the ground. The purpose of the memories is to correct for different arrival times of photons arriving at the BSM, as well as  coherently delay  the quantum states until the BSM results arrive at Bob. (a) The long term memory can hold the quantum entangled statues for the duration of several orbits, until contact with Bob's ground station is possible. b) The QM only has shorter time duration (approx. the signal time from Alice to Bob, and Charlie releases the photons just in time when Bob receives the BSM result from Alice. (c) Charlie is a photon receiver and performs a BSM between photons obtained from Alice and  Bob, and must hold these photons in memories in order to compensate different arrival times.}
    \label{fig:additional_telep_schemes_QM}
\end{figure}

\begin{figure}[ht]
    \centering
    \includegraphics[width=0.8\textwidth]{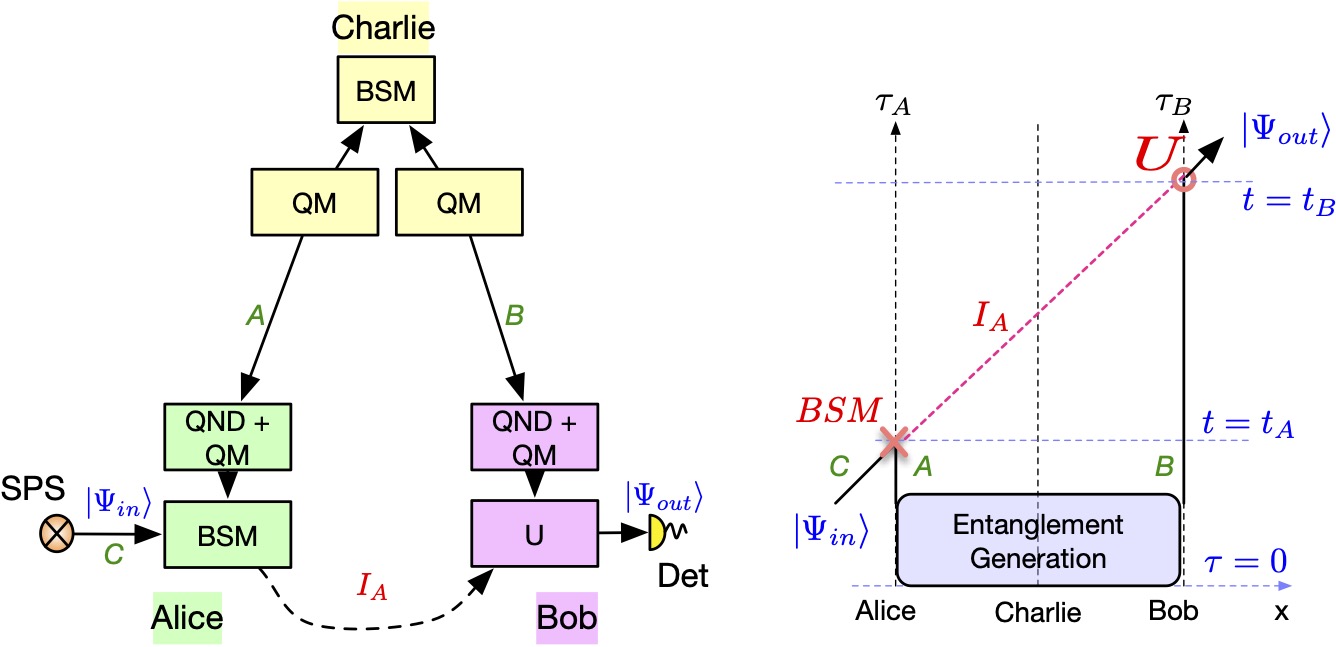}
    \caption{(left) The teleportation protocol uses an memory assisted entanglement distribution, where the photons can be released at Alice and Bob on demand. (right) Space-time arrangement of the signals. }
    \label{fig:teleportationScheme_mem}
\end{figure}

\clearpage

Figures~\ref{fig:additional_telep_schemes_QM} and \ref{fig:teleportationScheme_mem} show some further possible approaches for implementing the long-range quantum teleportation, now  with quantum memories on the satellite.

\subsection{Teleportation rate}
In the following, we estimate the teleportation rate and the minimum required time for a successful teleportation in two different schemes.

\begin{figure}[ht]  
    \centering\includegraphics[width=12cm]{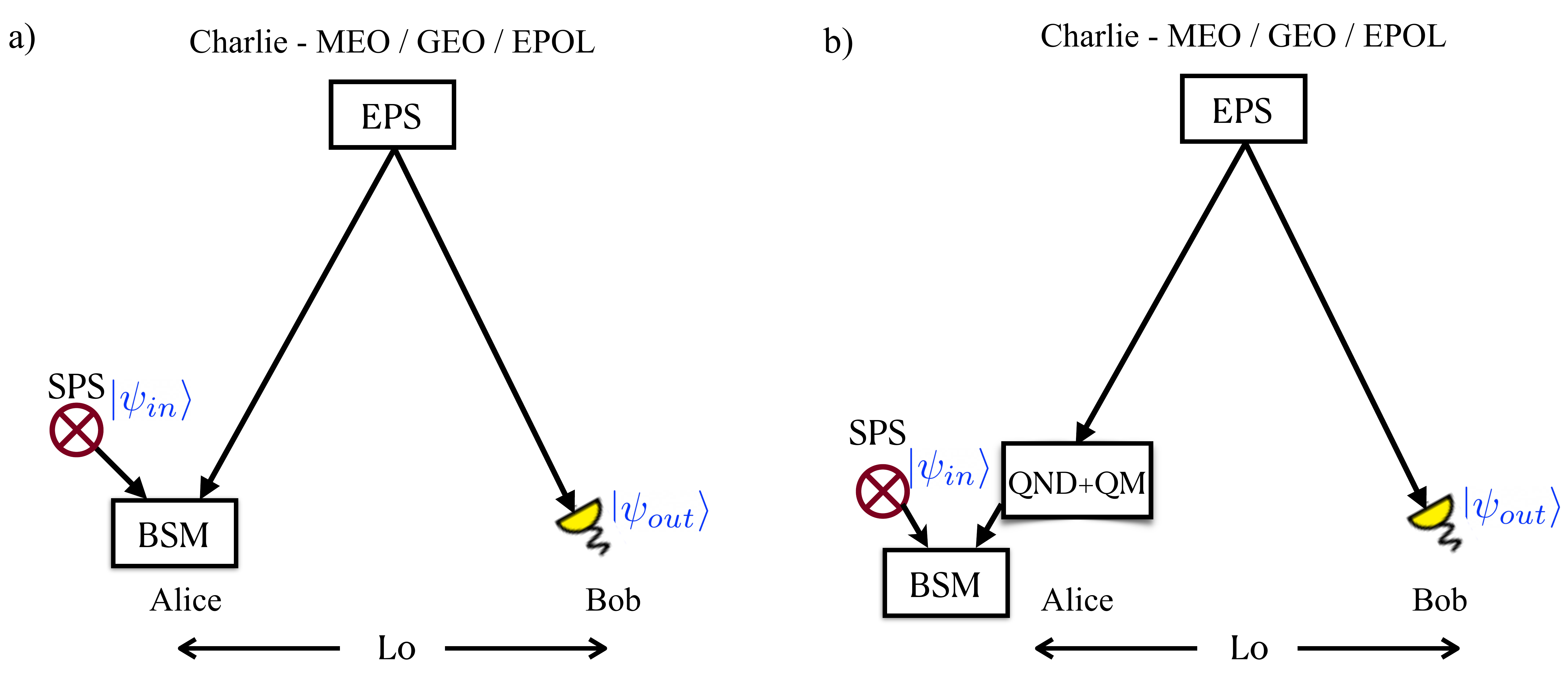}
    \caption{(a) Memory-less scheme, (b) On-demand scheme when there is one memory in the Alice station.}
    \label{fig:memoryless}
\end{figure}

\subsubsection{Memory-less scheme}\label{D11}
In the first scenario, a quantum satellite equipped with an entangled photon pair source (EPS), transfers a pair of entangled photons down to the ground stations (i.e., Alice and Bob). In the first ground station, Alice performs a BSM on her two photons as shown in Figure~\ref{fig:memoryless} (a). Without a memory, it is a matter of chance whether a photon pair will arrive. Therefore, we do not know which photon is going to be teleported. We look afterwards and pick the case that was successful.

In this situation, teleportation rate would be:
\begin{equation}
R_t=P_{ave} P_b R_{s} \eta_{eps},
\label{eq:memoryless}
\end{equation}
where  $P_{ave}$  is the average probability for the two-photon transmission, $ P_b $ is the success probability of the BSM given by $\eta_{sps}\eta_d^2/2$, $\eta_d$ is the detection efficiency, $\eta_{sps}$ is the single-photon source (SPS) efficiency, $\eta_{eps}$ is the EPS efficiency, and $R_{s}$ is the EPS repetition rate. Here, we assumed the same repetition rate for SPS and EPS. As the EPS is carried by the satellite, two downlinks are established for which the probability of successful transmission of a pair of photons is a function of systematic pointing error ($L_P$), Optical transmittance ($T_T,T_R$), atmospheric absorption ($A_{atm}$), and optical diffraction. Here we set $P_{ave}= 10 \log (2A)$, where $A$ is the attenuation of a single link (in dB) (see Appendix~\ref{Appendix:link_analysis}).

\subsubsection{On-demand scheme}\label{D12}
Quantum memories can be employed to improve the performance of quantum teleportation by making it on demand. 
The efficiency of a quantum memory can be decomposed into the storage and retrieval efficiencies $\eta_{st}$ and $\eta_r$, respectively. The BSM will be performed only after both memories are loaded with an entangled pair. The memory efficiency can decay exponentially during the storage time $\Delta t$ as $e^{-\Delta t/T_1}$ where $T_1$ is the memory lifetime. Note that here we ignore the reinitialization time of the memories as it is negligible compared to the transmission time.

\noindent
$\bullet\,\textbf{One-memory:}$

\noindent
We consider a situation when only Alice has a memory in her station (see Figure~\ref{fig:memoryless} (b)).
In this case, the teleportation rate is given by
\begin{equation}
R_t=P_{ave} P_b^\prime R_s \eta_{eps}
\eta_{st} \eta_r \eta_{Q\!N\!D} e^{-\Delta t_0/T_1} ,
\label{eq:one-memory-Alice}
\end{equation}
where $\eta_{Q\!N\!D}$ is the efficiency of the QND, $\Delta t_0=L_0/c$ is the communication time between Alice and Bob to inform each other about successful loading of the memory, $c$ is the speed of light, $L_0$ is the distance between Alice and Bob, and $P_b^\prime=\eta_{sps}\eta_d^2/4$.

On the other hand, there might be a situation when only Bob has a memory.
In this case, Alice can communicate the result of the BSM with Bob to inform him about the unitary operation he needs to apply on his photon. Therefore, an additional communication time 
should be considered.
For this case the rate would be 

\begin{equation}
R_t=P_{ave} P_b R_s \eta_{eps}
\eta_{st} \eta_r \eta_{Q\!N\!D} e^{-\Delta t_0/T_1} e^{-\Delta t_1/T_1}.
\label{eq:one-memory-Bob}
\end{equation}
Here $P_b=\eta_{sps}\eta_d^2/2$ is the BSM success probability, $e^{-\Delta t_1/T_1}$ is the decoherence rate of the Bob memory after the BSM where $\Delta t_1=L_0/c$.

\noindent
$\bullet\,\textbf{Two-memory:}$

\noindent
In this case, both stations have a QND and a quantum memory to store the photons (see Figure~\ref{fig:additional_telep_schemes_EPS} (a)). Here also, we consider a situation where Alice communicates the result of the BSM with Bob. 
Therefore, the teleportation rate is given by
\begin{equation}
R_t=P_{ave} P_b R_s \eta_{eps}
\eta_{st}^2 \eta_r^2 \eta_{Q\!N\!D}^2 e^{-2\Delta t_0/T_1}  e^{-\Delta t_1/T_1}.
\label{eq:two-memory}
\end{equation}
where the coefficient 2 in the exponential term is because both memory efficiencies decay during the communication time $\Delta t_0$.

\subsubsection{Repeater scheme}
In a more advanced case, satellites can be used to establish entanglement over elementary links of a quantum repeater \cite{boone2015entanglement} (see Figure~\ref{fig:additional_telep_schemes_EPS} (b)).
In a repeater scheme, before performing entanglement swapping between two neighboring links, entanglement should be established over them. 
Here we define $n_i$ as the number of attempts to successfully transmit an entangled photon pair over the $i$-th link. In a two-link repeater, the preparation time is defined by the entanglement generation time of the link that requires a larger number of attempts to be established. Therefore, the preparation time is given by $n_{max} T_0$ where $n_{max}=max\{n_i,n_{i+1}\}$, and $T_0$ is the time required for each attempt. Here this time is set by the repetition rate of the entangled photon pair source i.e., $T_0=1/R_s$. Loaded memories of the link that establishes the entanglement first, start to decay until entanglement generates in the neighboring link as well.  Hence, the decay time (waiting time) can be defined as $\Delta t=n_{dif} T_0$ where $n_{dif}=|n_i-n_{i+1}|$.
In this case,  the average entanglement swapping probability is given by \cite{wu2020near}

\begin{equation}
 \langle p_s\rangle
=\frac{1}{2}\eta_d^2 \eta_{st}^4 \eta_r^2 \eta_{QND}^4
 \langle exp(- 2n_{dif} T_0/T_1)\rangle.
\end{equation}
Here we assumed the time required to perform entanglement swapping is negligible compared to the entanglement generation time. Therefore, we neglected the decay time of the outer memories while performing entanglement swapping. To estimate $\langle exp(-2 n_{dif} T_0/T_1)\rangle$, we need to first calculate the probability distribution function of $n_{dif}$. Assume two independent events $a$ and $b$ that happens with probability $p$. The probability that the event $a$ happens until $n_a$th
attempt is 
\begin{equation}
    P_a=p(1-p)^{n_a-1}.
\end{equation}
Therefore, the joint probability distribution function for these two events is given by
\begin{equation}
    P_{ab}=p^2(1-p)^{n_a+n_b-2}.\label{pab}
\end{equation}
Using Eq.\ref{pab}, we can define the probability distribution function of $n_{dif}$ as
\begin{equation}
P(n_{dif})= \begin{cases}
\frac{p}{2-p} & n_{dif}=0\\
\frac{2p(1-p)^{n_{dif}}}{2-p} & n_{dif}\neq0 
\end{cases}
\end{equation}
Therefore, $\langle exp(- 2n_{dif} T_0/T_1)\rangle$ can be written as
\begin{equation}
\langle exp(- 2n_{dif} T_0/T_1)\rangle= \frac{p}{2-p}+\frac{2p(1-p)e^{- 2n_{dif} T_0/T_1}}{(2-p)(1-(1-p)e^{- 2n_{dif} T_0/T_1})},
\end{equation}
where in our case $p=P_{ave}\eta_{eps}$.
Following the approach of Ref \cite{wu2020near}, we can then estimate the  average entanglement distribution time of our two-link repeater protocol as 
\begin{equation}
T_0 \frac{\langle n_{min}\rangle}{\langle p_s\rangle} < \langle T_r\rangle < T_0 \frac{\langle n_{max}\rangle}{\langle p_s\rangle}.\label{R_r}
\end{equation}
Here, $n_{min}=min\{n_i,n_{i+1}\}$, and $\langle n_{min}\rangle$ and $\langle n_{max}\rangle$ are the expectation of $n_{min}$ and $n_{max}$. 

To estimate the overall teleportation rate, the same as for the on-demand scheme, we consider the situation where Alice informs Bob about the result of the BSM. Hence, the teleportation rate is given by 

\begin{equation}
\langle R_t\rangle=\langle R_r\rangle \eta_r^2 P_b \, e^{-2\Delta t_0/T_1}  e^{-\Delta t_1/T_1},
\label{eq:repeater}
\end{equation}
where the length of each elementary link is $L_0/2$, and $\langle R_r\rangle=1/\langle T_r\rangle$ is the repeater rate that is estimated using the average of the lower and upper bound of Eq.\ref{R_r}.

\subsubsection{Comparison for different satellite orbits}

In general, adding components to the system can increase the loss, since 100$\%$ efficiency is not physically possible. Therefore, on-demand schemes show a lower teleportation rate comparing to the memoryless case, and a higher rate comparing to the systems with repeaters. However, on-demand and repeater schemes assure a higher success probability and improve the practicality of implementing teleportation over
long distances. Employing quantum memories enables 'true' teleportation, as it allows Bob to perform a deterministic conditional rotation based on the result of the BSM at Alice station. Hence, the state can be teleported perfectly without the need for post-selection.

\begin{figure}[ht]
    \centering
\begin{subfigure}{0.4\textwidth}
\centering
\includegraphics[width=0.9\textwidth]{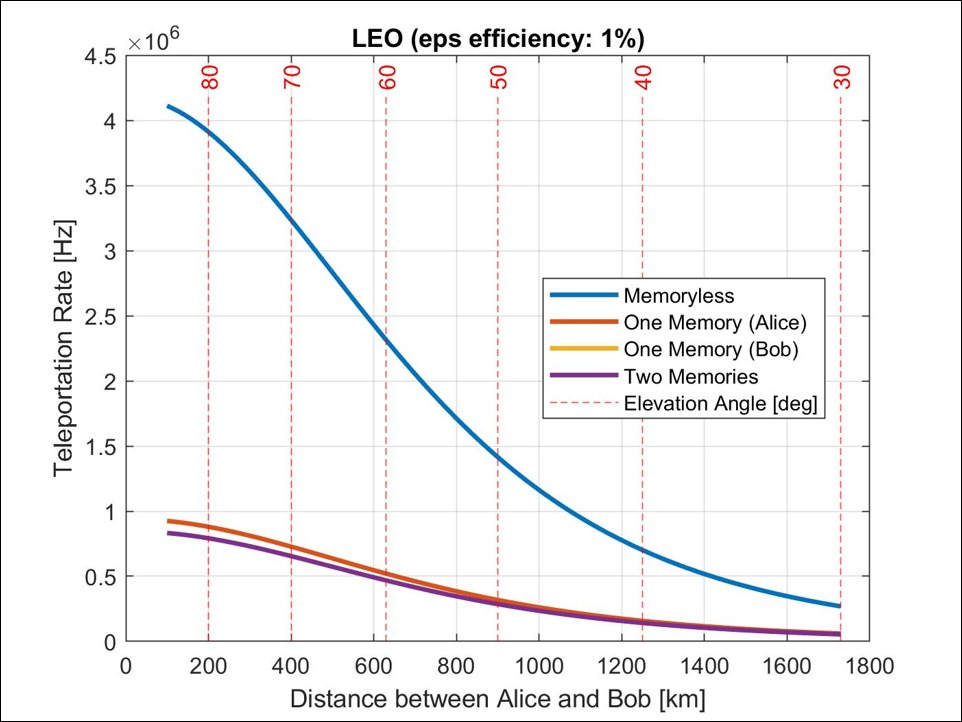} 
\caption{}
\end{subfigure}
\hspace{1cm}
\begin{subfigure}{0.4\textwidth}
\centering
\includegraphics[width=0.9\textwidth]{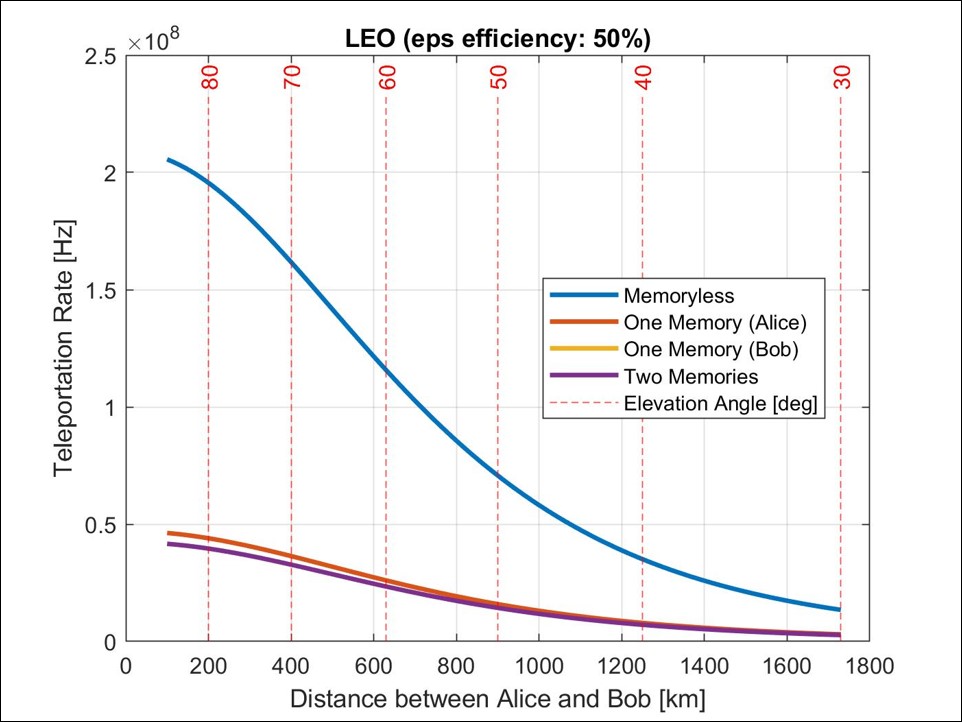} 
\caption{}
\end{subfigure}
\begin{subfigure}{0.4\textwidth}
\centering\includegraphics[width=0.9\textwidth]{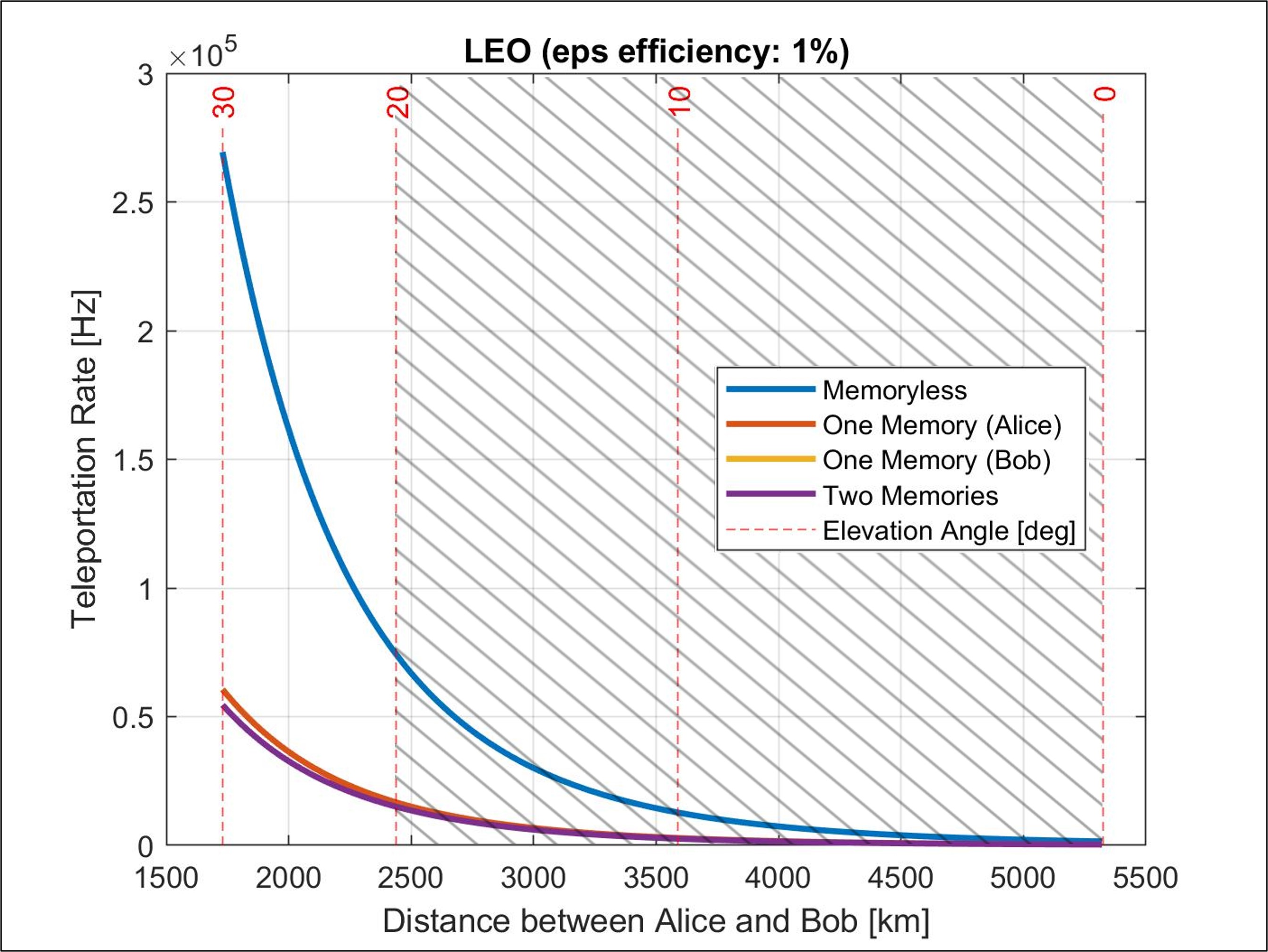} 
\caption{}
\end{subfigure}
\hspace{1cm}
\begin{subfigure}{0.4\textwidth}
\centering
\includegraphics[width=0.9\textwidth]{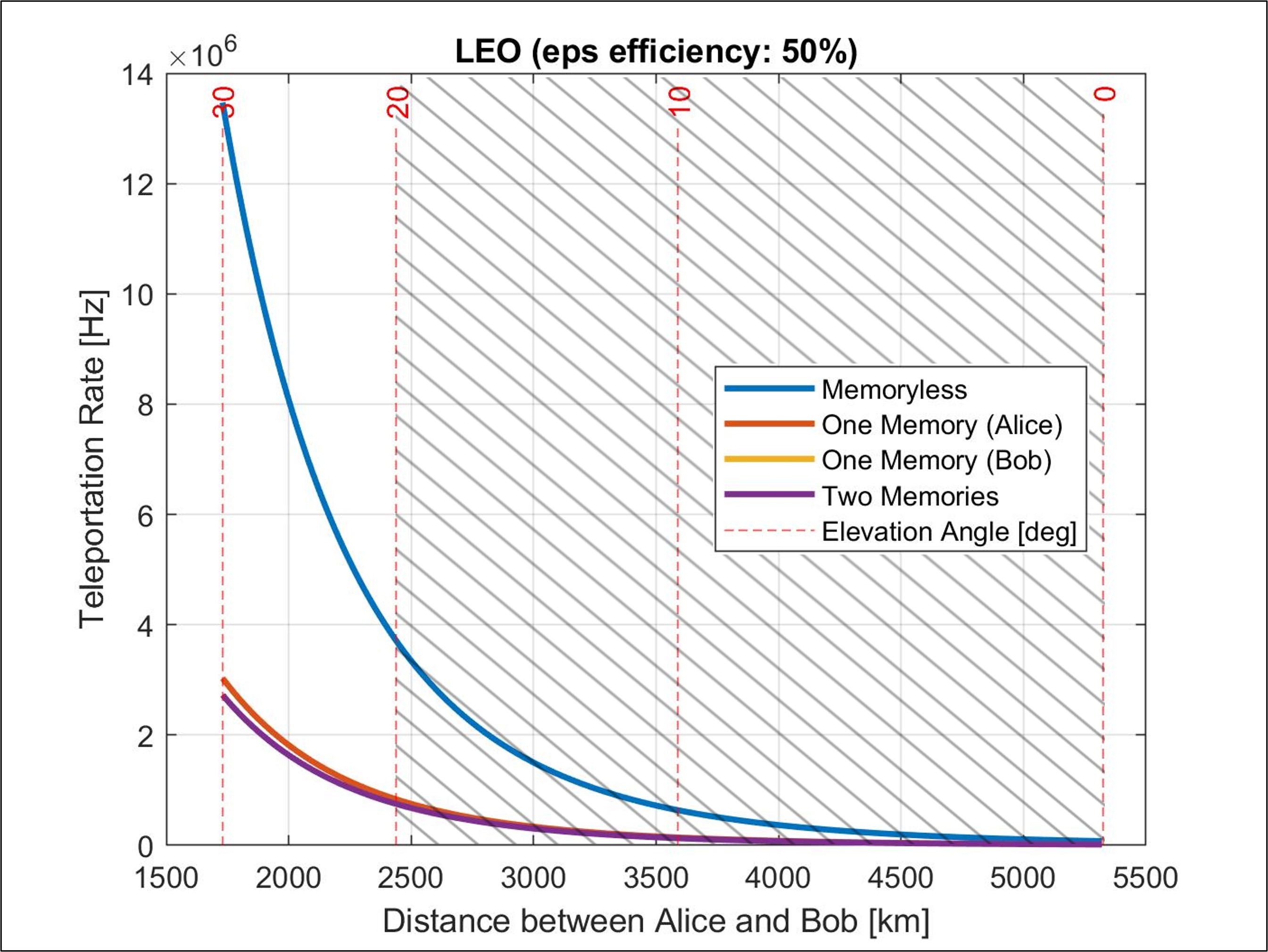} 
\caption{}
\end{subfigure}
\begin{subfigure}{0.4\textwidth}
\centering
\includegraphics[width=0.9\textwidth]{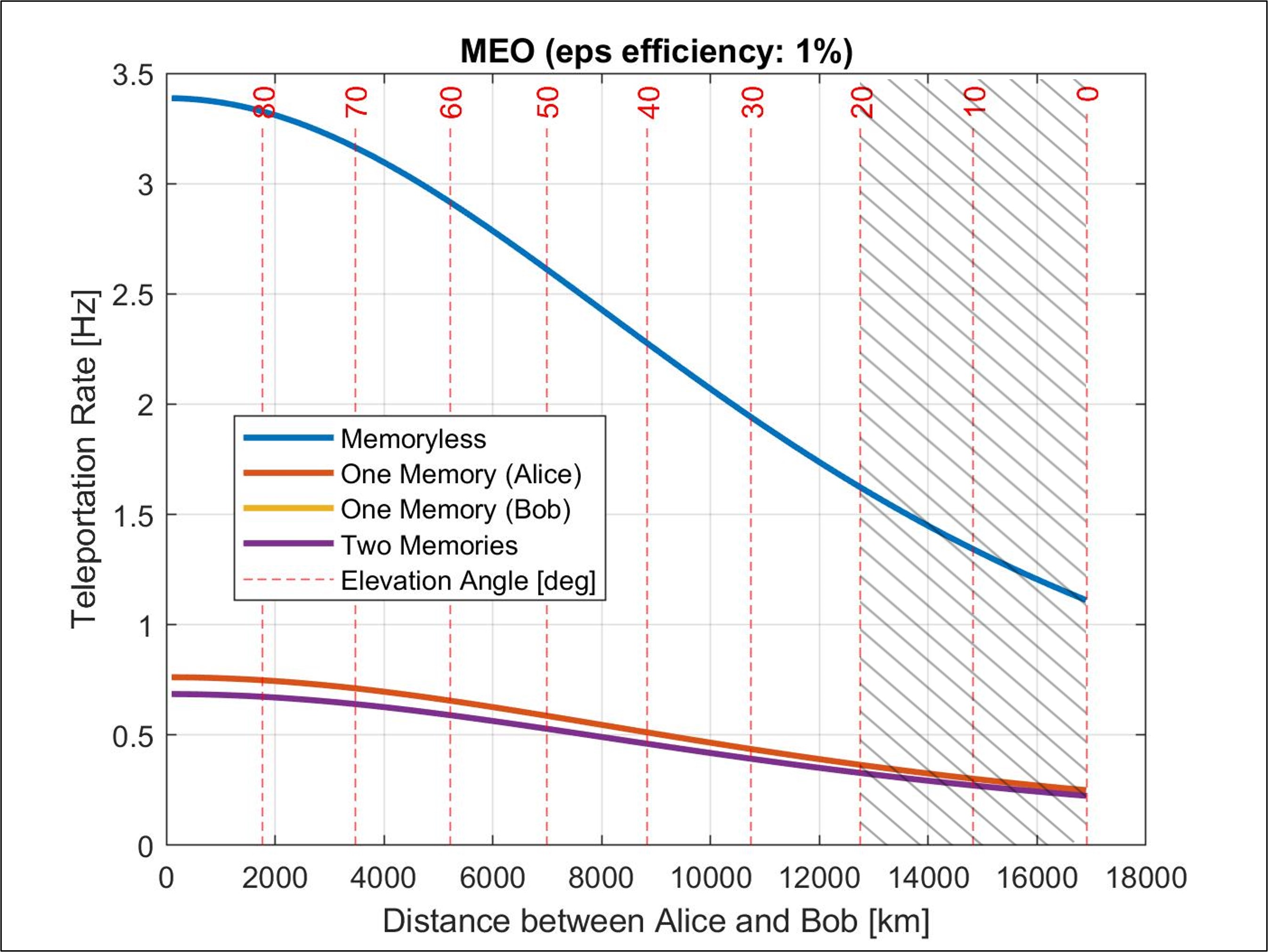} 
\caption{}
\end{subfigure}
\hspace{1cm}
\begin{subfigure}{0.4\textwidth}
\centering
\includegraphics[width=0.9\textwidth]{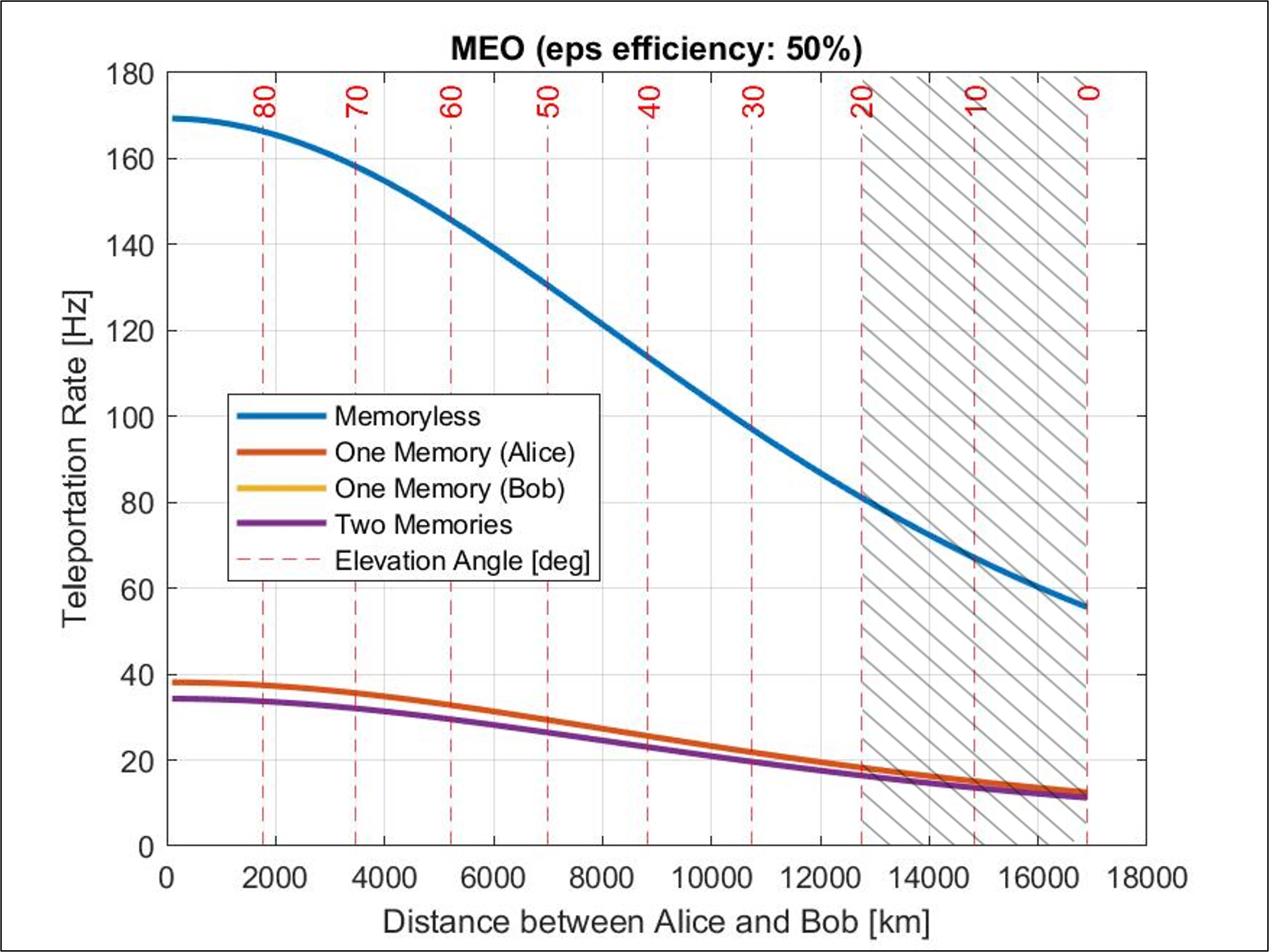} 
\caption{}
\end{subfigure}
\begin{subfigure}{0.4\textwidth}
\centering
\includegraphics[width=0.9\textwidth]{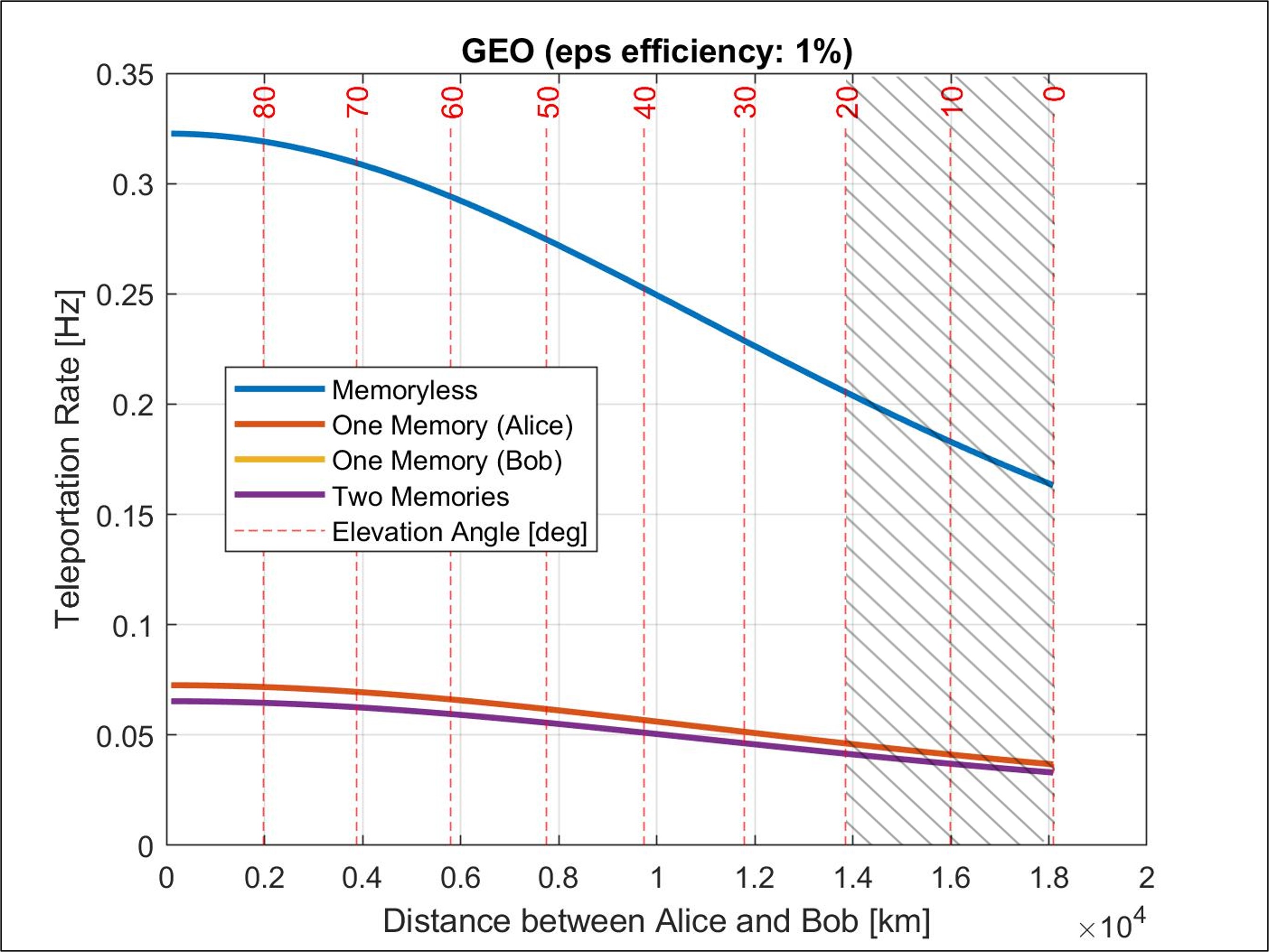} 
\caption{}
\end{subfigure}
\hspace{1cm}
\begin{subfigure}{0.4\textwidth}
\centering
\includegraphics[width=0.9\textwidth]{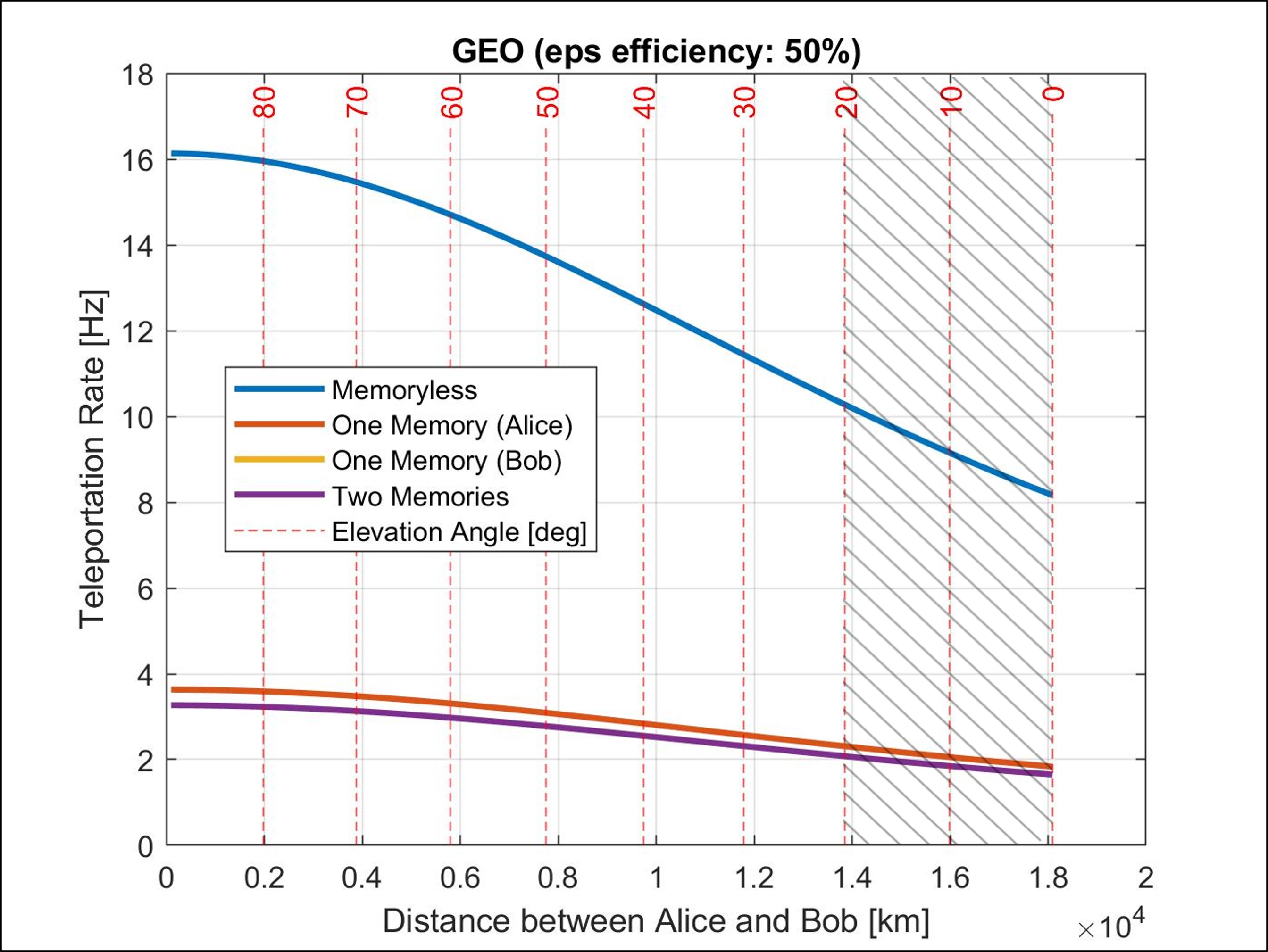} 
\caption{}
\end{subfigure}
    \caption{The plots show the teleportation rate for Alice and Bob as the distance between them increases (considering the curvature of the Earth), for two different photon source efficiencies, $\eta_{eps}=1\%$ on the left and $\eta_{eps}=50\%$ on the right. Each scenario is modelled for three orbits, 600 km (LEO), 20000 km (MEO) and 36000 km (GEO). LEO plots are split into $80^\circ$-to-$30^\circ$ and $30^\circ$-to-$0^\circ$ elevation angles to have a better visualisation of the results. The shaded area indicates that establishing a link might not be practical, due to the low elevation angle. In all plots \textit{One Memory (Bob)} is overlapped with \textit{Two Memories}. $\lambda =785$nm; $D_t = 40$cm; $D_r = 200$cm; Table~\ref{tab:Teleportation_assumptions}.}
    \label{fig:Teleportation_Rates}
\end{figure}

To compare the outcome of the mentioned scenarios in sections \ref{D11}, and \ref{D12}, in Figures~\ref{fig:Teleportation_Rates}, we estimated the teleportation rates considering a static satellite in three orbits: LEO (Low-Earth-orbit, 600 km), MEO (Medium-Earth-Orbit, 20,000 km), and GEO (Geostationary-Orbit, 36,000 km) that are located between Alice and Bob, such that the link lengths and elevation angles be the same at both sides.
As the link attenuation rises, the number of photons that can make it across falls, so the closer the EPS source is to the ground, the higher rate is expected.

\begin{figure}[ht]
    \centering
\begin{subfigure}{0.4\textwidth}
\centering
\includegraphics[width=0.9\textwidth]{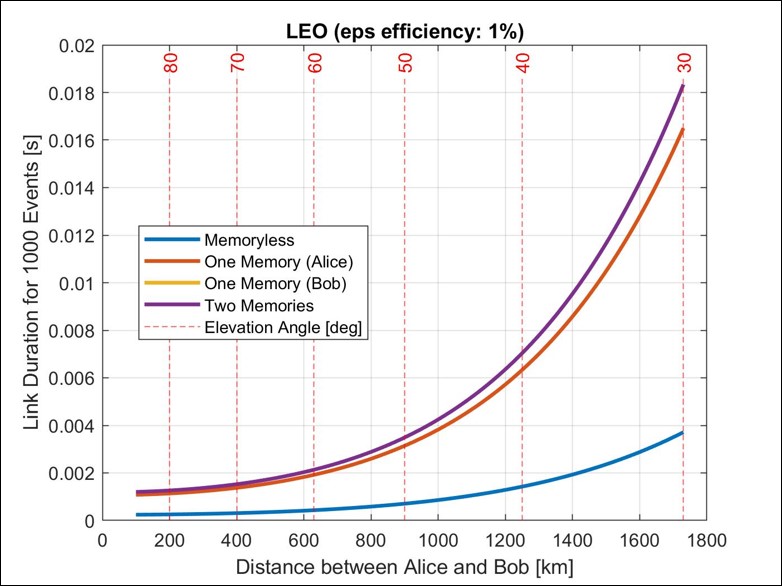} 
\caption{}
\end{subfigure}
\hspace{1cm}
\begin{subfigure}{0.4\textwidth}
\centering
\includegraphics[width=0.9\textwidth]{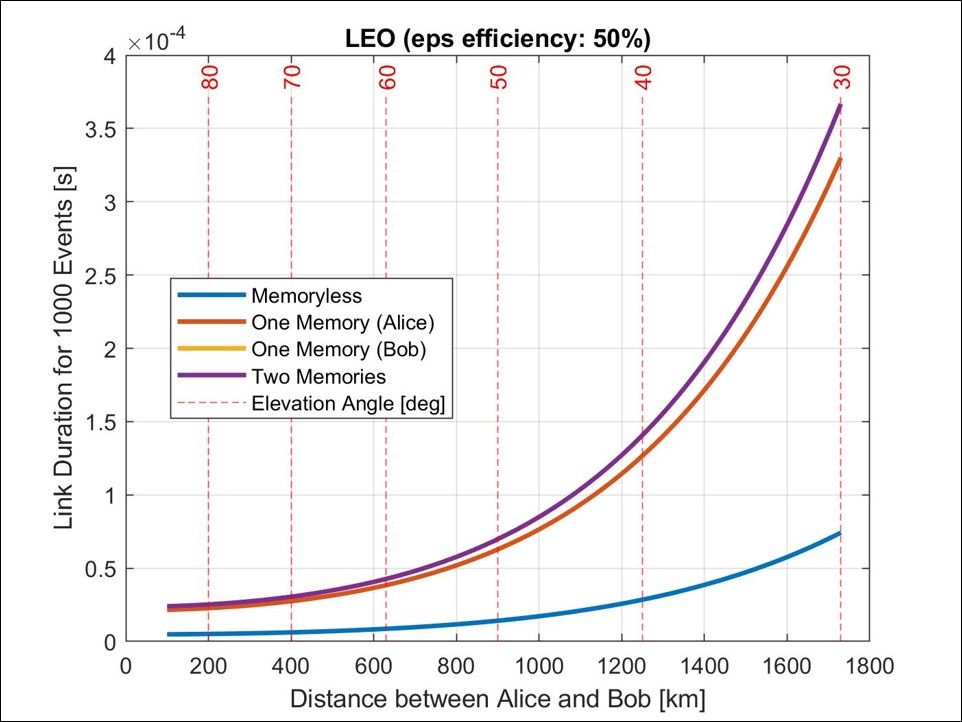} 
\caption{}
\end{subfigure}
\begin{subfigure}{0.4\textwidth}
\centering\includegraphics[width=0.9\textwidth]{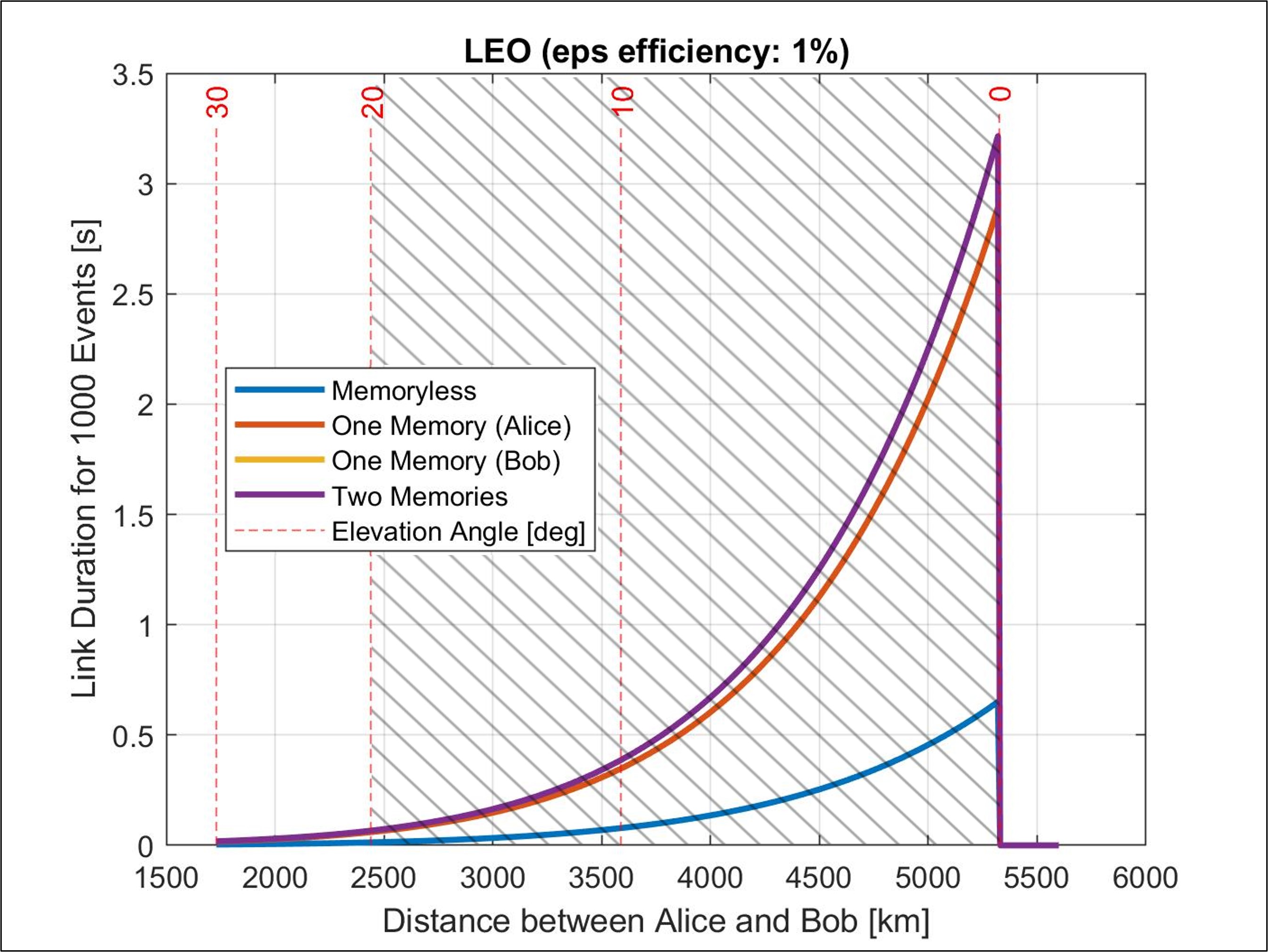} 
\caption{}
\end{subfigure}
\hspace{1cm}
\begin{subfigure}{0.4\textwidth}
\centering
\includegraphics[width=0.9\textwidth]{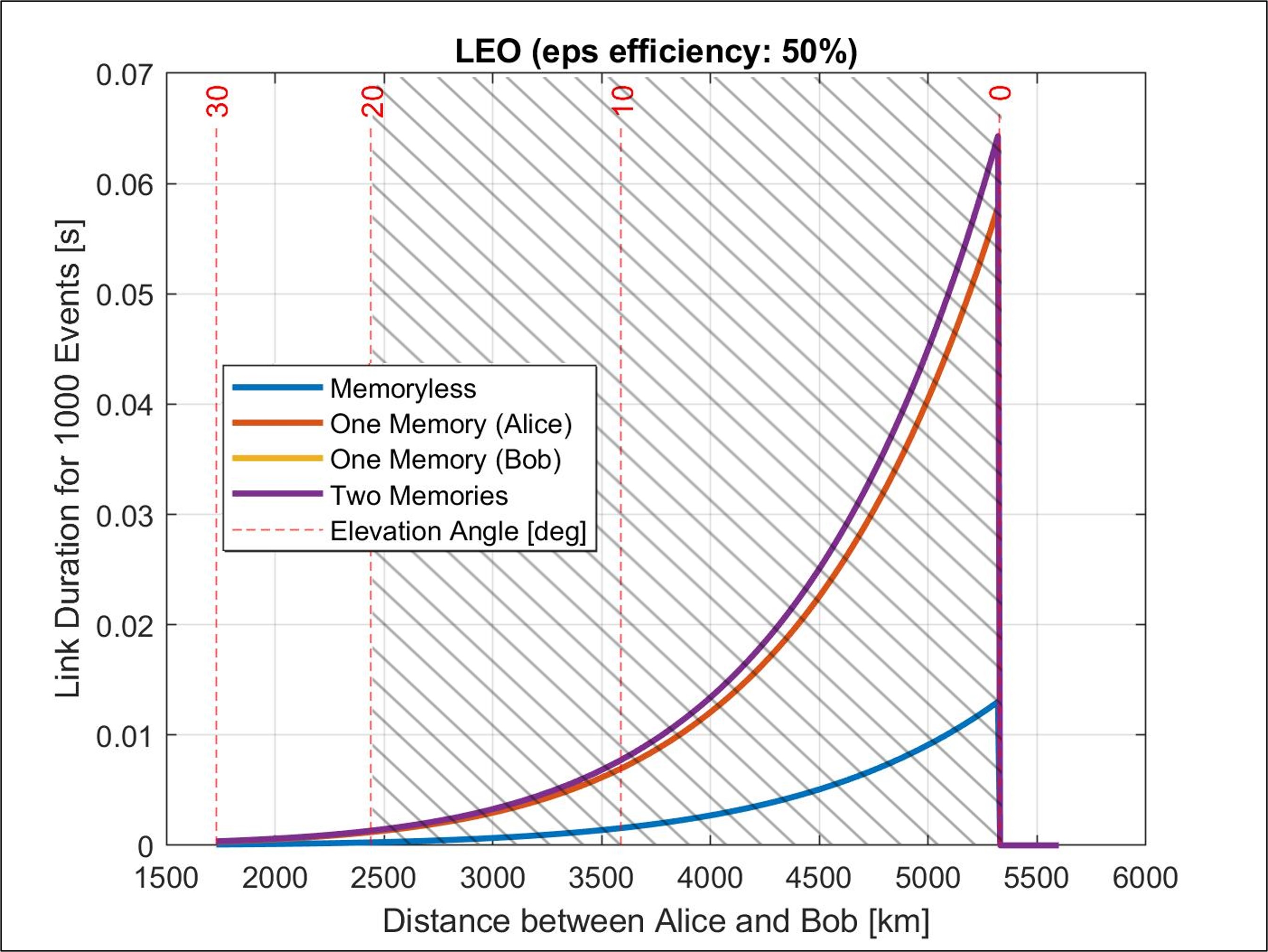} 
\caption{}
\end{subfigure}
\begin{subfigure}{0.4\textwidth}
\centering
\includegraphics[width=0.9\textwidth]{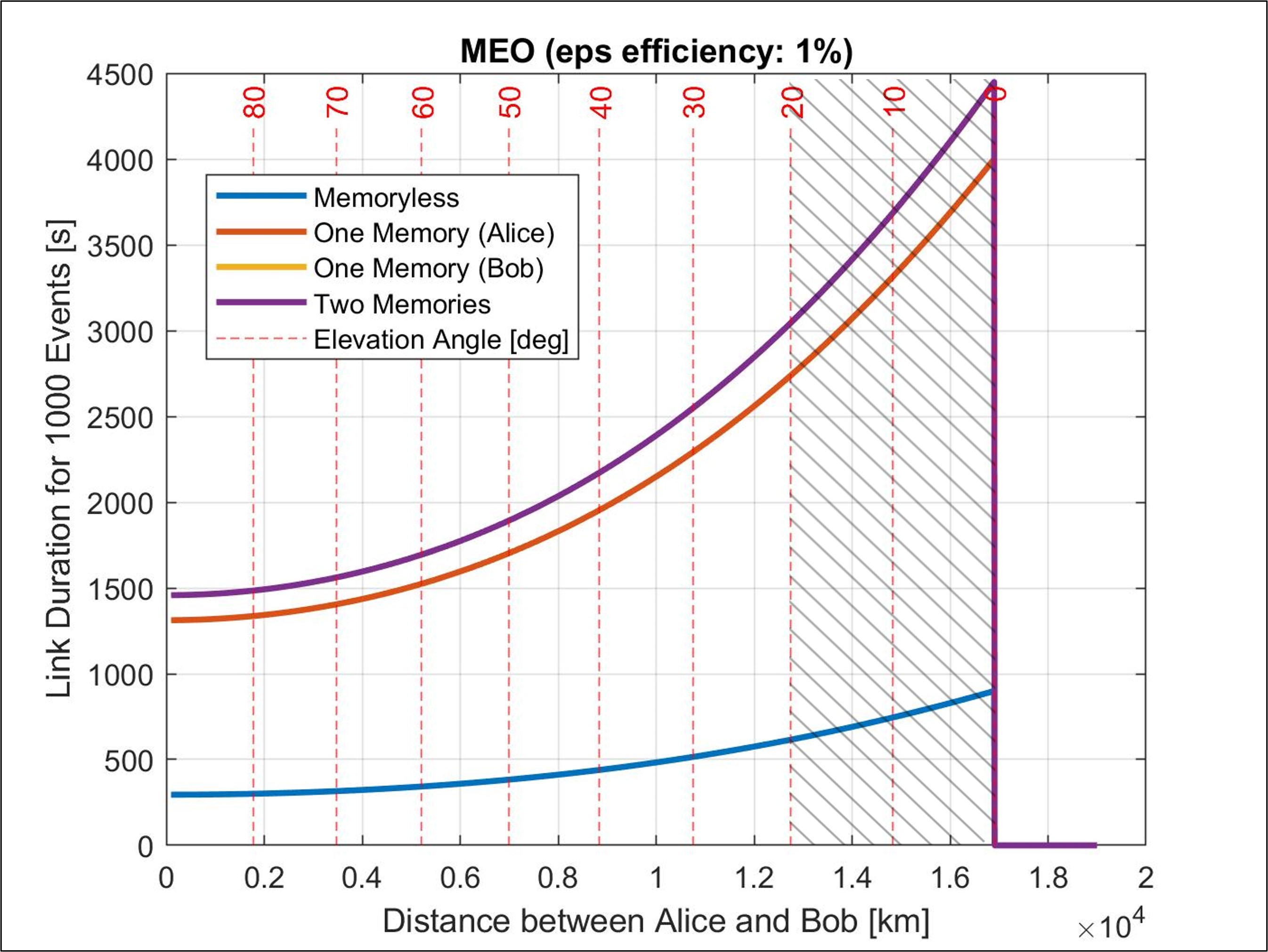} 
\caption{}
\end{subfigure}
\hspace{1cm}
\begin{subfigure}{0.4\textwidth}
\centering
\includegraphics[width=0.9\textwidth]{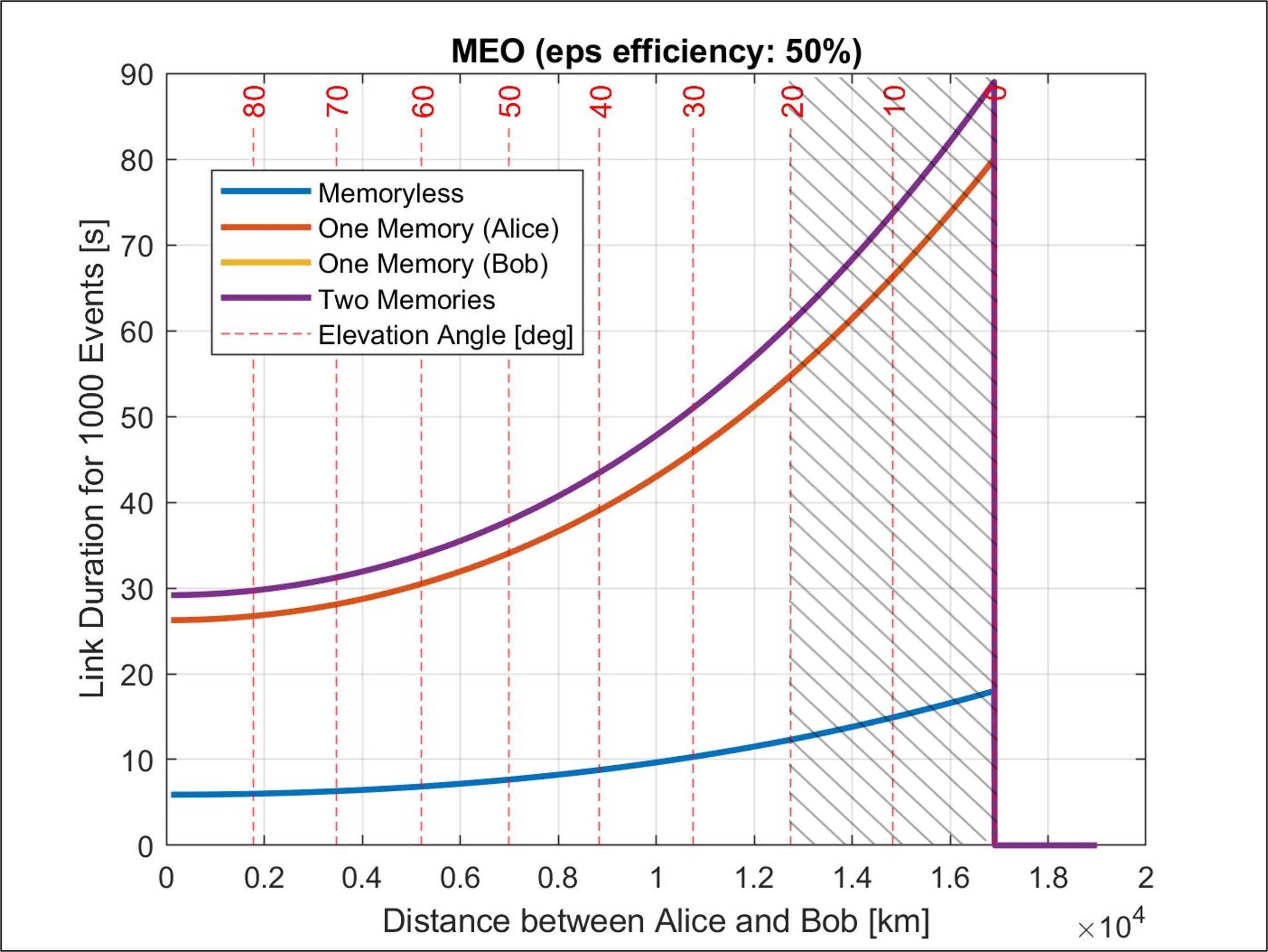} 
\caption{}
\end{subfigure}
\begin{subfigure}{0.4\textwidth}
\centering
\includegraphics[width=0.9\textwidth]{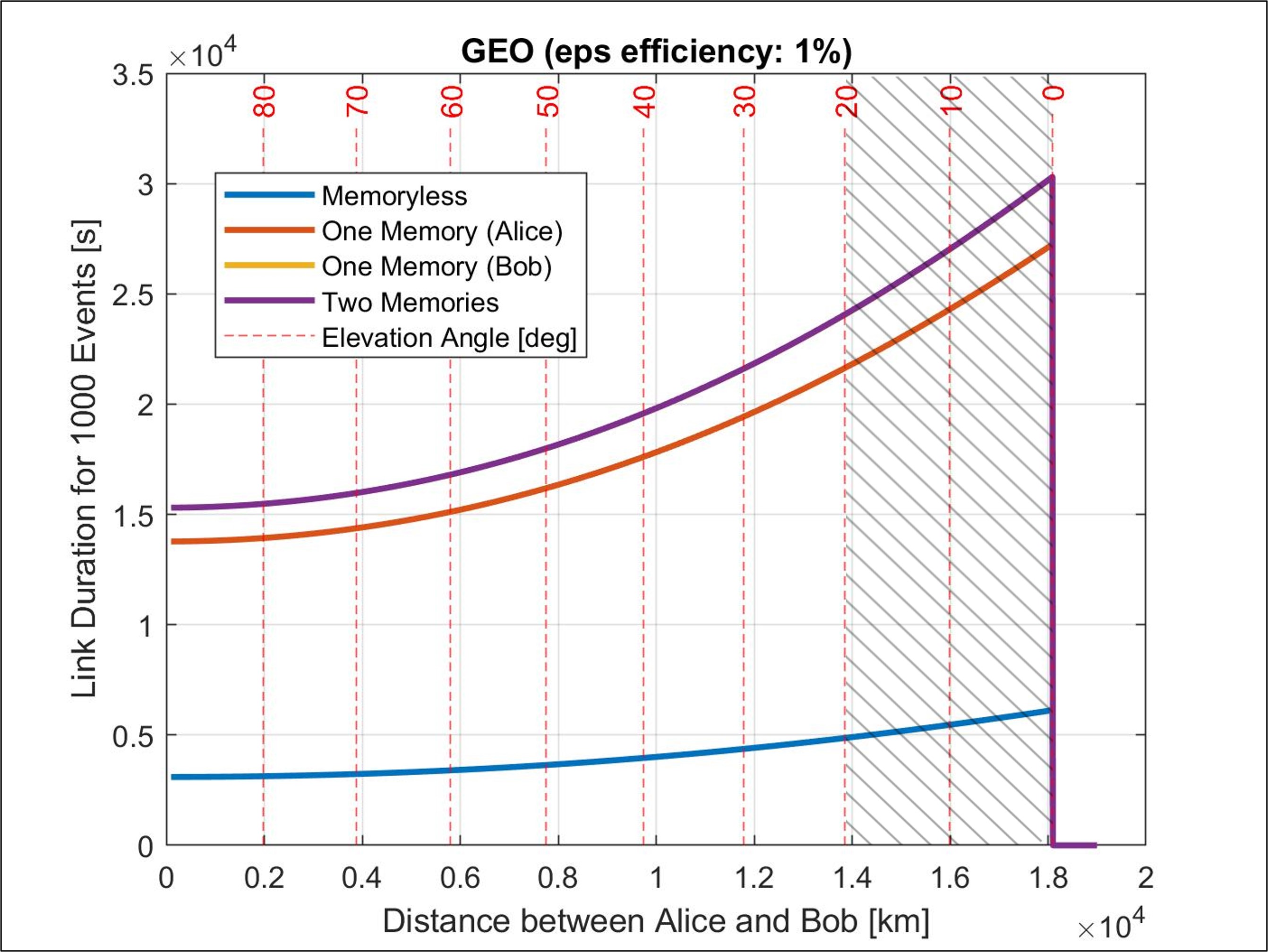} 
\caption{}
\end{subfigure}
\hspace{1cm}
\begin{subfigure}{0.4\textwidth}
\centering
\includegraphics[width=0.9\textwidth]{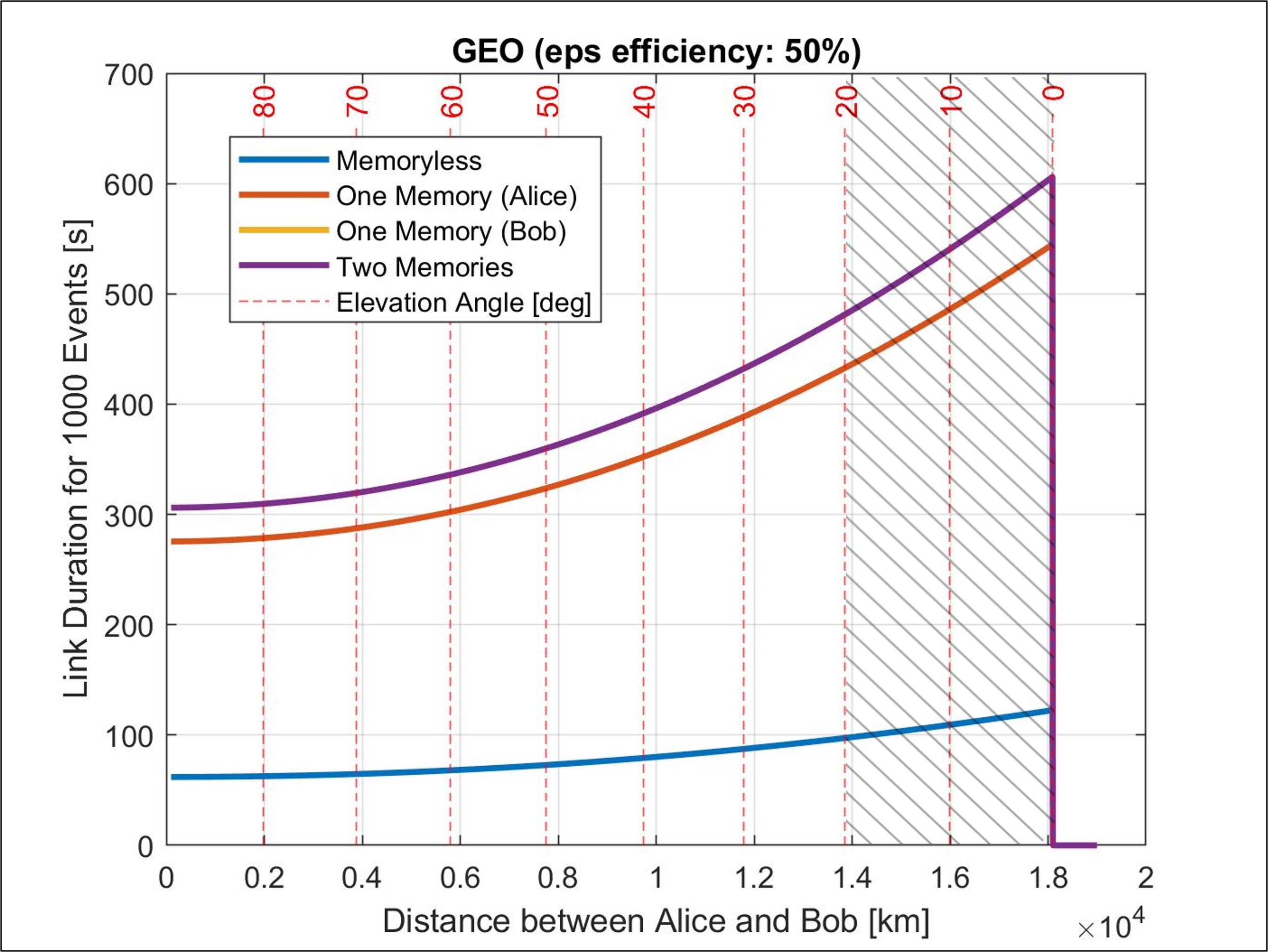} 
\caption{}
\end{subfigure}
    \caption{The plots show the required time for 1000 successful events at Alice and Bob as the distance between them increases (considering the curvature of the Earth), for two different photon source efficiencies, $\eta_{eps}=1\%$ on the left and $\eta_{eps}=50\%$ on the right. LEO plots are split into $80^\circ$ to $30^\circ$ and $30^\circ$ to $0^\circ$ elevation angles to have a better visualisation of the results. The shaded area indicates that establishing a link might not be practical, due to the low elevation angle. In all plots \textit{One Memory (Bob)} is overlapped with \textit{Two Memories}. $\lambda =785$nm; $D_t = 40$cm; $D_r = 200$cm; Table~\ref{tab:Teleportation_assumptions}.} 
    \label{fig:Teleportation_time}
\end{figure}

Note that, the distance between Alice and Bob is limited by the curvature of the Earth. For instance, If the photon source is orbiting the earth at 600 km (LEO) Alice and Bob links will be disrupted by the earth surface, if they are more than 5000 km apart. In other words, if the distance goes beyond this limit, the elevation angle at both sides must have negative values which is not physically possible. Figures~\ref{fig:Teleportation_Rates} and \ref{fig:Teleportation_time} are plotted from the zenith ($90^\circ$) to the horizon ($0^\circ$) for each orbit. However, below $20^\circ$ might not be practical, mostly because the beam has to travel a longer path in the turbulent atmosphere near the horizon. In 2007, an entanglement-based quantum communication was demonstrated over a 144 km horizontal link where the ground stations were at Roque de los Muchachos (2,392 m above sea level) on the island of La Palma, and Tenerife (2,410 m above sea level) \cite{UrsinNatPhys2007}. The high altitude of the ground stations helped with mitigating the severe atmospheric conditions near the Earth. Nevertheless, there are not many of such locations, especially in east of Canada. Therefore, $0^\circ$ to $20^\circ$ regions are shaded in the plots.

In Figures~\ref{fig:Teleportation_time}, we have estimated the minimum required time for a successful teleportation. Here, detection of 1000 photon pairs is the condition for a successful teleportation, hence the minimum link duration is achieved by dividing this number by the teleportation rate. 

\clearpage

\clearpage
\section{Wavelength selection for entangled photon sources}

\begin{figure}[ht]
\centering\includegraphics[width=0.8\linewidth]{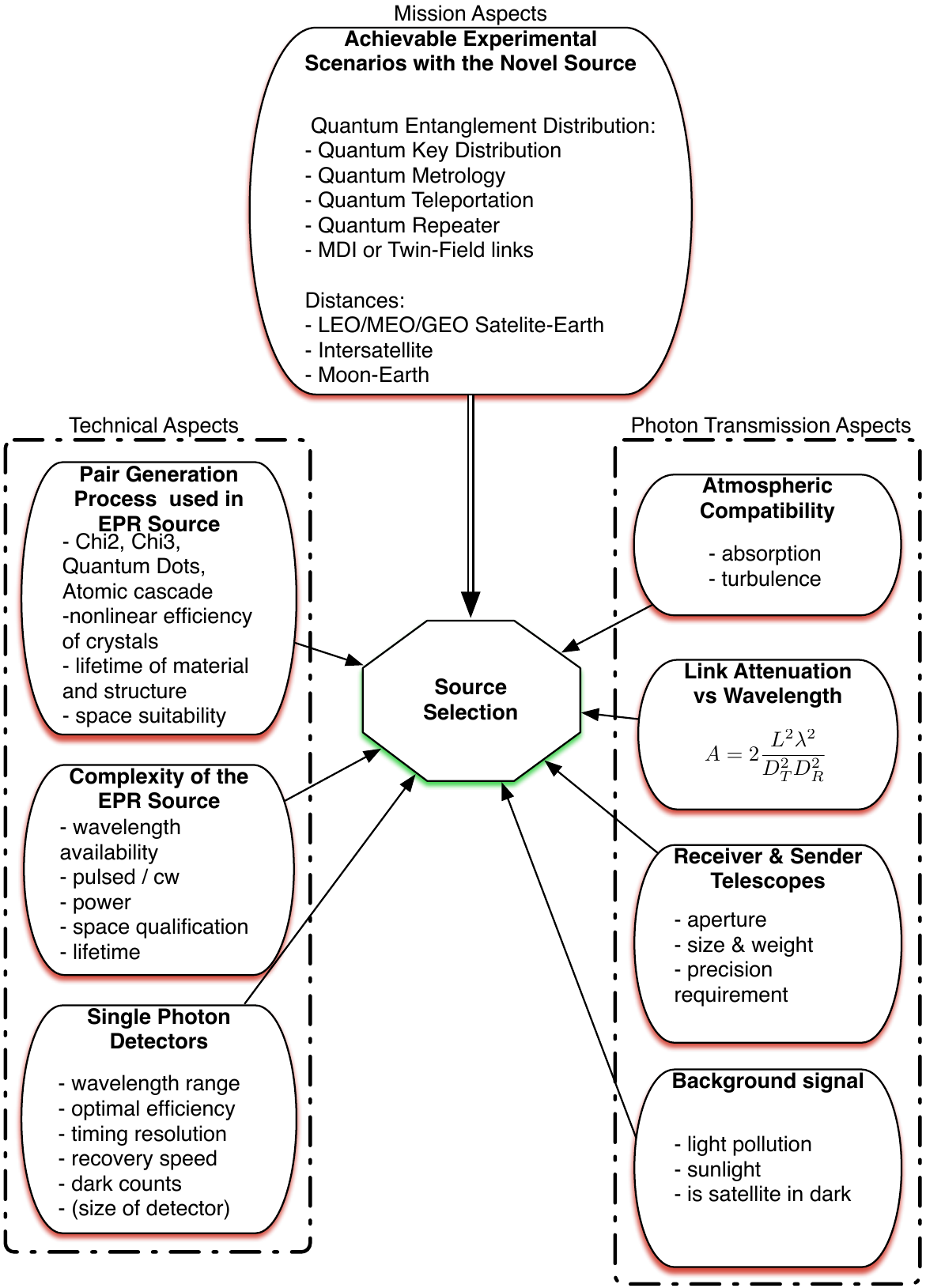}
\caption{Overview of the various considerations for selecting an EPS concept} \label{fig:Sources}
\end{figure}

Considerations and trade points for selecting a mechanism and type of EPS is given in Figure~\ref{fig:Sources}.

\clearpage

\section{Link analysis for different free-space scenarios} \label{Appendix:link_analysis}
A beam transmitted through an aperture and propagating in a medium such as the atmosphere gets attenuated and diffracted before reaching the receiver. Therefore, estimating the link loss in each free-space scenario is critical to argue the practicality of different QKD links. In downlink and intersatellite channels beam diffraction is the main cause of attenuation, whereas, in the uplink, atmospheric turbulence impacts the beam severely. In the following sections, we present a model to find the link attenuation in each free-space scenario, argue the feasibility of them with regard to today's technology and study the trade-off between the link loss and the link availability to suggest the optimal links for quantum communications.

\subsection{Link attenuation}
 If the receiver is at far field of the transmitter ($H\gg \frac{D_T^2}{\lambda}$) and the telescope is diffraction-limited, the link attenuation is defined as,
 \begin{equation}
        A= \frac{H^2(\theta_{T}^2+\theta_{atm}^2)}{D_{R}^2}  \frac{1}{T_{T}(1-L_{P})T_{R}} + A_{atm} +A_{add},
    \label{attenuation}
\end{equation}
where $\theta_{T}=\frac{\lambda}{D_{T}}$ and $\theta_{atm}=\frac{\lambda}{r_{0}}$ are the divergence angles due to the transmitter aperture and the turbulence ($r_0$). Here $\lambda$ is the wavelength of the beam transmitted from a telescope with $D_{T}$ aperture diameter and $T_{T}$ transmission factor and received by a satellite with $D_{R}$ and $T_{R}$; $L_P$ is the pointing loss, and $A_{atm}$ is the atmosphere transmittance which varies with wavelength. Figure~\ref{fig:Atmospheric_Transmittance} shows an average transmittance of 0.8 at 785nm and 0.9 at 1550nm, which adds $\sim$1dB (at 785nm) and $\sim$0.5dB (at 1550nm) to the total loss.
$A_{add}$ is the additional loss due to the imperfections of the components or coupling efficiencies which is considered as 6dB in total, for simplicity.
This work uses Eq.~\ref{attenuation} to estimate the link loss in different link configurations (uplink, downlink, intersatellite link) with satellites in different orbits. It is assumed that the satellites pass over the ground station at zenith and all orbits are in the same plane as the ground station. Moreover, all orbits except the HEO (Highly Elliptical Orbit) are circular orbits with similar inclination angles. To model HEO, the properties of Molniya orbits \cite{molniya} have been chosen as an example. Molniya has a perigee of approximately 600km and an apogee of about 40,000km with an eccentricity of 0.74 and a period of 718 min. Since this orbit spends a noticeable portion of its orbit above the northern hemisphere, it can be a good candidate for free-space commutations in countries such as Canada. 

\subsubsection{Intersatellite links}
The link attenuation in intersatellite links is mostly dependent on the wavelength and aperture size of the transmitter and receiver since there is no atmospheric loss. Therefore, $A_{atm}$ and $\theta_{atm}$ are both zero and the attenuation results from the diffraction of the transmitted beam and the portion of it received by the receiver satellite. choosing shorter wavelengths such as 405nm that diffract less and Using larger telescope apertures help with minimizing the link loss. 
Note that, the pointing error is increased to $30\%$ for intersatellite links.

\subsubsection{Downlinks}
In a downlink configuration, the propagated beam only goes through the atmosphere at the end of its path and as a result, it has a higher spatial coherence compared to the up-link. Therefore, the turbulence effect on the wavefront distortion can be neglected, $\theta_{atm}=0$. However, it causes a beam-wander that makes the fibre coupling of the large beam quite challenging. In the case of using multi-mode fibre, the ratio of the focused beam diameter to the fibre core size, including the image jitter, indicates the power loss. If using single-mode fibre, the mode of the coupled light must be taken into consideration. This loss is assumed as a part of $A_{add}$ in this work. 

\subsubsection{Uplinks}
Atmospheric turbulence can be detrimental to an up-link. Therefore it is crucial to study the atmosphere structure. For this purpose, different models are presented such as the H-V model, HAP model, AFGL AMOS Night model, and SLC Day and Night models \cite{HV}. In this work, we are looking at small zenith angles (less than $60^{\circ}$ or $45^{\circ}$ in case of strong ground-level turbulence) and using weak fluctuations theory based on the Rytov approximation. Based on Hufnagel-Valley (H-V) model, the atmosphere structure parameter ($C_{n}^{2}$) of the atmosphere can be defined as \cite{HV}\cite{LaserBeamPropagation}, 
\begin{equation}
        C_{n}^{2}(h)= 0.00594 (\frac{V}{27})^{2} (10^{-5}h)^{10} exp (-\frac{h}{1000}) + 2.7\times10^{-16} exp (-\frac{h}{1500}) + A' exp (-\frac{h}{100}).
        \label{Cn}
\end{equation}
Here, V is RMS wind speed which is normally 21 m/s and $A'=1.7\times10^{-14} m^{-\frac{2}{3}}$ is the nominal value of $C_{n}^{2}$ on the ground. The optical wave experiences loss of spatial coherence and fluctuations of its intensity as it goes through turbulence. The atmospheric coherence diameter $r_{0}$ (also known as Fried's parameter) indicates how long the spatial coherence of the propagating beam can be preserved over a path of length H. 
\begin{equation}
\begin{split}
         r_{0} &= (0.423 \mu_{0} k^2 \sec \zeta)^{-\frac{3}{5}}, \\
        \mu_{0} &= \int_{h_{0}}^{H} C_{n}^{2}(h) dh.
\end{split}
    \label{r0}
\end{equation}
Then, Eq.~\ref{attenuation} is used to estimate the link attenuation. This equation considers both the diffraction from the transmitter aperture ($\theta_{T}=\frac{\lambda}{D_{T}}$) and the diffraction from the imaginary apertures with $r_0$ diameter in the atmosphere ($\theta_{atm}=\frac{\lambda}{r_{0}}$).

\begin{figure}[ht]
    \centering
    \includegraphics[width=\linewidth]{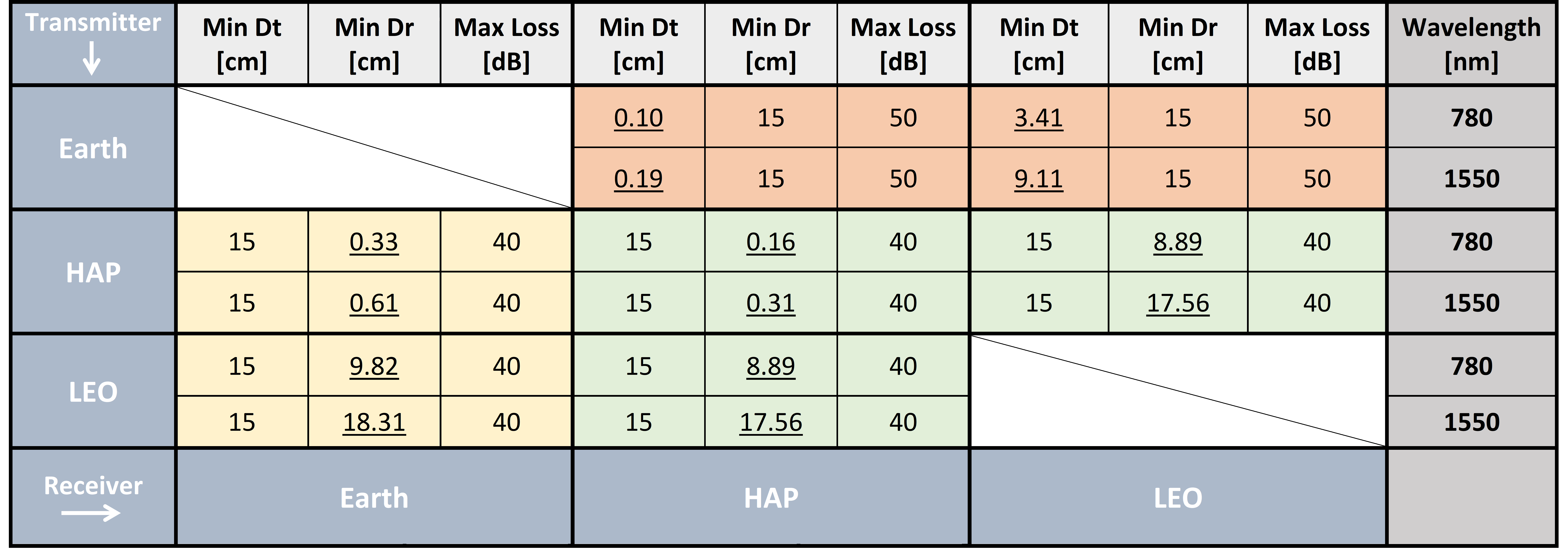}
    \caption{The table indicates the scenarios for which a fixed link attenuation of 40dB in downlink and 50dB  in up-link is achievable with apertures smaller than 25 cm in the space and 200cm on the ground. The underlined numbers are obtained from our link analysis and represent the minimum aperture sizes required to satisfy the mentioned conditions. The results are reported for both 785nm and 1550nm wavelengths. Orange cells: up-links; yellow cells: downlinks; green cells: intersatellite links. The parameters used for this modelling are listed in Table~\ref{tab:Teleportation_assumptions}.}
    \label{fig:Link_Analysis_Static}
\end{figure}

\subsection{Static satellite model}
Recent research on free-space QKD, suggests that quantum links can tolerate up to 50dB loss in an uplink, and 40dB loss in a downlink to successfully perform QKD. In a static satellite model, the satellite is assumed in the zenith of the other station and Eq.~\ref{attenuation} is used to estimate the minimum required apertures size according to the mentioned tolerance for both a 785nm and a 1550nm beam. The results are listed in Figure~\ref{fig:Link_Analysis_Static}. In higher orbits such as MEO and GEO, the beam divergence increases the link loss which can be slightly compensated by expanding the aperture sizes. However, on one hand, telescopes with larger than 1m or 1.5m apertures cannot operate fast enough to track LEO satellites; on the other hand, satellite apertures are limited to 25cm with current technology. Therefore, Figure~\ref{fig:Link_Analysis_Static} only includes the links that experience 50dB loss in up-link and 40dB loss in down-link, considering the mentioned constraints.

\subsection{Dynamic satellite model}
\begin{figure} [ht]
\centering
    \includegraphics[width=\linewidth]{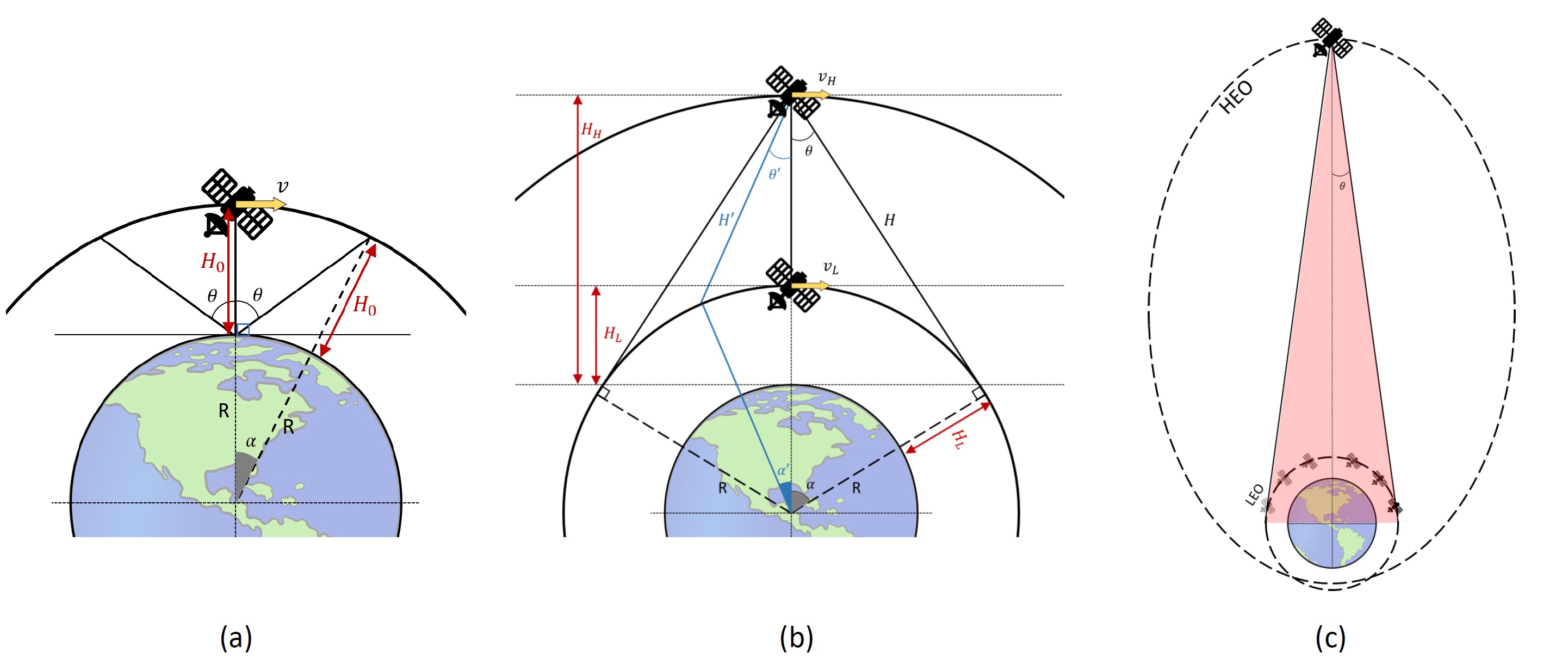}
    \caption{ (a) Uplink/downlink and (b) intersatellite link for circular orbits; (c) HEO-LEO intersatellite link. All orbits are assumed to be in the same plane as the ground station with satellites orbiting in the same direction. $H_0$, orbit altitude; $\theta$, zenith angle; $v$, satellite velocity; R, Earth's radius; $H_L$, $v_L$ lower orbit altitude and velocity; $H_H$, $v_H$ higher orbit altitude and velocity.}
    \label{fig:Slant_Path}
\end{figure}

\begin{figure}[ht]
    \centering
    \includegraphics[width=\linewidth]{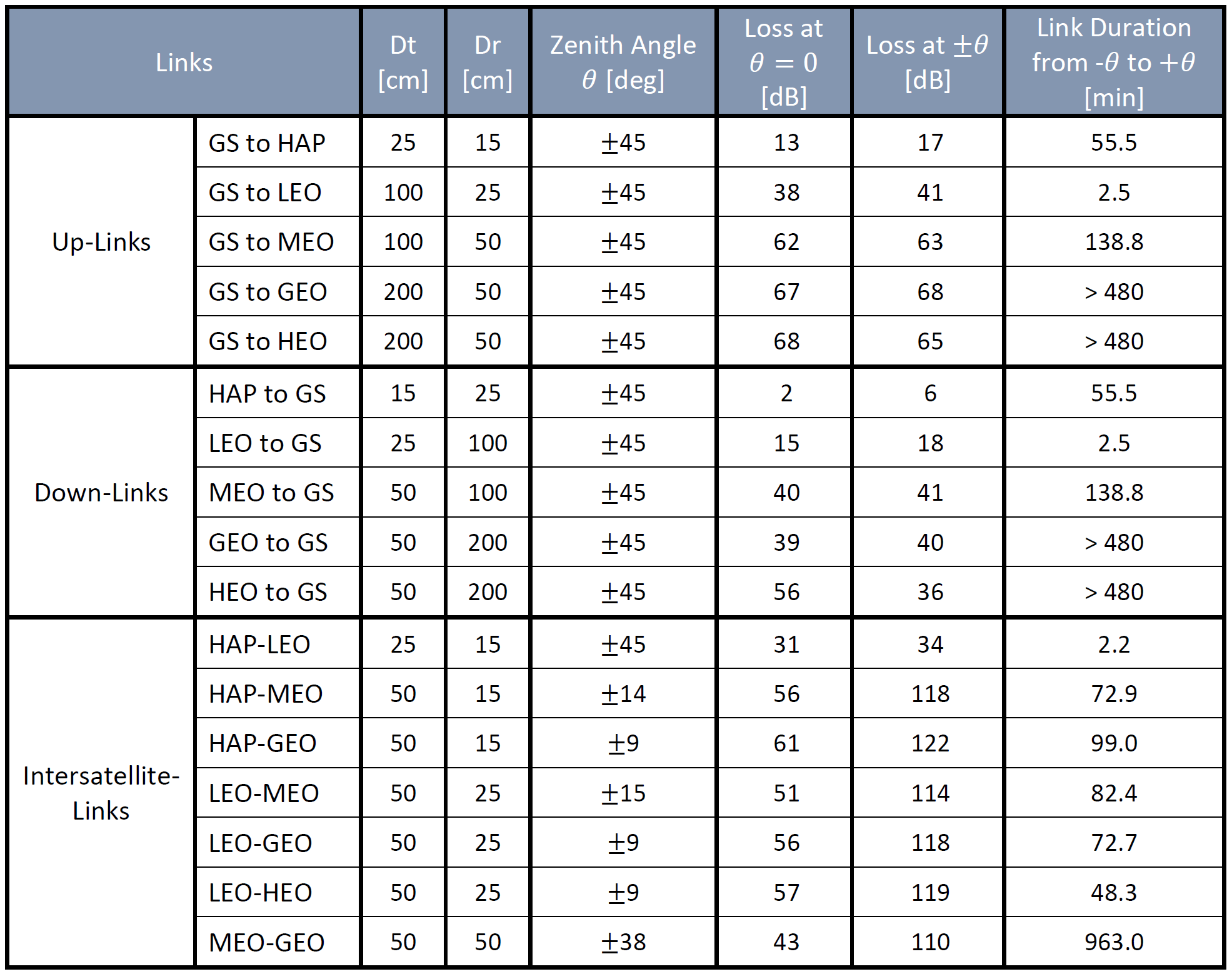}
    \caption{An overview of the free-space scenarios considering their estimated link duration and link attenuation for a certain flyby path; $D_t$ and $D_r$ are the ideal apertures for the transmitter and receiver according to their location (HAP: 15cm, LEO: 25cm, MEO/GEO/HEO: 50cm). The trade-off between the link loss and link duration at different orbits must be taken into consideration in defining an optimal communication channel. Table~\ref{tab:Teleportation_assumptions} has a list of the assumptions used in this analysis. Earth's rotation is ignored in the link duration calculations.}
    \label{fig:Link_Analysis_Dynamic}
\end{figure}

Even though higher obits experience more link loss, their slower speed allows for having the link for a longer time. For instance, the LEO satellite's orbital period is 90min to 120min, whereas GEO has almost the same orbital period as the earth, or HEO satellites e.g. Molniya have a period of 12 hours, that spends about 10 hours over the northern hemisphere. Hence, there is a trade-off between link attenuation and link duration. In Figure~\ref{fig:Link_Analysis_Dynamic}, we considered almost all the possible scenarios and estimated the link duration and the link loss for a satellite flying from $-45^\circ$ to $+45^\circ$ zenith angle over the ground station in uplinks and downlinks. At $\theta=45^\circ$ the orbital velocity of the satellites in circular orbits $v=\sqrt{GM/(H_0+R)}$, and $\alpha$ (as shown in Figure~\ref{fig:Slant_Path}) are calculated to determine the pass-time of the satellite. M and R are the mass and the radius of the Earth and G is the gravitational constant. Earth's rotation is not considered in the following analysis. 
\begin{equation}
    Link Duration = \frac{2\alpha(H_0+R)}{v}.
\end{equation}
In elliptical orbits, the flying time of a satellite from the perigee to any point on the orbit can be calculated as below.\cite{OrbitalMechanics}
\begin{equation}
    t_{\phi} = \sqrt{\frac{a^3}{\mu}}M_e,
\end{equation}
\begin{equation}
    M_e = E-e\sin{E},
\end{equation}
\begin{equation}
E = 2\tan^{-1}\Bigr[\sqrt{\frac{1-e}{1+e}}\tan(\phi/2)\Bigr],
\end{equation}
where, $\phi$ is the true anomaly; $E$ is the eccentric anomaly; $M_e$ is the mean anomaly; and e is the eccentricity of the ellipse. Thus, these equations can be used to find the flying time of a satellite from the zenith of the ground station to any point in the orbit. 
As a result, our model shows that a HEO satellite with Molniya orbit properties (except the inclination), can stay within $\pm45^\circ$ of the ground station for more than 8 hours. This link duration can be an asset to free-space communication across northern countries such as Canada.

As the beam travels through the atmosphere on a slant path, the atmospheric loss increases. For a vertical path, the transmittance is $e^{-\tau}$, where $\tau$ is the optical depth. For zenith angles below $70^\circ$, this parameter can be approximated as $e^{\frac{-\tau}{cos\theta}}$. Hence the additional loss at each angle can be estimated as $sec\theta \times A_{atm_{vertical}}$.

The same approach is valid for intersatellite links. Considering that both satellites are in the same plane and orbit the Earth in the same direction, their relative speed ($v_H-v_L$) is used to find the passing time. In these cases, the time required for the lower orbit satellite to fly $45^\circ$ away from the higher orbit satellite is used to compare the link duration in each scenario. Note that, when the orbits are far apart, the maximum angle between the flying satellites is limited. In other words, the link line tangent to the lower orbit corresponds to the maximum $\theta$ (Figure~\ref{fig:Slant_Path}~(b)), which can be less than $45^\circ$. Therefore, in Figure~\ref{fig:Link_Analysis_Dynamic} the maximum angle is reported as the satellite coverage if $\pm45^\circ$ is not possible.

In HEO-LEO link it is assumed that the HEO satellite is far enough that the link lasts for at least half of the LEO period (Figure~\ref{fig:Slant_Path}~(c)). With this assumption, the new position of the HEO satellite is found to study the change in the link loss. It is worth noting that, in this scenario, a quarter of the LEO satellite's revolution from zenith, equals $2.6^\circ$ displacement of the HEO satellite. Therefore, it is safe to assume that the HEO satellite is constant relative to LEO for 48.3 minutes (half of the orbital period of LEO at 600km).

\begin{table}[]
\centering
\resizebox{\textwidth}{!}{%
\begin{tabular}{|c|c|c|}
\hline
$\lambda$   & wavelength                    & 785 nm, 1550 nm   \\ \hline
$R_s$       & eps repetition rate           & 1 GHz    \\ \hline
$\eta_{sps}$  & sps efficiency                & 75\%      \\ \hline
$\eta_d$    & detector efficiency           & 90\%      \\ \hline
$\eta_{st}\eta_r$    & memory efficiency             & 50\%      \\ \hline
$\eta_{QND}$  & QND efficiency                & 90\%      \\ \hline
$\eta_{dif}$ & $|n_i-n_{i+1}|$                      & 1         \\ \hline
$T_1$       & memory lifetime               & 100 ms  \\ \hline
$T_T, T_R$ & \begin{tabular}[c]{@{}c@{}}receiver and transmitter \\ transmittance\end{tabular} & 80\% \\ \hline
$L_p$       & pointing erorr                & 20\%      \\ \hline
\multirow{3}{*}{\begin{tabular}[c]{@{}c@{}}Additional\\ Loss\end{tabular}} &
  optical loss &
  6 dB \\ \cline{2-3} 
 &
  \begin{tabular}[c]{@{}c@{}}atmospheric absorption ($A_{atm}$) \\ (vertical path)\end{tabular} & 
  \begin{tabular}[c]{@{}c@{}}785 nm : 1 dB\\ 1550 nm : 0.5 dB\end{tabular} \\ \hline
$r_0$  & Fried's parameter      & 7.5 cm    \\ \hline
$D_t$       & transmitter aperture diameter & Variable     \\ \hline
$D_r$       & receiver aperture diameter   & Variable    \\ \hline
$H_{HAP}$   & High-Altitude-Platform       & 20-30 km    \\ \hline
$H_{LEO}$   & Low-Earth-Orbit       & 600 km    \\ \hline
$H_{MEO}$   & Medium-Earth-Orbit    & 20,000 km \\ \hline
$H_{GEO}$   & Geostationary-Orbit   & 36,000 km \\ \hline
$H_{HEO}$   & Highly-Elliptical Orbit   & 600-40,000 km \\ \hline
\end{tabular}%
}
\caption{List of the parameters used to model the free-space links.}

\label{tab:Teleportation_assumptions}
\end{table}

\end{appendix}

\newpage
\bibliography{main}{}
\bibliographystyle{apsrev4-1}

\end{document}